\documentclass[]{aa}  
\usepackage{amsmath}
\usepackage{amssymb}
\usepackage{natbib}
\usepackage{graphicx}
\usepackage{subcaption}
\usepackage{txfonts}
\usepackage[english]{babel}
\usepackage{hyperref}
\usepackage{units}
\usepackage{footnote}
\usepackage{threeparttable}
\usepackage{longtable}
\usepackage{xcolor}
\usepackage{morefloats}
\begin{document}
\title{The ARCiS framework for Exoplanet Atmospheres}
\subtitle{Modelling Philosophy and Retrieval}

\author{
	Michiel Min\inst{1}
		\and
	Chris W. Ormel \inst{2}
		\and
	Katy Chubb \inst{1}
		\and
	Christiane Helling \inst{1,3,4}
		\and
	Yui Kawashima \inst{1}
}

\offprints{M. Min, \email{M.Min@sron.nl}}

\institute{
\inst{1} SRON Netherlands Institute for Space Research, Sorbonnelaan 2, 3584 CA Utrecht, The Netherlands\\
\inst{2} Department of Astronomy, Tsinghua University, Beijing 100084, China\\
\inst{3} Centre for Exoplanet Science, University of St Andrews, Nort Haugh, St Andrews, KY169SS, UK \\
\inst{4} SUPA, School of Physics \& Astronomy, University of St Andrews, St Andrews, KY16 9SS, UK
}

   \date{Last edit: \today}

  \abstract
   {}
   {ARCiS, a novel code for the analysis of exoplanet transmission and emission spectra is presented. The aim of the modelling framework is to provide a tool able to link observations to physical models of exoplanet atmospheres.}
   {The modelling philosophy chosen in this paper is to use physical and chemical models to constrain certain parameters while keeping free the parts where our physical understanding is still more limited. This approach, in between full physical modelling and full parameterisation, allows us to use the processes we understand well and parameterise those less understood. A Bayesian retrieval framework is implemented and applied to the transit spectra of a set of 10 hot Jupiters. The code contains chemistry and cloud formation and has the option for self consistent temperature structure computations.}
   {The code presented is fast and flexible enough to be used for retrieval and for target list simulations for e.g. JWST or the ESA Ariel missions. We present results for the retrieval of elemental abundance ratios using the physical retrieval framework and compare this to results obtained using a parameterised retrieval setup.}
   {We conclude that for most of the targets considered the current dataset is not constraining enough to reliably pin down the elemental abundance ratios. We find no significant correlations between different physical parameters. We confirm that planets in our sample with a strong slope in the optical transmission spectrum are the planets where we find cloud formation to be most active. Finally, we conclude that with ARCiS we have a computationally efficient tool to analyse exoplanet observations in the context of physical and chemical models.}


   \maketitle
%

\section{Introduction}

The study of exoplanet atmospheres is of great interest to put the rich variety of planets we find into the context of their formation \citep[e.g.][]{2011ApJ...743L..16O}. The composition of the atmosphere of an exoplanet contains important information on where the atmosphere was accreted, how it evolved and how it was polluted with solid materials. This information on the dynamics of the planet forming environment will help us to put the Solar System into context \citep[see also][]{2014Life....4..142H}.

Over the past decade the field of exoplanet research has developed from finding and first order characterising planets into a much more quantitative analysis of the characteristics of exoplanet atmospheres. The workhorse for this effort has been for a large fraction HST WFC3, which provides a great window on the presence of water vapour \citep[e.g.][for an example on the current state of the art]{2018AJ....155..156T}. Future instrumentation, most notably the instruments on board the James Webb Space Telescope (JWST), will provide us with a much more in depth view of the content of exoplanet atmospheres. The wavelength coverage of JWST extend into the mid-infrared, allowing us to probe solid state features of proposed cloud materials \citep[see e.g.][]{2015A&A...573A.122W, 2017A&A...600A..10M}.

Analysis of the rich data that will be coming our way calls for robust analysis tools. These tools have to transform the observational data and the attached uncertainties into characteristics of the atmosphere and the attached uncertainties. In order to obtain a set of atmospheric parameters from observed exoplanet spectra one has to use so-called retrieval methods \citep[see e.g.][]{2019A&A...627A..67M, 2018MNRAS.481.4698F, 2017ApJ...834...50B, 2015ApJ...802..107W, 2013ApJ...775..137L, 2012ApJ...753..100B, 2009ApJ...707...24M}. The meaning of the term retrieval is often debated in the context of the difference between retrieval and model fitting. Classical retrieval methods for atmospheres come from the analysis of Solar System objects and Earth observations. Here the process is used to obtain directly the likelihood distribution of a (set of) parameter(s) with a minimum number of assumptions on the underlying physical system while employing relatively well known, tight priors. For example for the Earth this is easily done since we have exquisite datasets that allow for detailed retrieval of molecular abundances. This is different for the case of exoplanets where we have to explore a very large parameter space, rely on little prior knowledge and are dealing with limited observational data. Many studies in the literature on exoplanet retrieval focus on figuring out which level of complexity is required -- in the form of for example clouds \citep[see e.g][]{2017ApJ...834...50B} or 3D structure of the planet \citep{2019A&A...623A.161C} -- to reliably retrieve atmospheric information.

In this paper we present a new modelling framework in the form of the ARCiS code (where ARCiS stands for ARtful modelling Code for exoplanet Science). 
We aim to provide a framework which is both fast and flexible enough to allow robust retrieval methods to be used. Such a retrieval framework that can take into account constraints on physics and chemistry of the atmosphere in a non-parameterised manner is needed to 1) gain insight into the physical and chemical processes like cloud formation, and 2) make sure that our results are not biased by oversimplified assumptions on the structure of the atmosphere.
The idea behind this is not new in the literature and is basically a generalisation with additional physical processes of what one could refer to as the \emph{chemically consistent retrieval} \citep[see e.g.][]{2017ApJ...847L...3O}.
In \citet{2019A&A...622A.121O} we have already introduced a physical transport model for the vapour and condensate species. 
Here we add chemical processes to further constrain the molecular abundances \citep{2018A&A...614A...1W} and different cloud species to make the code applicable to a wider variety of environments.

In Section\ref{sec:modelling} we describe the modelling philosophy and setup as well as the retrieval part. We apply the retrieval part of the code to the transit spectra of ten exoplanets in Section~\ref{sec:retrieval results}. We discuss the results and conclude in Section~\ref{sec:conclusions}.

\section{Modelling setup}
\label{sec:modelling}
\begin{figure*}[!tp]
\centerline{\resizebox{0.8\hsize}{!}{\includegraphics{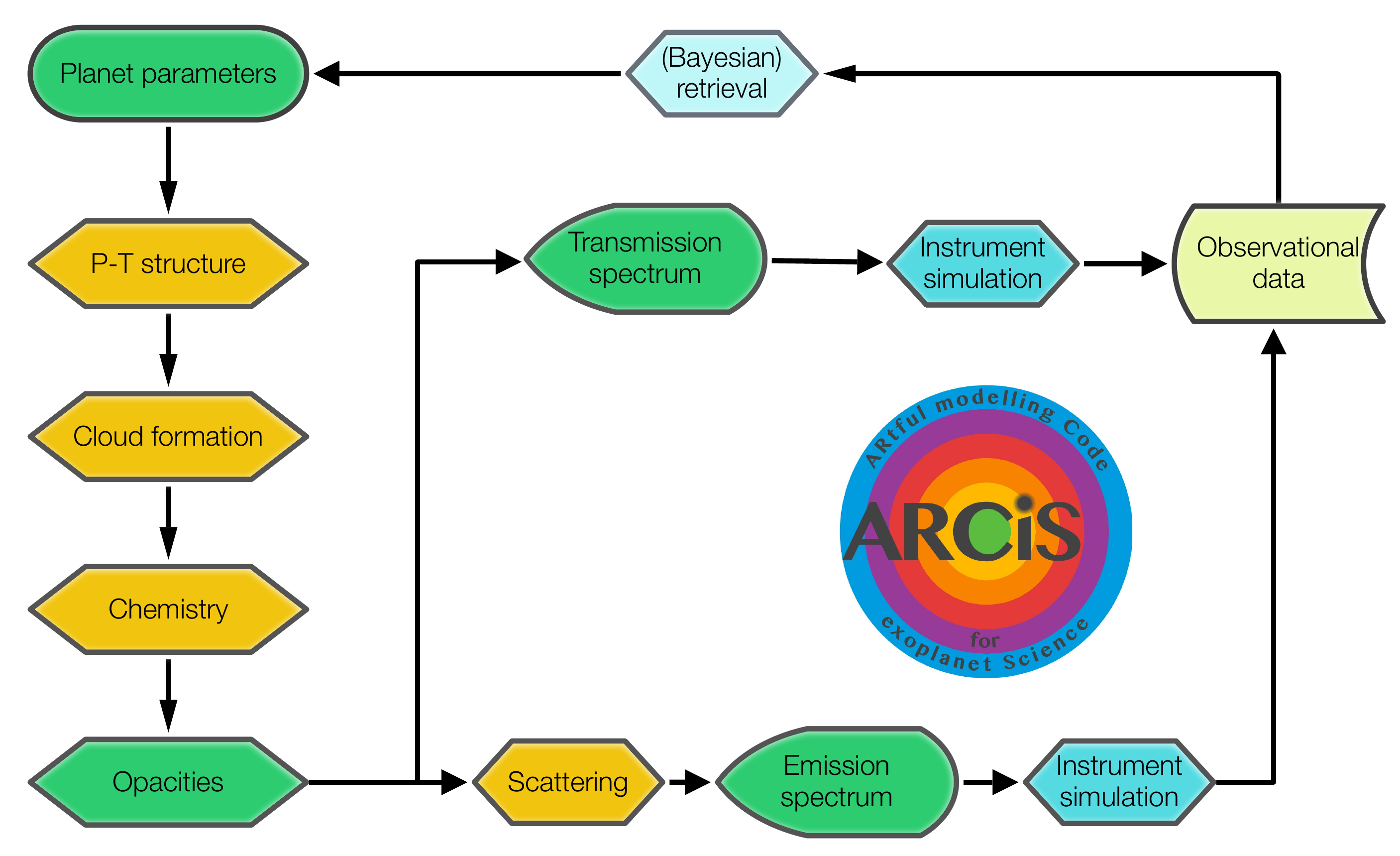}}}
\caption{Schematic layout of the modelling chain.}
\label{fig:scheme}
\end{figure*}

The modelling in the ARCiS (ARtful modelling Code for exoplanet Science) framework aims at using physical parameters from planet formation and evolution to predict observable quantities. Furthermore, we aim to develop a framework with a mild computational load so that robust retrieval algorithms can be employed to link these parameters to observations. The modelling chain is schematically outlined in Fig.~\ref{fig:scheme} and described briefly below. We mainly refer the reader to the relevant papers discussing the various modules in more detail. We note, however, that the concepts for most of the modules are under continuous development. Numerical implementations of some of the modules can be found on \url{http://www.exoclouds.com}.

\subsection{Atomic abundances}
\label{sec:elements}

Setting up an exoplanet atmosphere starts from the abundances of the atoms that form the molecules and condensed species in the atmosphere. The abundances of these elements can, in principle, be predicted from the theory of planet formation and evolution \citep[e.g.][]{2011ApJ...743L..16O}. In many studies the oxygen and/or carbon abundances are scaled with respect to solar values to obtain different chemical conditions. The C/O ratio is one of the most important parameters determining the gas phase chemistry. However, in this paper we also wish to include cloud formation. For this, also the ratios of other elements are important. 

In this paper we consider the ratios C/O, Si/O and N/O. We start from the Solar abundances as derived by \citet{2009ARA&A..47..481A}. First, for C/O ratios above Solar we adjust the oxygen abundance (while keeping C at Solar abundances), for ratios below Solar we adjust the carbon abundance (while keeping O at Solar abundances). Next we adjust the nitrogen abundance to match the N/O abundance. All other elements are scaled with the same value such that we match the Si/O abundance.

After the relative abundances of all elements heavier than He are computed using the above scheme, we scale the H and He abundances with the metallicity parameter, $\left[Z\right]$. In our definition of the metallicity this is the added number density of all elements heavier than He divided by the added number density of H and He.
\begin{equation}
\left[Z\right]=\log_{10}\left[\left(\frac{\Sigma_{i\neq H,He} X_i}{\Sigma_{i=H,He} X_i}\right) \cdot \left(\frac{\Sigma_{i=H,He} X_i}{\Sigma_{i\neq H,He} X_i} \right)_{\rm{Solar}}\right],
\end{equation}
with $X_i$ the number density of element $i$.

\subsection{Cloud formation}

When clouds condense, the element abundances available for the molecules changes. Thus, the first step in our modelling procedure is to compute cloud formation. In many studies, cloud formation is either highly parameterised \citep[see e.g.][]{2017ApJ...834...50B} or treated in complete and detailed coupling to the atmospheric chemistry \citep[see e.g.][]{2008A&A...485..547H, 2017A&A...608A..70J, 2019arXiv190608127H}. Here we employ a model of intermediate complexity that still catches the essential physics of condensation and transport of the cloud particles. The method is described in \citet{2019A&A...622A.121O}. Here, we expand the method for multiple cloud species for completeness. This is especially important for higher values of the C/O ratio (where it becomes important to include carbon based cloud species) or high temperature planets (where it becomes important to include high temperature condensates). For the saturation vapour pressure of the species we follow \citet{2016A&A...596A..32V},
\begin{equation}
P_V=\exp{(-\mathcal{A}/T+\mathcal{B})} \,\mathrm{dyn}\,\mathrm{cm}^{-2}.
\end{equation}
Here $\mathcal{A}$ and $\mathcal{B}$ are constants given in Table~\ref{tab:clouddata}, $T$ is the temperature in K, and $P_V$ is the saturation vapour pressure in dyn\,cm$^{-2}$. 
At low temperatures, below 680\,K, metallic Fe reacts with S to form FeS[s] \citep{1994ApJ...421..615P}. Therefore, we convert part of the metallic iron below 680\,K into FeS[s] such that all S condenses at these temperatures.

When computing which elements are available for the cloud formation we remove gas phase CO. We do this by removing the maximum amount of CO that can form. This implies that when C/O$<1$, SiC[s] and C[s] will not condense while for C/O$>1$ all the oxides will not condense. 
The basic assumption behind this is that CO is a very stable molecule that locks away carbon and oxygen for the formation of solids \citep[see also][for a detailed discussion on cloud formation in carbon and oxygen rich environments]{2017A&A...603A.123H}.

\begin{table*}[!tb]
\begin{center}
\begin{threeparttable}[b]
\caption{References for the laboratory data for the cloud species. For the complete list of references for the $\mathcal{A}$ and $\mathcal{B}$ values, see \citet{2016A&A...596A..32V}.}
\begin{tabular}{l|ll|l}
\hline
\hline
Solid state species	&	$\mathcal{A}$	&	$\mathcal{B}$		&	refractive index ref.\\
\hline
(NaAl)$_x$Mg$_y$SiO$_3$[s]\tnote{(a)}	&	68908		&	38.1				&	{\cite{1996ApJS..105..401S, 1998A&A...339..904J}}	\\
SiO$_2$[s]							&	69444		&	33.1				&	{\cite{1976JNCS...20..153G}}	\\
Fe[s] \& FeS[s]\tnote{(b)}					&	48354		&	29.2				&	{\cite{1996A&A...311..291H}}	\\
Al$_2$O$_3$[s]						&	77365		&	39.3				&	{\cite{1995Icar..114..203K}}	\\
C[s]								&	93646		&	36.7				&	{\cite{1996MNRAS.282.1321Z}}	\\
SiC[s]								&	78462		&	37.8				&	{\cite{1993ApJ...402..441L}}	\\
TiO$_2$[s]\tnote{(c)}					&	\multicolumn{2}{c|}{same as Al$_2$O$_3$}	&	{\cite{2003ApJS..149..437P}}	\\
VO[s]\tnote{(c)}						&	\multicolumn{2}{c|}{same as Al$_2$O$_3$}	&	opacities not computed\tnote{d}\\
\hline
\hline
\end{tabular}
   \begin{tablenotes}
	\item[(a)] Values for $x$ and $y$ are adjusted to condense all of the Mg and Na. Vapour pressure parameters and opacities are taken to be those for MgSiO$_3$[s].
	\item[(b)] FeS[s] is only considered for temperatures below 680\,K.
	\item[(c)] For TiO$_2$[s] and VO[s] we adopt the values for Al$_2$O$_3$[s].
	\item[(d)] For VO[s] no optical constants were available, so VO[s] is ignored in the opacity calculation.
   \end{tablenotes}
\label{tab:clouddata}
\end{threeparttable}
\end{center}
\end{table*}

\begin{table}[!tb]
\caption{References for the line-lists used in the gas opacity computations.}
\begin{center}
\begin{tabular}{l|l}
\hline
\hline
Molecule							&	reference\\
\hline
H$_2$		& HITRAN;~\cite{HITRAN_2016} \\
CO, CO$_2$	& HITEMP;~\cite{HITEMP}	\\
H$_2$O		& ExoMol;~\cite{ExoMol_H2O}\\
C$_2$H$_2$	& ExoMol;~\cite{ExoMol_C2H2}\\
CH$_4$		& ExoMol;~\cite{ExoMol_CH4}\\
NH$_3$		& ExoMol;~\cite{jt771}\\
PH$_3$		& ExoMol;~\cite{ExoMol_PH3}\\
SiO			& ExoMol;~\cite{ExoMol_SiO}\\
SO$_2$		& ExoMol;~\cite{ExoMol_SO2}\\
HCN			& ExoMol;~\cite{ExoMol_HCN}\\
H$_2$S		& ExoMol;~\cite{ExoMol_H2S} \\
Na, K 		& NIST;~\cite{NISTWebsite}\\
\hline
\hline
\end{tabular}
\end{center}
\label{tab:linelists}
\end{table}

\subsection{Chemistry}

For this study we consider equilibrium chemistry for the gas phase. The element abundances are computed after the cloud formation process. Thus, element depletion or enrichment by cloud formation is taken into account. Some of the elements will have been removed by the cloud particles and rained down to deeper layers. This will create elemental abundances that vary throughout the atmosphere \citep[similar to what was shown in e.g.][]{2008A&A...485..547H, 2019A&A...626A.133H}. We compute the molecular abundances from this elemental abundance structure using the equilibrium chemistry code GGchem by \citet{2018A&A...614A...1W}.

In the upper layers of the atmosphere, the chemical timescales may become longer than the mixing timescale. This is where so-called quenching happens. In the uppermost layers, the chemistry may also be affected by UV radiation from the central star \citep[for both effects see e.g.][]{2012A&A...546A..43V, 2018ApJ...853....7K}. Both these effects are currently not taken into account in the model but will be the subject of future studies.

\subsection{Molecular opacities}

Once the molecular abundances are computed, we have to compute the opacities of the molecules. This is done using the available line-lists in combination with a method to compute the line profiles. ARCiS uses correlated-k tables for the radiative transfer computations \citep{1991JGR....96.9027L, 1989JQSRT..42..539G}. Using accurate line-lists is crucial for obtaining the right transmission and emission spectra. The most complete line-lists for molecules at high temperatures are provided by the ExoMol project \citep{ExoMol_gen}. The problem with this is that the large number of lines makes direct computation of the line-profiles numerically cumbersome. A solution for this is given in \cite{2017A&A...607A...9M} using a statistical line-sampling technique. However, here we use the standardised precomputed correlated k-tables (at R~=~$\lambda/\Delta \lambda$~=~1000 for the 0.3~-~50$\mu$m wavelength region) as presented in \citet{Chubb} which have been computed using the ExoCross program \citep{2018A&A...614A.131Y}. The molecules we use in this paper are listed in Table~\ref{tab:linelists} together with the line-list data used. The pressure and temperature broadened profiles for the resonance doublets of Na and K are computed using \cite{16AlSpKi.broad} and \cite{19AlSpLe.broad}.

\subsection{Cloud opacities}

The opacities of the cloud particles depend on their size, shape, and composition. The composition and average size are computed from the cloud formation procedure. For the shape we have to make assumptions. A frequently used assumption is that the particles are homogeneous spheres, so Mie-theory can be applied \citep{Mie}. However, even small deviations from a perfect homogeneous sphere can cause large differences with Mie-theory \citep[see e.g.][]{2003A&A...404...35M}. Therefore, here we employ the DHS method \citep[Distribution of Hollow Spheres][]{2005A&A...432..909M}. This method has been successfully applied to the study of cosmic dust in the ISM, protoplanetary disks, and AGB winds \citep[see e.g][]{2007A&A...462..667M, 2016A&A...586A.103W, 2012A&A...544L..18L}. The spectroscopic effects of using DHS particles or perfect spheres was investigated by \citet{2017A&A...600A..10M} showing that the difference could be observable with future telescopes.

The refractive index of homogeneous particles is available through measurements in the laboratory for various species (see also Table~\ref{tab:clouddata}). The cloud particles we are considering are inhomogeneous, different materials condense on the same particle. For this we can use effective medium theory to construct an effective refractive index from the refractive indices of the various materials \citep{BohrenHuffman}. We use the Bruggeman mixing rule for this \citep[similar to the standard adopted in][]{2016A&A...586A.103W}.

Note that, like in \cite{2019A&A...622A.121O}, we assume that the nuclei on which the clouds condense have negligible mass and also do not contribute to the opacity of the clouds. These nuclei basically form what is usually referred to as haze. We realise that the assumption that the only opacity is caused by the condensation species and not by the hazes/nuclei can be significant in certain corners of parameter space. The influence of this assumption will be investigated in a future study.

\subsection{Computation of temperature structure}

In principle the temperature structure of the atmosphere can be computed using radiative transfer. This is implemented in the ARCiS code. To compute the temperature structure of the atmosphere we use the Variable Eddington Factor method as also explained in \cite{2015ApJ...813...47M}. The radiative transfer is a strong function of the opacities of the atmosphere. Thus we have to iterate cloud formation, chemistry and computation of the opacities with the radiative transfer to converge to a consistent solution. While currently implemented in the code, we will not use this mode to self-consistently compute the temperature in this paper and leave the discussion of this to a future study. Here we use a parameterised pressure-temperature profile.

\subsection{Scattering}

The scattered flux is important when considering emission spectra. Scattering can be important for both the thermal flux from the planet itself as well as for scattering of the flux coming in from the central star. Details on the scattering will be presented in a future paper where we will also consider the emission spectra, here we only briefly mention the methods implemented. There are currently two ways scattering is implemented in ARCiS. The first is Monte Carlo scattering where the scattered flux is computed by emitting a number of photon packages and tracing them trough the atmosphere \citep[see][]{2017A&A...607A..42S}. The photon packages are emitted as thermal emission from the atmosphere itself as well as from the central star. At random positions, determined by the optical depth traveled by the photon package, the packages interact with the local scatterers and change direction. This iterates until a photon package leaves the system and is collected in one of the bins placed around the planet. This method allows to take into account the full anisotropic scattering matrix including polarisation effects. However, the drawback is that it can become slow and noisy and is therefore not suitable for implementation in retrievals. For details on Monte Carlo radiative transfer in exoplanet atmospheres see \citet{2017A&A...607A..42S, 2019MNRAS.487.2082L}. The second method we implemented is isotropic scattering using a direct matrix inversion of the scattered flux. Mathematical details on this method will be presented in a future study. In this paper we only consider transmission spectra. Although scattering can influence the transmission depth in specific cases of e.g. highly forward scattering clouds \citep{2017ApJ...836..236R}, we do not include these scattering effects in the computations presented here.

\subsection{Computation of transmission and emission spectra}

Once the opacity and thermal structure is setup and the contribution of scattered light computed, the transmission and emission spectra can be constructed. For the transmission spectra this is simply done by computing the optical depth through the atmosphere at different impact parameters and at all wavelengths considered. This way we compute how much light is blocked by the planet at each wavelength. This can in turn be converted in an effective radius of the planet at each wavelength. For the emission spectra we have to compute the formal solution of the radiative transfer equation along various rays over the planet. The computation of observables have been benchmarked against petitCODE in \cite{2019A&A...622A.121O}. Also ARCiS was part of the code comparison challenge conducted in the framework of the Ariel mission comparing the forward modelling and retrieval outputs of five different codes \citep{TechNote}. All five codes in this comparison showed excellent agreement on retrieved properties and accompanying uncertainties.

\section{Retrieval methods}
\label{sec:retrieval}

As mentioned in the introduction, there are significant differences between classic retrieval methods designed for Earth observations and retrievals for exoplanet atmospheres. In the retrievals for Earth and Solar System atmospheres we often have high signal to noise ratio datasets and a relatively good understanding of the system. This allows for a retrieval approach using relatively well constrained priors and few assumptions on the physics and chemistry going on.
For exoplanets, here we run into some issues. As the data is not as good and complete as typical for Earth or Solar System planetary science we always have to make some assumptions. With the data currently available, and also with data expected from future instruments, we cannot obtain the full horizontal and vertical distribution of temperature, density and molecular abundances. When applied to transmission spectra the most important commonly used set of assumptions is roughly:
\begin{enumerate}
\item 1D spherical geometry
\item Isothermal temperature structure 
\item Vertical hydrostatic equilibrium; ideal gas law
\item Homogeneous molecular abundances
\item Grey or smoothly varying cloud opacities
\end{enumerate}
It is essential to keep in mind, that all parameters derived by means of a retrieval method are only valid within the assumptions applied. It is therefore that many studies in the literature are focussing on pinpointing which assumptions can be made under which circumstances to still get reliable results \citep[see e.g.][]{2016ApJ...833..120R, 2017MNRAS.469.1979M, 2019ApJ...886...39C, 2019A&A...623A.161C}. Many of these assumptions can be debated, meaning also the outcome of the retrieval in this way is not necessarily the best representation of the actual parameters of the atmosphere and the interpretation in the context of more complex models becomes very important. Also, for future instrumentation it will be necessary to include more physically correct model description into retrieval models, like for example pressure dependent molecular abundances and temperatures. We will refer to methods using the above set of assumptions as the \emph{classical retrieval method}.

\subsection{Model constrained retrieval}

With the ARCiS code we have the ability to perform the classical retrieval. However, the added physics and chemistry also allows for what we will refer to as a \emph{model constrained retrieval method}. In principle any set of model assumptions can be made. In this paper we will compare the classical retrieval to the model constrained retrieval using the following set of assumptions:
\begin{enumerate}
\item 1D spherical geometry
\item Parameterised temperature structure
\item Vertical hydrostatic equilibrium; ideal gas law
\item Molecular abundances from chemical equilibrium
\item Cloud opacities computed from cloud formation
\end{enumerate}
Note that the number of parameters in the model constrained retrieval method is usually a bit lower than for the classical retrieval method since the molecular abundances and cloud properties are computed from only a limited set of parameters. For the temperature structure in the retrieval procedure we do not use the radiative transfer iterations but chose to adopt a parameterised structure following \citet{2010A&A...520A..27G}
\begin{multline}
    \label{eq:Guillot}
    T^4 = \frac{3T_\mathrm{int}^4}{4} \left( \frac{2}{3}+\tau \right) \\ +\frac{3T^4_\mathrm{irr}f_\mathrm{irr}}{4}\left[ \frac{2}{3} +\frac{1}{\gamma\sqrt{3}} +\left( \frac{\gamma}{\sqrt{3}} -\frac{1}{\gamma\sqrt{3}} \right) e^{-\gamma\tau\sqrt{3}} \right]
\end{multline}
with $T_\mathrm{int}$ the internal temperature of the planet. The irradiation from the host star is computed using $T_\mathrm{irr}=T_\star (R_\star/r_p)^2$ where $T_\star$ is the stellar temperature, $R_\star$ the stellar radius and $r_p$ the distance to the host star. Here $\tau$ is the optical depth at IR wavelengths, which we compute using a constant value for $\kappa_\mathrm{IR}$, and $\gamma = \kappa_\mathrm{vis}/\kappa_\mathrm{IR}$ the ratio between the opacity at visual (irradiated) and IR (outgoing) wavelengths. The parameter $f_\mathrm{irr}$ specifies the distribution of the incoming flux over the planet. 

For the sampling of parameter space we apply the MultiNest algorithm \citep{2008MNRAS.384..449F, 2009MNRAS.398.1601F, 2013arXiv1306.2144F}. This algorithm searches parameter space and iteratively zooms in on the most promising area of parameter space, i.e. the area with maximum likelihood. The algorithm provides accurate posterior distributions for all parameters. Usually at least $10^5$ model evaluations are needed for a parameter space which is roughly $10$ dimensional (see next section for more details on which parameters are used).

\subsection{Retrieval validation}

\begin{figure*}[!tp]
\centerline{\resizebox{\hsize}{!}{\includegraphics{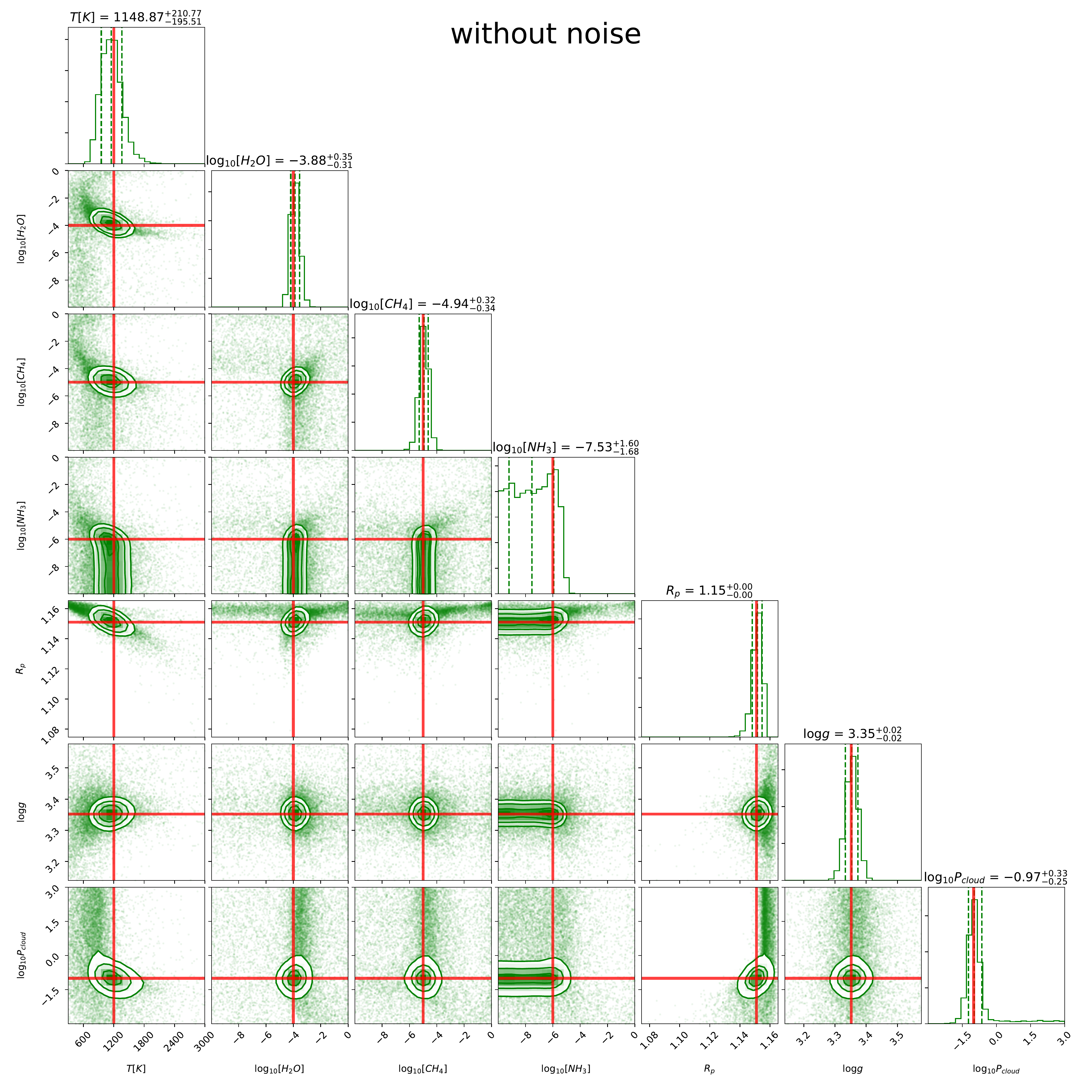}\includegraphics{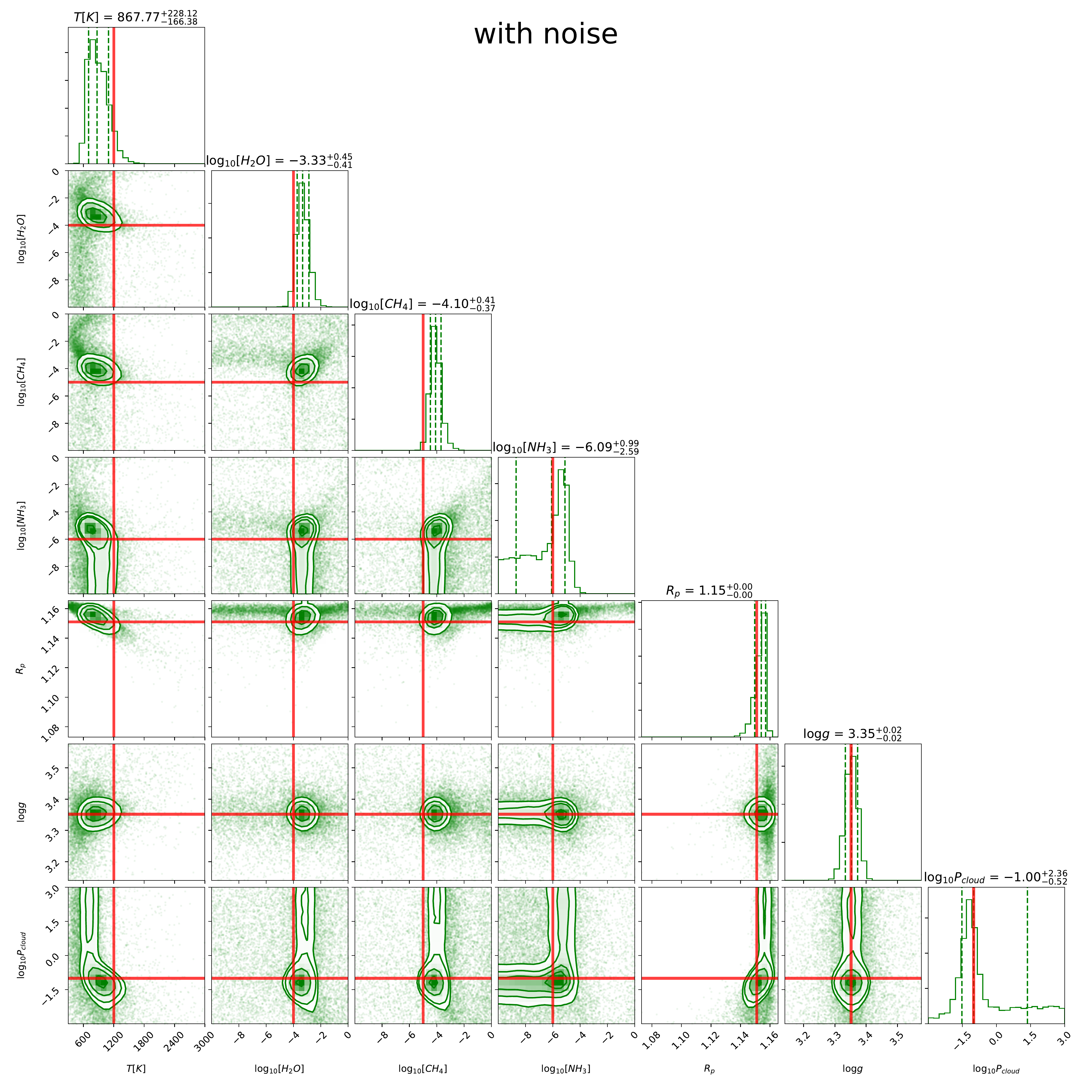}}}
\caption{Results from the retrieval validation tests. Left panel shows the posterior distributions when the retrieval is performed on the mock data without applying Gaussian noise to the simulated observations, right panels are the posterior distributions when Gaussian noise is applied to the simulated observations. Red lines indicate the input values in the forward model. The green contour lines display the 1, 2, 3$\sigma$ uncertainties.}
\label{fig:HD189733b_mock}
\end{figure*}

First we need to validate the retrieval method as implemented in our code. We have already validated the retrieval algorithm in the framework of the Ariel mission, comparing our retrieval outcome to that of four other codes. Here we present a simple test where we use ARCiS to simulate an observed spectrum and subsequently use the retrieval to retrieve the input parameters. We generate a planet with the basic properties of HD189733b. We use an isothermal profile of 1200\,K. The atmosphere consists only of H$_2$O, CH$_4$ and NH$_3$ with volume mixing ratios $10^{-4}, 10^{-5}, 10^{-6}$ respectively. The remaining atmosphere consists of H$_2$ and He in a ratio $0.85/0.15$. We add an optically thick grey cloud deck at $P=0.1\,$bar. We simulate a spectrum using the observational wavelength points and accompanying uncertainties currently available (see also next section). The resulting posterior distributions from the retrieval of this mock spectrum are presented in Fig.~\ref{fig:HD189733b_mock}. The left plot represents the retrieval performed on the exact data-points as computed in the forward model, the right panel shows the results of the retrieval when we apply Gaussian noise to the data with amplitude of the uncertainties in the observations. As can be seen, the input parameters (represented by the red lines) are retrieved consistently.

\subsection{Retrieval applied to ten Hot Jupiters}
\label{sec:retrieval results}

In this section we apply the retrieval framework described in Section \ref{sec:retrieval} to the transmission spectra from 10 hot Jupiters as presented in \cite{2016Natur.529...59S}. For easy comparison and consistency we use this dataset in full as it can be downloaded from \href{https://pages.jh.edu/~dsing3/David_Sing/Spectral_Library.html}{the website of D. Sing} with only two exceptions. For WASP-6b we replace the observations with the homogeneously reduced dataset from \citet{2020MNRAS.494.5449C}. For WASP-12b we add the WFC3 spectrum as presented in \citet{2015ApJ...814...66K}. This spectrum has a slight offset with respect to the observations in  \cite{2016Natur.529...59S}, so during the retrieval we allow for an additional parameter scaling the WFC3 spectrum to match the remaining observational data. This scaling parameter is not treated in the same way as the other retrieval parameters but determined for each forward model computed by using a linear least square fit determining its optimal value, therefore it is not represented in the Bayesian posterior distributions.
We apply both the classical and the constrained retrieval and compare the results.

\begin{table}[!tb]
\caption{References for the observations used in this study.}
\begin{center}
\begin{tabular}{l|l|l}
\hline
\hline
\#	&	Planet		&	references	\\
\hline
1	&	HD189733b 	&	\cite{2013MNRAS.432.2917P} \\
	&				&	\cite{2014ApJ...791...55M} \\
	&				&	\cite{2016Natur.529...59S} \\
2	&	HAT-P-12b 	&	\cite{2016Natur.529...59S}  \\
3	&	WASP-12b	&	\cite{2013MNRAS.436.2956S} \\
	&				&	\cite{2015ApJ...814...66K} \\
	&				&	\cite{2016Natur.529...59S}  \\
4	&	WASP-6b 		&	\cite{2020MNRAS.494.5449C} \\
5	&	WASP-31b 	&	\cite{2015MNRAS.446.2428S} \\
	&				&	\cite{2016Natur.529...59S} \\
6	&	HAT-P-1b 		&	\cite{2013MNRAS.435.3481W} \\
	&				&	\cite{2014MNRAS.437...46N} \\ 
7	&	WASP-39b 	&	\cite{2016ApJ...827...19F} \\
	&				&	\cite{2016Natur.529...59S} \\
	&				&	\cite{2018AJ....155...29W} \\
8	&	WASP-19b 	&	\cite{2013MNRAS.434.3252H} \\
	&				&	\cite{2016Natur.529...59S} \\
9	&	HD209458b 	&	\cite{2016Natur.529...59S} \\
10	&	WASP-17b 	&	\cite{2016Natur.529...59S} \\ 
\hline
\hline
\end{tabular}
\end{center}
\label{tab:references}
\end{table}

We use the nested sampling algorithm as described in Section~\ref{sec:retrieval} for both retrieval setups. Though the molecular opacities were computed at a spectral resolution of $R=300$, for the retrieval the correlated-k tables are remapped onto the observational grid given the spectral width of each observational point. Typically between $7\cdot10^{4}$ and $1.5\cdot10^5$ model evaluations were needed for the multinest algorithm to converge. An exception to this was the retrieval for HD189733b which needed $10^6$ model runs to converge. These were run on a single desktop computer with 8 CPU cores. The classic retrieval method takes on the order of 30 minutes per target to converge ($\sim50$ model evaluations per second), while the model constrained retrieval took around 6 hours per target ($\sim3$ model evaluation per second). Exact computation times depend on CPU model and even differ from model to model depending on the input parameters and choices on atmospheric setup, thus these numbers should be taken only as order of magnitude. The most time-consuming part in the constrained retrieval setup is the computation of cloud structure and opacities consuming together almost 90\% of the computation time. Only a small part of the time is spend in the equilibrium chemistry module since the GGchem code used is extremely efficient.

For the classic retrieval method we use seven species in the gas (H$_2$O, CO, CO$_2$, CH$_4$, NH$_3$, Na, K) and two parameters for the cloud formation (pressure level of the opaque grey cloud deck and the strength of a diffuse, Rayleigh scattering haze). Together with the reference radius (defined at 10\,bar), the planet $\log g$ value and isothermal temperature of the planet, this results in 12 parameters to retrieve. For the model constrained retrieval we have four parameters determining the chemistry (C/O, Si/O, N/O and metallicity), four parameters for the temperature structure ($T_\mathrm{int}$, $f_\mathrm{irr}$, $\gamma$ and $\kappa_\mathrm{vis}$) and two cloud formation parameters \citep[the diffusion coefficient, $K_{zz}$, and the nucleation rate, $\dot\Sigma$;][]{2019A&A...622A.121O}. Together with the radius and $\log g$ of the planet this results in 12 parameters to retrieve as well. All parameters are given in Table~\ref{tab:parameters}. For the model constrained retrieval we include the opacities of all molecules given in Table~\ref{tab:linelists}.

\begin{table*}[!tb]
\caption{Parameters used in our retrievals.}
\begin{center}
\begin{tabular}{l|c|c|l}
\hline
\hline
Parameter				&	symbol			&	range									& prior	\\
\hline
\multicolumn{4}{c}{Classic retrieval} \\
\hline
planet radius			&	$R$				&	$5\sigma$ around lit. value					&	flat linear \\
planet $\log(g)$			&	$\log(g)$			&	$5\sigma$ around lit. value					&	gaussian prior \\
atmospheric temperature	&	$T$				&	300 -- 3000\,K								&	flat linear \\
molecular abundances	&	$X_i$			&	$10^{-10}$ -- $1$							&	flat log \\
cloud pressure			&	$P_\mathrm{cloud}$	&	$10^{-3}$ -- $10^{3}$bar						&	flat log \\
Rayleigh haze			&	$f_\mathrm{mix}$	&	$10^{-3}$ -- $10^{6}$ $\times$ H$_2$ scattering	&	flat log \\
\hline
\multicolumn{4}{c}{Constrained retrieval} \\
\hline
planet radius			&	$R$				&	$5\sigma$ around lit. value					&	flat linear \\
planet mass			&	$M$				&	$5\sigma$ around lit. value					&	gaussian prior \\
diffusion coefficient		&	$K_{zz}$			&	$10^{5}$ -- $10^{12}$\,cm$^2$\,s$^{-1}$			&	flat log \\
nucleation rate			&	$\dot\Sigma$		&	$10^{-17}$ -- $10^{-7}$\,g\,cm$^{-2}$\,s$^{-1}$		&	flat log \\
C/O					&	C/O				&	0.1 -- 1.3									&	flat linear \\
Si/O					&	Si/O				&	0 -- 0.3									&	flat linear \\
N/O					&	N/O				&	0 -- 0.3									&	flat linear \\
metallicity				&	$\left[Z\right]$		&	-1 -- 3									&	flat linear \\
irradiation parameter		&	$f_\mathrm{irr}$	&	0 -- 0.25									&	flat linear \\
internal temperature		&	$T_\mathrm{int}$	&	10 -- 3000	\,K								&	flat log \\
infrared opacity			&$\kappa_\mathrm{IR}$	&	$10^{-4}$ -- $10^{4}$\,cm$^2$\,g$^{-1}$			&	flat log \\
visible to IR opacity		&	$\gamma$		&	$10^{-2}$ -- $10^{2}$						&	flat log \\
\hline
\hline
\end{tabular}
\end{center}
\label{tab:parameters}
\end{table*}

We ran the model constrained retrieval also with 1) TiO and VO \citep[using the lines-lists from][]{2019MNRAS.488.2836M, 2016MNRAS.463..771M}, 2) without Na and K, and 3) constraining the value of $f_\mathrm{irr}$ to be smaller than 0.25. These resulted in slightly different sets of retrieved parameters, all within 2 sigma of each other. The only notable exception to this is the fit for HD189733b with restricted $f_\mathrm{irr}$. Since for this source we find a temperature which is much above the expected value from the irradiation (i.e. $f_\mathrm{irr}>>0.25$), the retrieval was pushed into a very different corner of parameter space. The results with unrestricted and restricted values for $f_\mathrm{irr}$ for the metallicity were $7\sigma$ different flipping from sub-solar to super-solar metallicity. Fortunately, the results for all other sources were not significantly affected by these changes. Performing the retrieval with slightly varying model constraints and viewing the results of them all, makes the interpretation more robust. The results presented below and the correlations discussed are found for all four sets of retrieved parameters. Here we present the results without TiO and VO, with Na and K, and with $0<f_\mathrm{irr}<0.25$.

In Table~\ref{tab:H2O} we compare the water abundance we find using the classic retrieval to the results found in \cite{2019MNRAS.482.1485P} for the same sources. We note here that we used different data for WASP-6b, WASP-39b and HAT-P-12b compared to that study. Also, we use a different model setup, especially for the temperature and cloud structure, which can introduce systematic differences in the derived abundances. In the right most column in Table~\ref{tab:H2O} we give the difference between the two studies in terms of how many sigma they deviate. Here we see that statistically speaking the two studies seem consistent (0 to 1 source with more than $2\sigma$ difference, 2 to 3 sources between 1 and $2\sigma$ and 6 to 7 sources below $1\sigma$ difference).

Interestingly, we find three sources where, using the classic retrieval, we find constraints on the NH$_3$ abundance; WASP-31b, WASP-39b and HD209458b. The abundance of NH$_3$ we find in WASP-31b is in excellent agreement with that found by \cite{2017ApJ...850L..15M}, who also report a tentative detection of nitrogen bearing species in HD209458b \citep[see also][]{2017MNRAS.469.1979M}. Nitrogen chemistry is for most sources not yet very well constrained by the current data, but this will change with the additional precision and broader wavelength coverage of JWST and Ariel.

\begin{table}[!tb]
\caption{Retrieved values of the H$_2$O abundance in the classic retrieval as compared to the results from \cite{2019MNRAS.482.1485P}.}
\begin{center}
\begin{tabular}{l|l|c|c|c}
\hline
\hline
\#	&	Planet		&	$\log_{10}$[H$_2$O]				&	$\log_{10}$[H$_2$O]		&	$\sigma$\\
			&	 \cite{2019MNRAS.482.1485P}		&	this study				&	difference\\
\hline
1	&	HD189733b 	&	$-5.04^{+0.46}_{-0.30}$			&	$-3.29^{+0.57}_{-0.83}$	&	1.84 \\
2	&	HAT-P-12b 	&	$-3.91^{+1.01}_{-1.89}$			&	$-6.09^{+2.07}_{-2.52}$	&	0.78 \\
3	&	WASP-12b	&	$-3.16^{+0.66}_{-0.69}$			&	$-2.48^{+0.76}_{-1.00}$	&	0.57 \\
4	&	WASP-6b 		&	$-6.91^{+1.83}_{-2.07}$			&	$-3.35^{+0.47}_{-1.17}$	&	1.64 \\
5	&	WASP-31b 	&	$-3.97^{+1.01}_{-2.27}$			&	$-3.27^{+1.44}_{-2.18}$	&	0.29 \\
6	&	HAT-P-1b 		&	$-2.72^{+0.42}_{-0.56}$			&	$-1.23^{+0.78}_{-3.12}$	&	0.47 \\ 
7	&	WASP-39b 	&	$-4.07^{+0.72}_{-0.78}$			&	$-1.85^{+1.23}_{-0.58}$	&	2.41 \\
8	&	WASP-19b 	&	$-3.90^{+0.95}_{-1.16}$			&	$-2.68^{+1.15}_{-1.60}$	&	0.66 \\
9	&	HD209458b 	&	$-4.66^{+0.39}_{-0.30}$			&	$-4.68^{+0.53}_{-0.20}$	&	0.03 \\
10	&	WASP-17b 	&	$-4.04^{+0.91}_{-0.42}$ 			&	$-1.99^{+0.75}_{-1.20}$	&	1.37 \\ 
\hline
\hline
\end{tabular}
\end{center}
\label{tab:H2O}
\end{table}

\begin{figure*}[!tp]
\centerline{\resizebox{\hsize}{!}{\includegraphics[page=1]{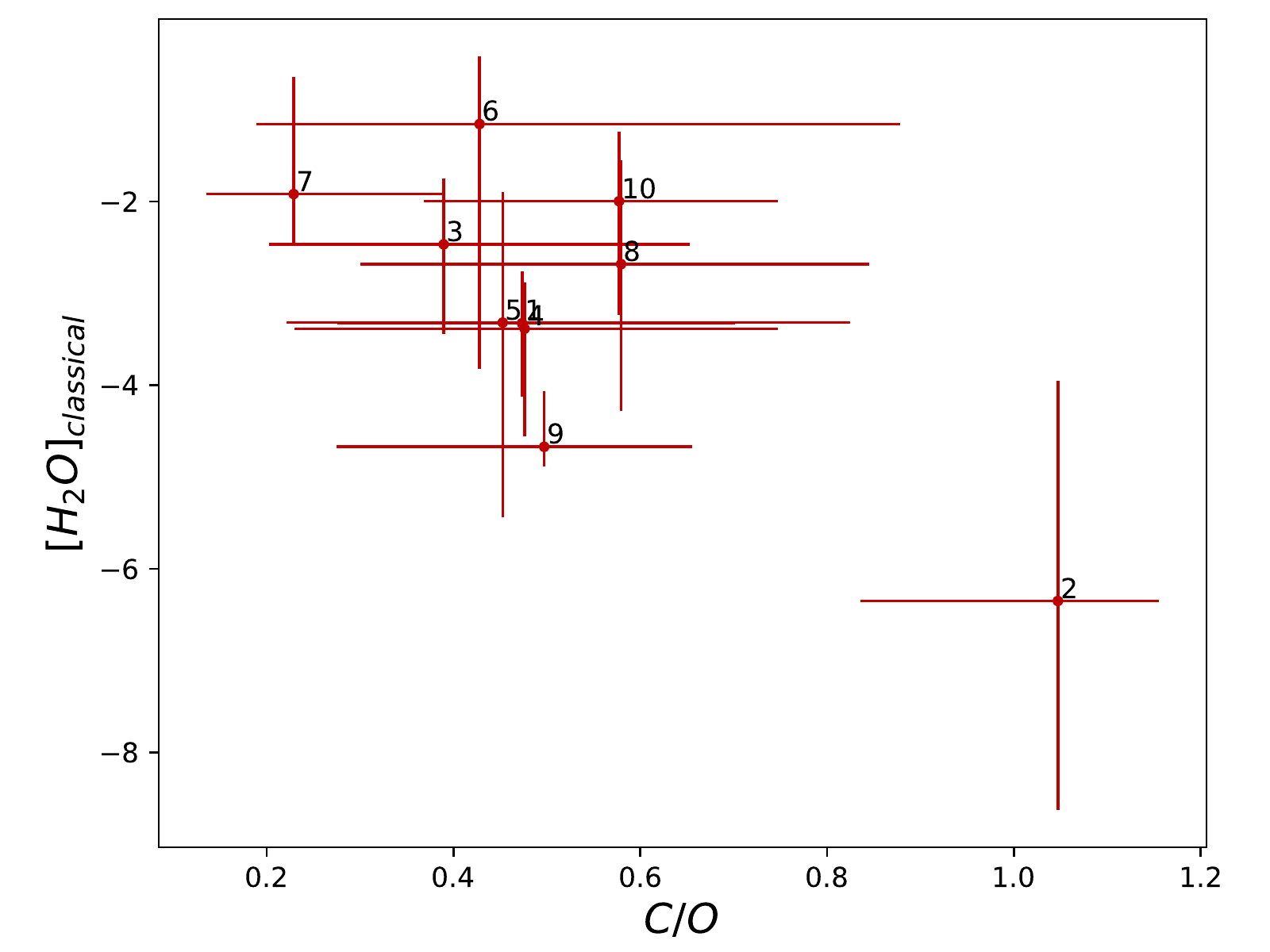}\includegraphics[page=2]{scatter}}}
\caption{The $\log_{10}$ of the water abundance as derived from the classical retrieval plotted versus the C/O ratio (left) and the metallicity (right) derived from the constrained retrieval. The annotated numbers correspond to numbers for the planets as given in Table~\ref{tab:H2O}.}
\label{fig:H2O_Z_vs_COratio}
\end{figure*}

\subsection{Comparing classic and constrained retrieval results}

The water abundance in classic retrieval results is often used as an indicator of either the C/O ratio of the atmosphere or its metallicity as chemical models predict these parameters are linked. Here we can test this assumption by comparing the H$_2$O abundance as derived from the classic retrieval to the C/O ratio and the metallicity obtained from the model constrained retrieval. 
In the left panel of Fig.~\ref{fig:H2O_Z_vs_COratio} we plot the H$_2$O mixing ratio as obtained from the classic retrieval versus the C/O ratio obtained from the model constrained retrieval. We see no significant correlation between these two parameters. The correlation between the water abundance and the metallicity (right panel of Fig.~\ref{fig:H2O_Z_vs_COratio}), shows a tentative trend that sources with high metallicity have a high water abundance. For both the metallicity and the C/O ratio we conclude (unsurprisingly) that the data used here is not of sufficient quality and quantity to assess the validity of using the water abundance as a tracer for either the C/O ratio or the metallicity. We note that the way we defined the carbon and oxygen abundances is an important factor here. For sub-solar C/O ratios, the oxygen abundance is for example only influenced by the metallicity. A better understanding on how the elemental abundances are coupled from the perspective of planet formation would be very helpful here.

We further compare the atmospheric temperatures obtained with both retrieval approaches. In Fig.~\ref{fig:PT} in Appendix \ref{app:PT} we show the pressure temperature structure for the classic (in red) and constrained (in blue) retrieval methods. 
The local temperatures appear to agree reasonably well within the error contours derived, especially around the pressures that contribute most to the observed spectrum (i.e. around $10^{-1} - 10^{-3}\,$bar). The temperature structure at very low and very high pressures as derived from both retrieval methods is inaccurate and suffers significantly from the assumptions on the shape of the temperature structure (either isothermal or given by Eq.~\ref{eq:Guillot}).

The Bayesian algorithm we employ allows to compare the classic and constrained retrieval in a formal way. The Bayes factor, which is the ratio between the Bayesian evidence of two models, can tell us if one of the two models is significantly preferred over the other (which is the case when $\ln B_{01}>5$). The number of free parameters is the same between the two models. We find that for all sources the Bayesian evidence is higher or equal for the classic retrieval. However, the difference is only significant, i.e. the natural logarithm of the Bayes factor $\ln B_{01}>5$,  for HAT-P-12b, WASP-39b and HD209458b. The logarithm of the Bayes factor is given for all sources in Table~\ref{tab:fBe}. The relatively large value of $\ln B_{01}$ for HAT-P-12b, WASP-39b and HD209458b hint at that for these sources we have enough information to indicate that there might be physical or chemical processes going on that are not captured by our constrained retrieval model. For HAT-P-12b this is already shown in the exceptional value for the C/O ratio in the constrained retrieval and the low H$_2$O abundance in the classic retrieval. This is partly caused by the lack of a clear water feature in the WFC3 spectrum of this source. Future observations will have to tell us if this is real.

\begin{table}[!tb]
\caption{Natural log of the Bayes factor, $\ln B_{01}$,  between the classical and constrained retrievals and the retrieved value for $K_{zz}$ in the constrained retrievals.}
\begin{center}
\begin{tabular}{l|l|c|r}
\hline
\hline
\#	&	Planet		&	$\ln B_{01}$ 	&	$\log_{10}K_{zz}$\\
\hline
1&      HD189733b   &           2.0   &  $11.2_{-1.2}^{+0.5}$\\
2&      HAT-P-12b   &   \textbf{6.0}  &  $10.7_{-3.2}^{+0.9}$\\
3&       WASP-12b   &           2.9   &  $10.0_{-1.3}^{+1.2}$\\
4&        WASP-6b   &           0.1   &  $ 7.6_{-0.7}^{+0.8}$\\
5&       WASP-31b   &           2.8   &  $ 7.6_{-0.9}^{+1.8}$\\
6&       HAT-P-1b   &           1.1   &  $ 7.5_{-0.9}^{+1.7}$\\
7&       WASP-39b   &   \textbf{5.7}  &  $ 6.6_{-0.3}^{+0.4}$\\
8&       WASP-19b   &          -0.2   &  $ 6.1_{-0.7}^{+2.0}$\\
9&      HD209458b   &   \textbf{5.1}  &  $ 5.9_{-0.2}^{+0.1}$\\
10&       WASP-17b   &           0.4   &  $ 5.7_{-0.4}^{+0.5}$\\
\hline
\hline
\end{tabular}
\end{center}
\label{tab:fBe}
\end{table}

\subsection{Model constrained retrieval results}

We analysed the results of the model constrained retrievals by looking at correlations between the different derived parameters. Due to the large uncertainties on the parameters (caused by the small wavelength coverage of the observations and the large errorbars on the data), no significant correlations have been found. Nevertheless, in this section we summarise our retrieval results for a set of 10 giant gas planets all treated with the same retrieval method.

\begin{figure}[!tp]
\centerline{\resizebox{\hsize}{!}{\includegraphics[page=8]{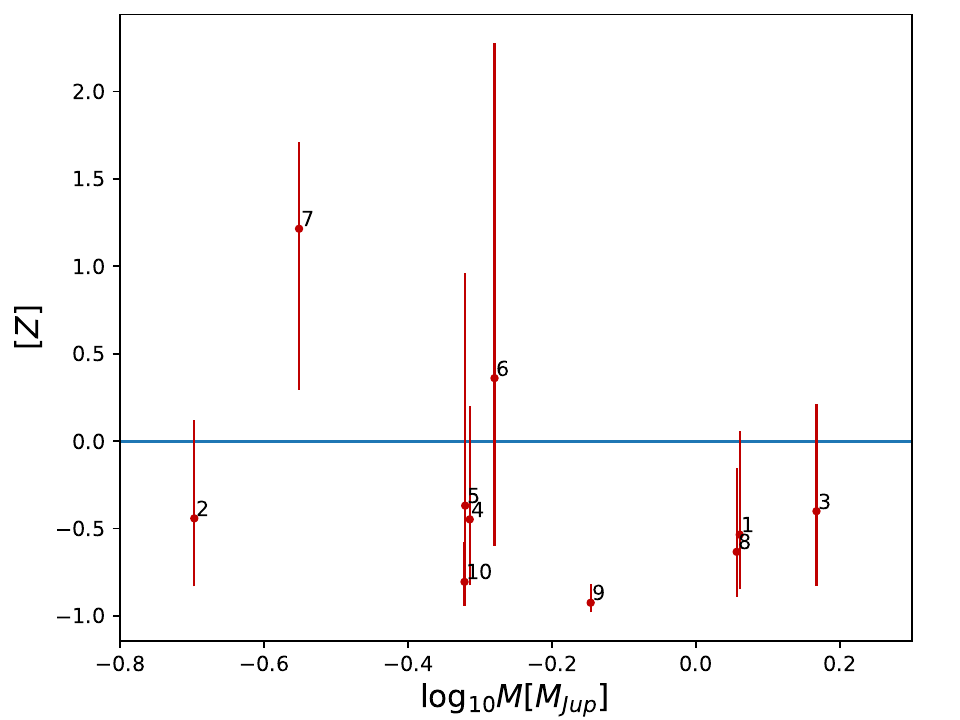}}}
\caption{The metallicity derived by the model constrained retrieval versus the mass of the exoplanets considered. Mass of the exoplanets is given in Jupiter masses.}
\label{fig:Z_vs_M}
\end{figure}

\begin{figure}[!tp]
\centerline{\resizebox{\hsize}{!}{\includegraphics[page=9]{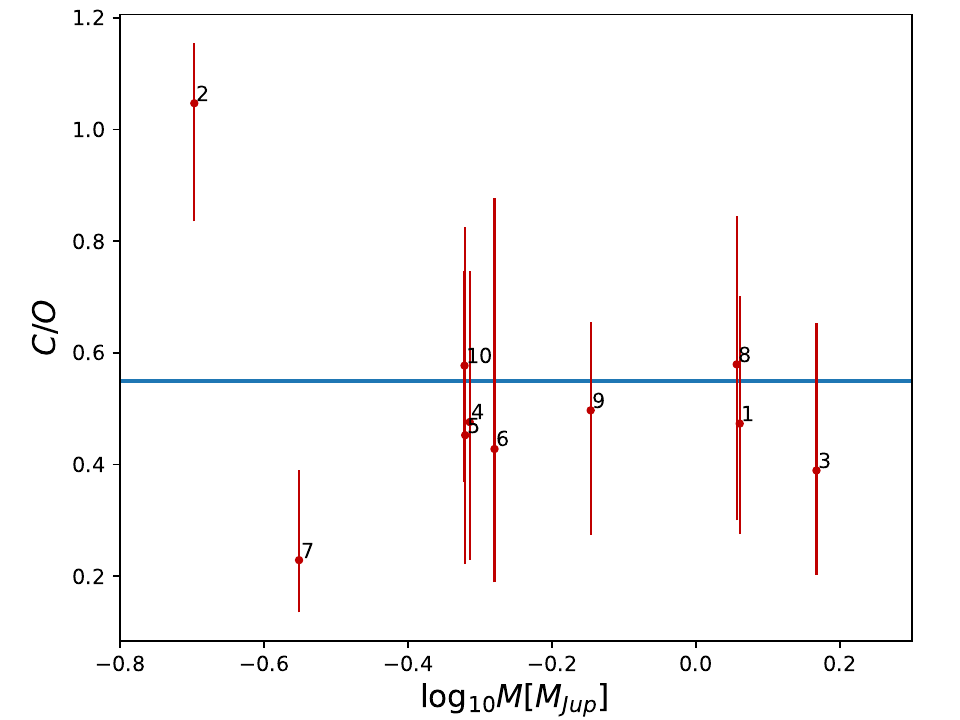}}}
\caption{The C/O ratio derived by the model constrained retrieval versus the mass of the exoplanets considered.}
\label{fig:COratio_vs_M}
\end{figure}

\begin{figure}[!tp]
\centerline{\resizebox{\hsize}{!}{\includegraphics[page=18]{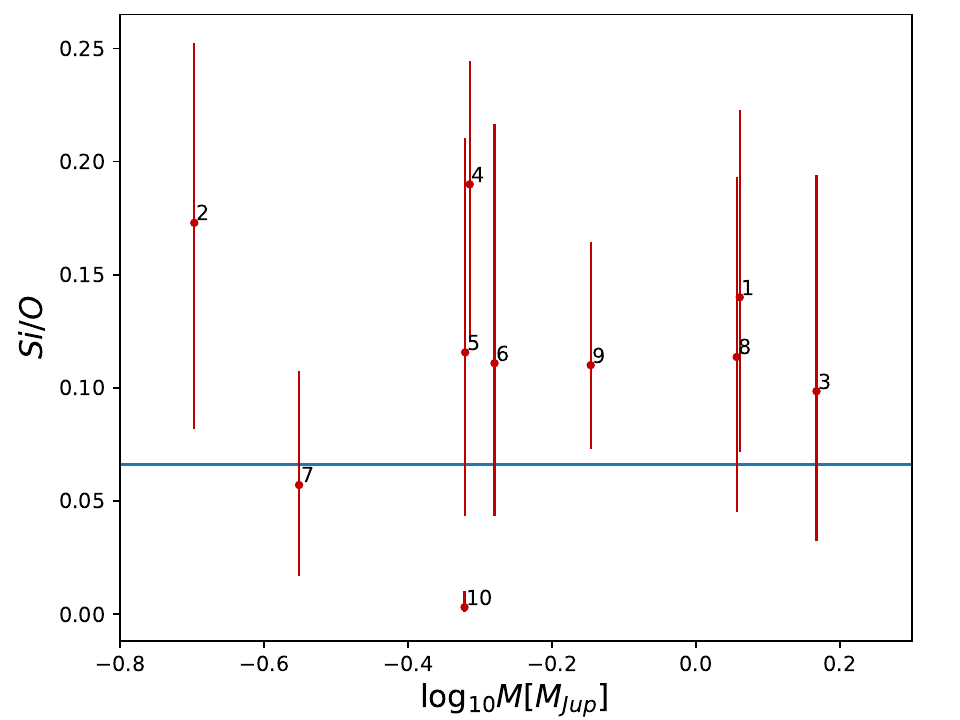}}}
\caption{The Si/O ratio derived by the model constrained retrieval versus the mass of the exoplanets considered.}
\label{fig:SiOratio_vs_M}
\end{figure}

In Figs.~\ref{fig:Z_vs_M}, \ref{fig:COratio_vs_M} and  \ref{fig:SiOratio_vs_M} we plot the metallicity, the C/O ratio, and the Si/O ratio versus the mass of the exoplanet. \citet{2017Sci...356..628W} find a correlation between metallicity and mass. We find that eight of the ten planets are within 2$\sigma$ of the solar metallicity. The exceptions to this are WASP-17b and HD209458b. The C/O ratio shows no correlation at all with the planet mass. Interestingly, we find one planet with a C/O ratio significantly higher than the solar value (HAT-P-12b), and one with a lower C/O ratio (WASP-39b). We remind the reader here that these two planets also have $\ln B_{01}>5$ compared to the classic retrieval, which indicates we have to apply extra caution interpreting the retrieval results. Finally, the Si/O ratio is poorly constrained by the current dataset. Only WASP-17b seems to be an outlier here with a significantly low Si/O value. We find that overall most planets are within $2\sigma$ consistent with the solar ratio. \citet{2019MNRAS.482.1485P} found that most planets show a sub-solar H$_2$O abundance which can be caused by sub-solar metallicity or super-solar C/O ratio. We confirm the sub-solar O abundance either with high values for C/O or low metallicity. When we look at the so-called corner-plots, visualising the posterior distributions of the Bayesian sampling (see Appendix~\ref{app:cornerplots}), we see that for some sources there is a correlation between the C/O and Si/O. This indicates that for these sources the O abundance is actually constrained and not the abundance ratios with C and Si.

\begin{figure}[!tp]
\centerline{\resizebox{\hsize}{!}{\includegraphics[page=5]{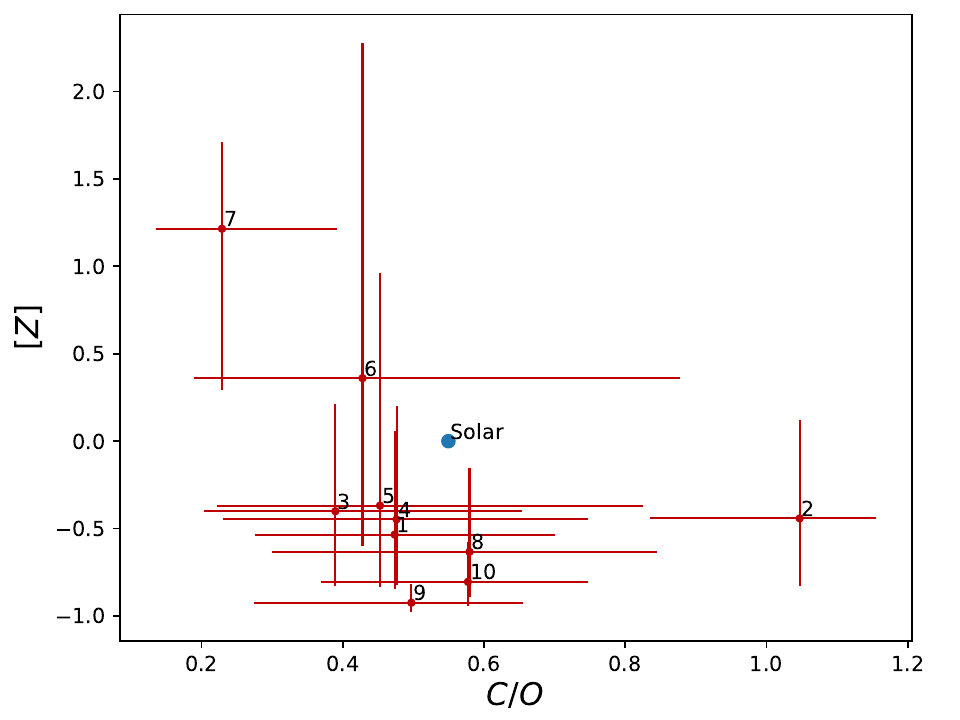}}}
\caption{The metallicity versus the C/O ratio of the exoplanets considered derived by the model constrained retrieval.}
\label{fig:Z_vs_COratio}
\end{figure}

\begin{figure}[!tp]
\centerline{\resizebox{\hsize}{!}{\includegraphics[page=3]{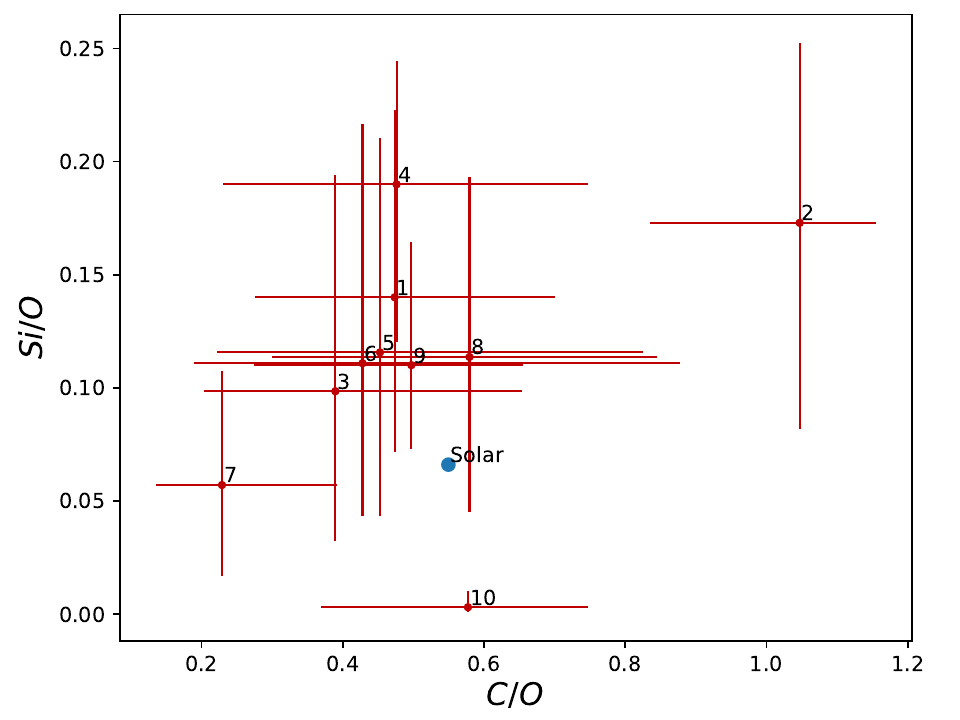}}}
\caption{The Si/O ratio versus the C/O ratio of the exoplanets considered derived by the model constrained retrieval.}
\label{fig:SiOratio_vs_COratio}
\end{figure}

When a planet forms, it accretes mass from different regions in the protoplanetary disk. These different regions have different elemental abundances. It is therefore expected that the abundance ratios can be used as indicators for the formation location. Especially, from formation simulations we may be able to predict correlations between the C/O ratio, the Si/O ratio and the metallicity of the accreted atmosphere. In Figs.~\ref{fig:Z_vs_COratio} and \ref{fig:SiOratio_vs_COratio} we plot these parameters as derived from our constrained retrieval. As can be clearly seen, there are no significant correlations present, mostly due to the large uncertainties on the derived parameters. Future observations can be used to reduce these uncertainties and look for correlations between these parameters.

\begin{figure*}[!tp]
\centerline{\resizebox{\hsize}{!}{\includegraphics[page=10]{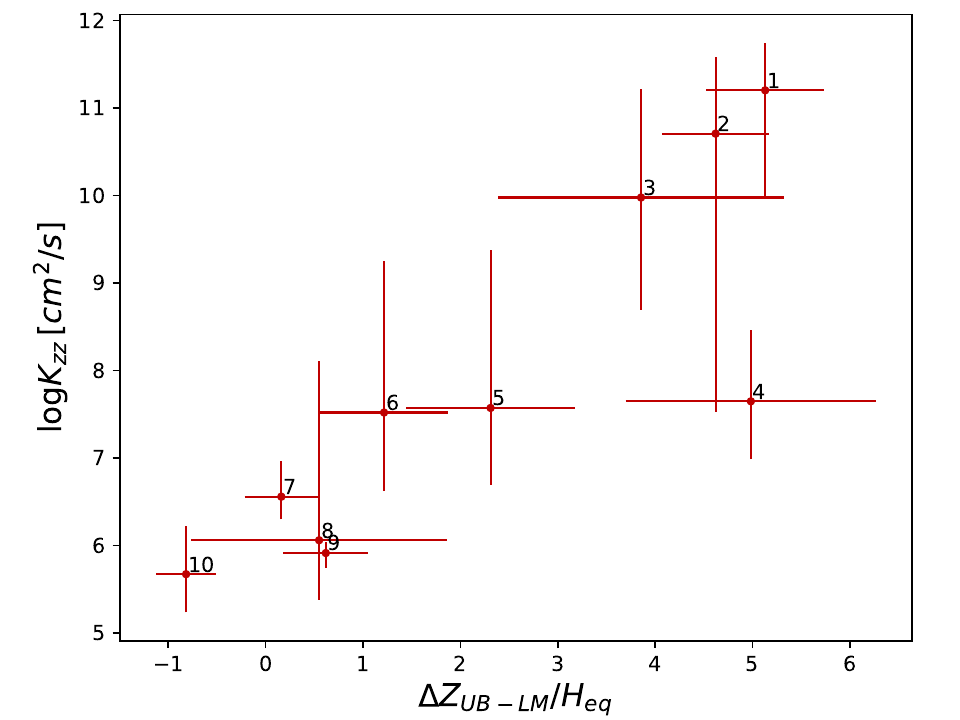}\includegraphics[page=11]{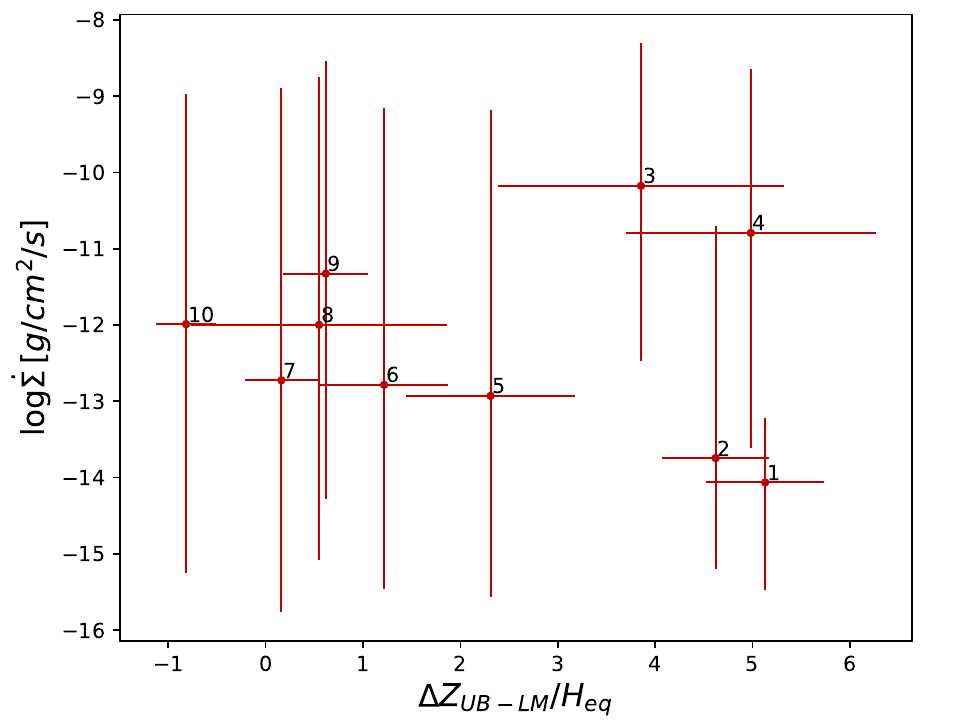}}}
\caption{The value for K$_{zz}$ (left) and $\dot\Sigma$ (right) versus the spectral index as introduced by \citet{2016Natur.529...59S} of the exoplanets considered.}
\label{fig:Cloud_vs_SingIndex}
\end{figure*}

We can also compare our results directly to the spectral index as computed by \citet{2016Natur.529...59S}. In that paper the authors use the spectral index of the transmission spectrum as a direct indicator for clouds. In Fig.~\ref{fig:Cloud_vs_SingIndex} we plot the cloud parameters, i.e. the values we derive for the diffusion strength $K_{zz}$ and the nucleation rate $\dot\Sigma$, versus the slope of the transmission spectrum $\Delta Z_{UB-LM}/H_{eq}$ as defined in \citet{2016Natur.529...59S}. There is a strong correlation between the slope and the value of $K_{zz}$. The sources we find that have a high diffusion strength, and thus strong cloud formation, also are the sources flagged by \citet{2016Natur.529...59S} as cloudy based directly on the observed spectrum. This is a strong sanity check for both the observational method in \citet{2016Natur.529...59S} and the theoretical cloud formation model considered here. However, we do expect that extremely cloudy objects will have a flat spectrum and populate the upper left corner of Fig.~\ref{fig:Cloud_vs_SingIndex}. For the nucleation rate, we do not find a significant correlation. However, this could be due to the large errors on the derived values for the nucleation rate. The values we derived for the nucleation rate are consistent with the range expected from detailed nucleation rate computations using the models from \cite{2018A&A...614A.126L}. In this study we have assumed that the nuclei on which the clouds condense contain negligible mass and thus also do not contribute to the opacity \citep[like in][]{2019A&A...622A.121O}. This is an assumption that can break down for atmospheres with low values of $K_{zz}$ and inefficient coagulation such that the nuclei can accumulate high in the atmosphere and form a haze layer. We leave investigation of the effect of these hazes to a future study.

We made quite specific choices on the elemental abundance ratios that we include and how to compute the elemental abundances from this. These choices can be debated and should in fact be related to how we expect the elemental abundances to change from planet formation theory. This is a study we are currently conducting. The influence of these choices is beyond the scope of this paper and is actually expected to become really relevant with more accurate and complete data coming available from future missions (especially JWST).

We have checked many other correlation between physical properties of the atmosphere and the planet. However, none of them turned out to be significant.

\section{Predictions for future missions}

\begin{figure*}[!tp]
\centerline{\resizebox{\hsize}{!}{\includegraphics{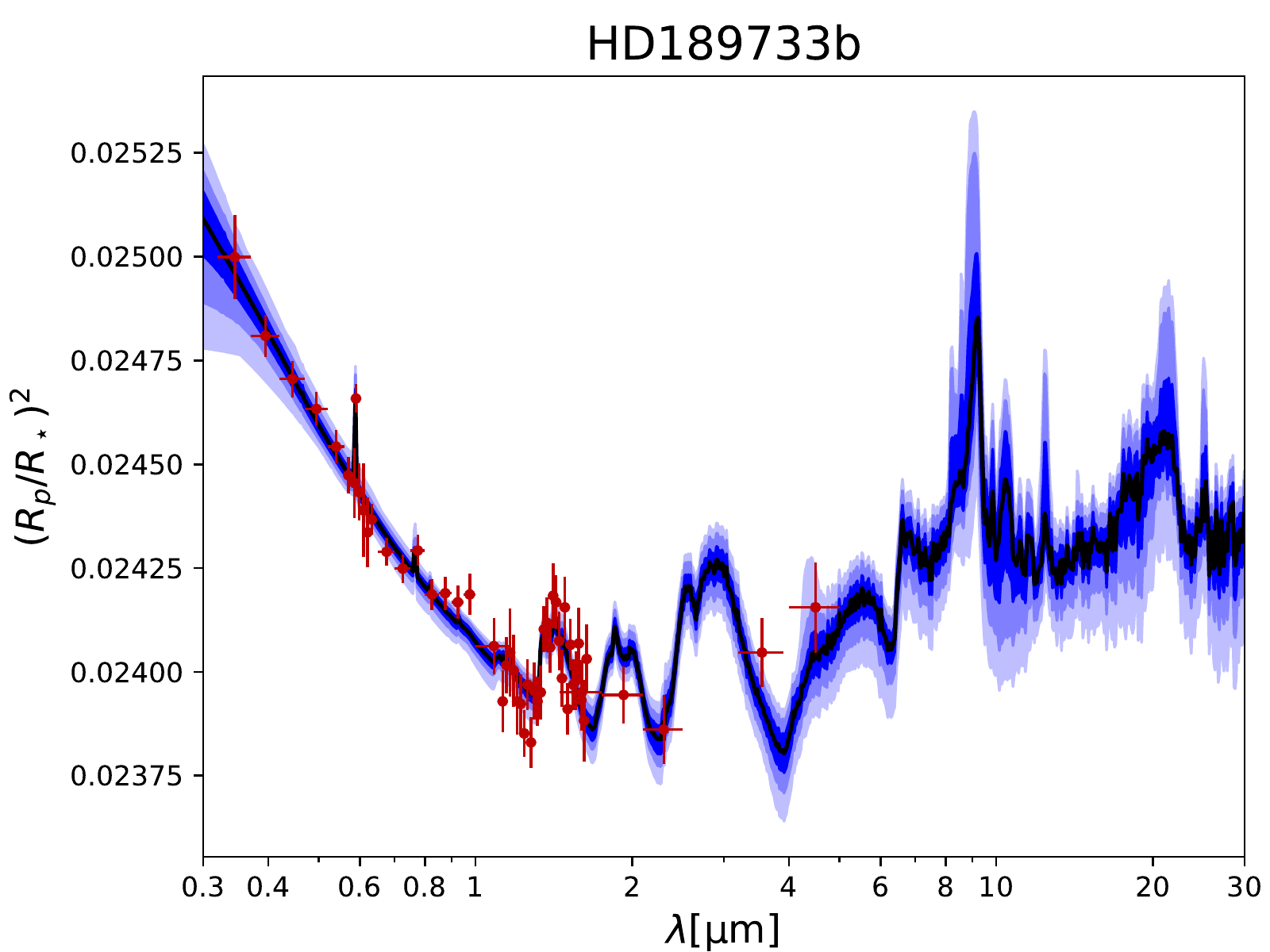}\includegraphics{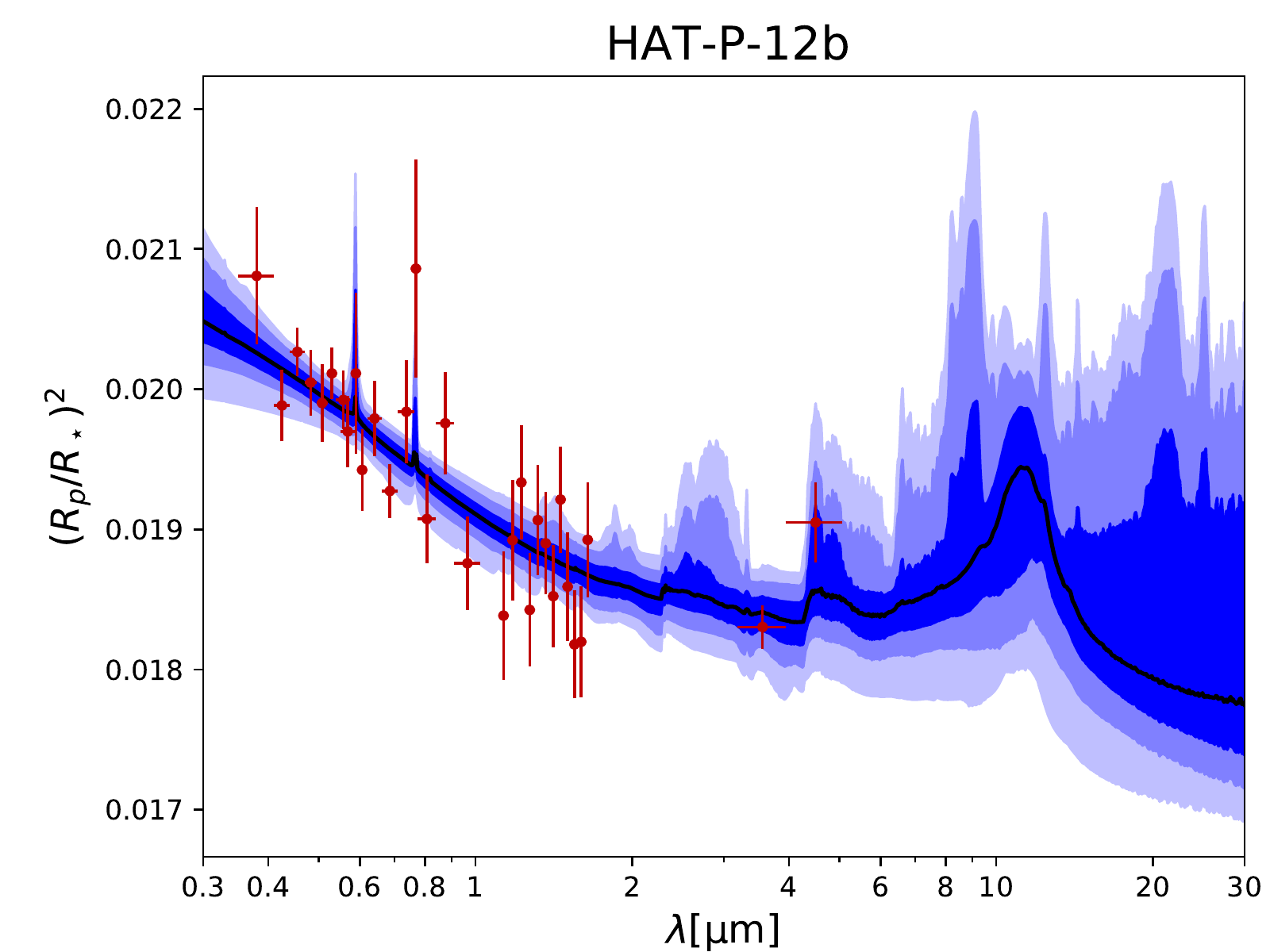}}}
\centerline{\resizebox{\hsize}{!}{\includegraphics{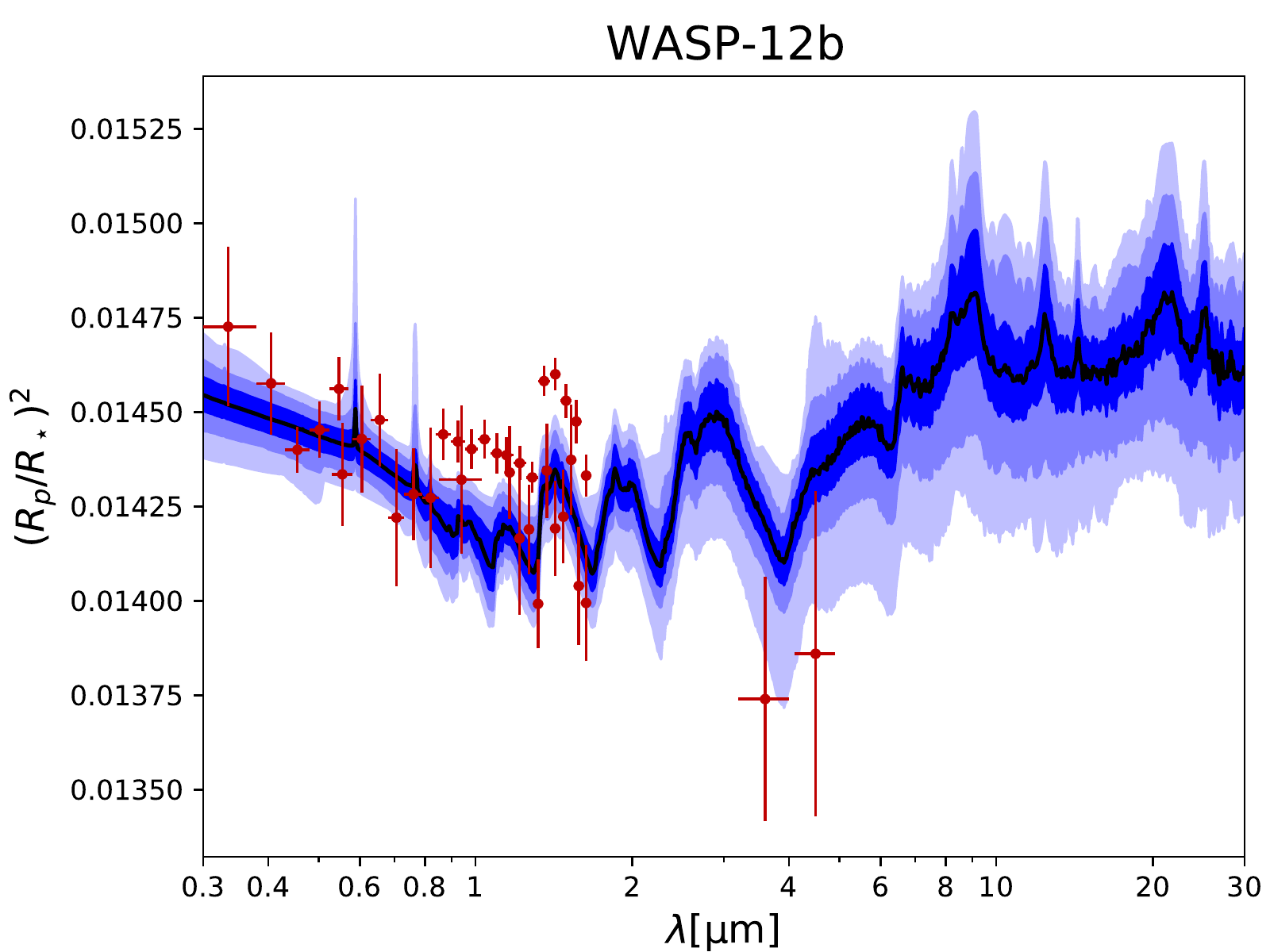}\includegraphics{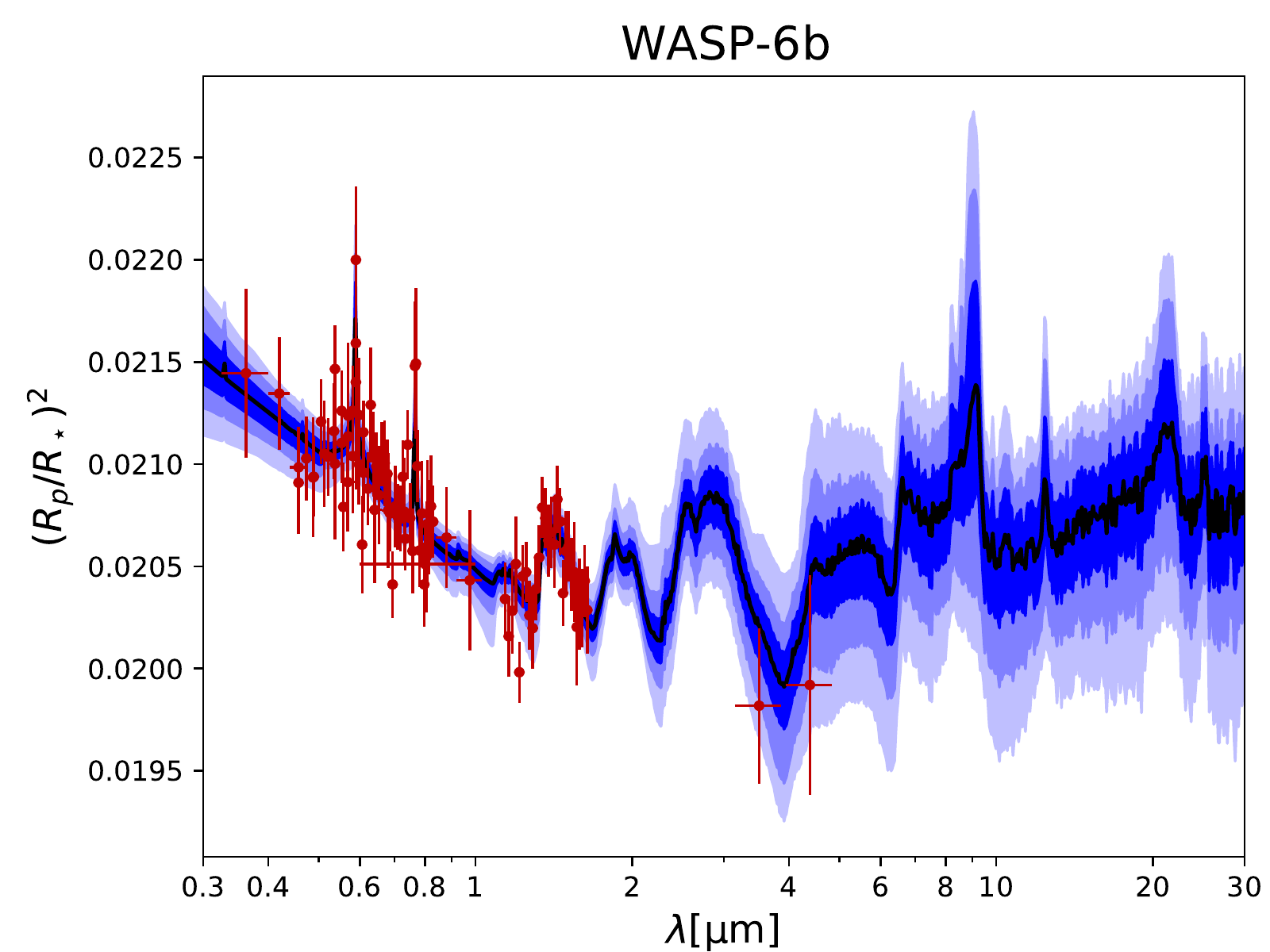}}}
\centerline{\resizebox{\hsize}{!}{\includegraphics{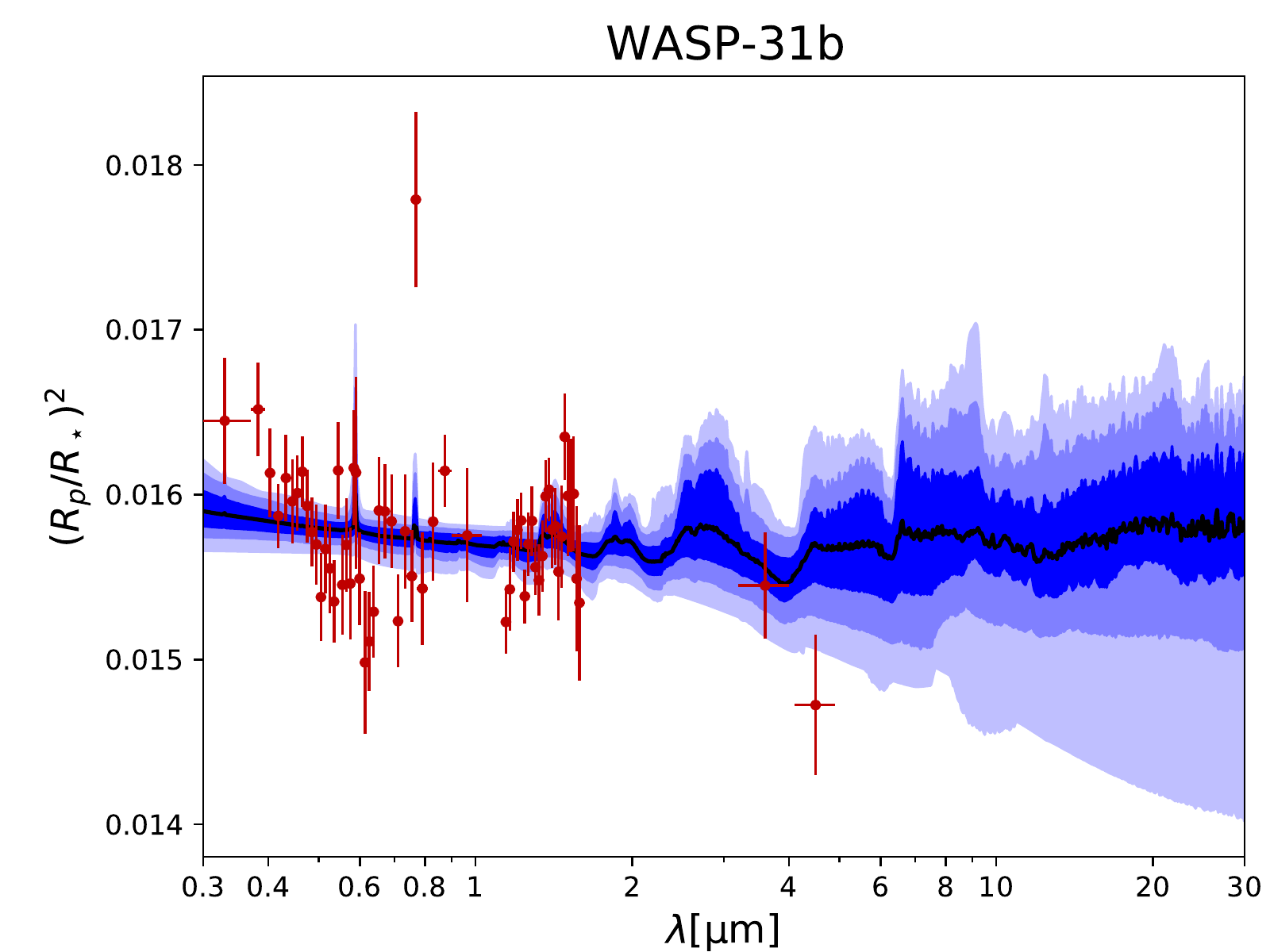}\includegraphics{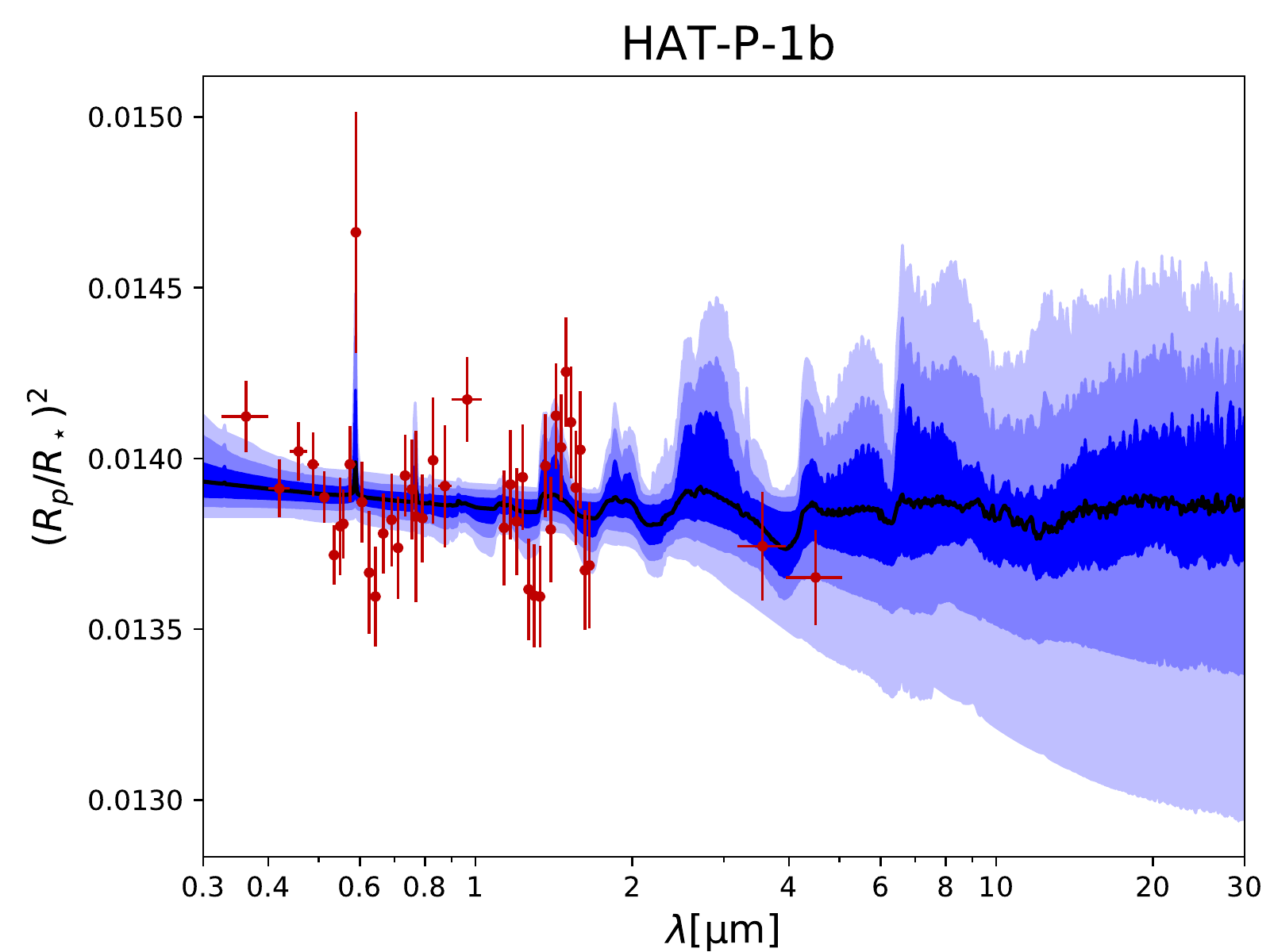}}}
\caption{Extrapolated transmission spectrum into the JWST wavelength range for all 10 planets considered. The planets are ordered according to the retrieved value for $K_{zz}$ with highest value at the top left and lowest at the bottom right corner. Data points are taken from \citet{2016Natur.529...59S}. The blue shaded regions denote the 1, 2, and 3 $\sigma$ confidence intervals.}
\label{fig:spectra}
\end{figure*}
\begin{figure*}[!tp]
\ContinuedFloat
\centerline{\resizebox{\hsize}{!}{\includegraphics{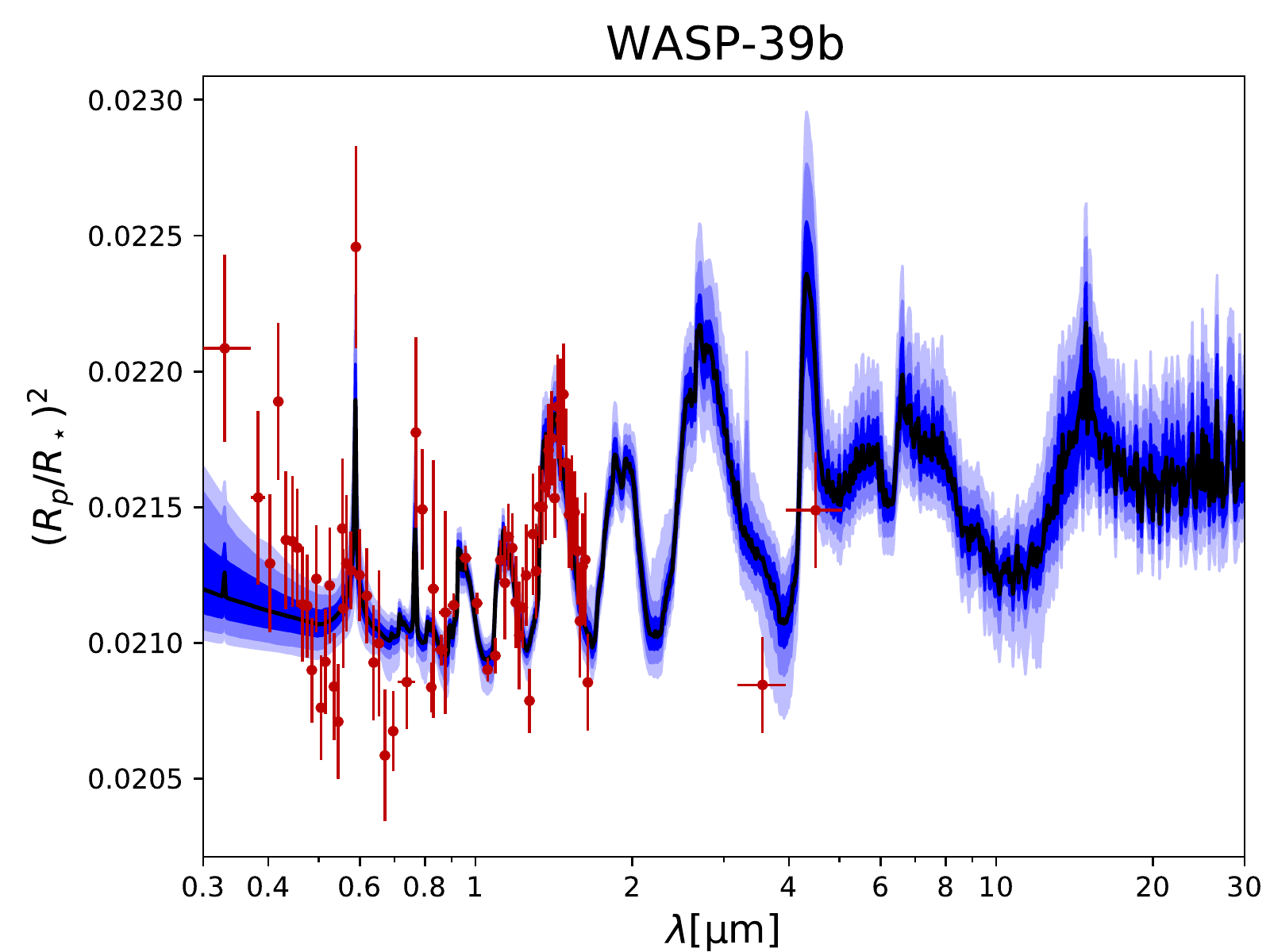}\includegraphics{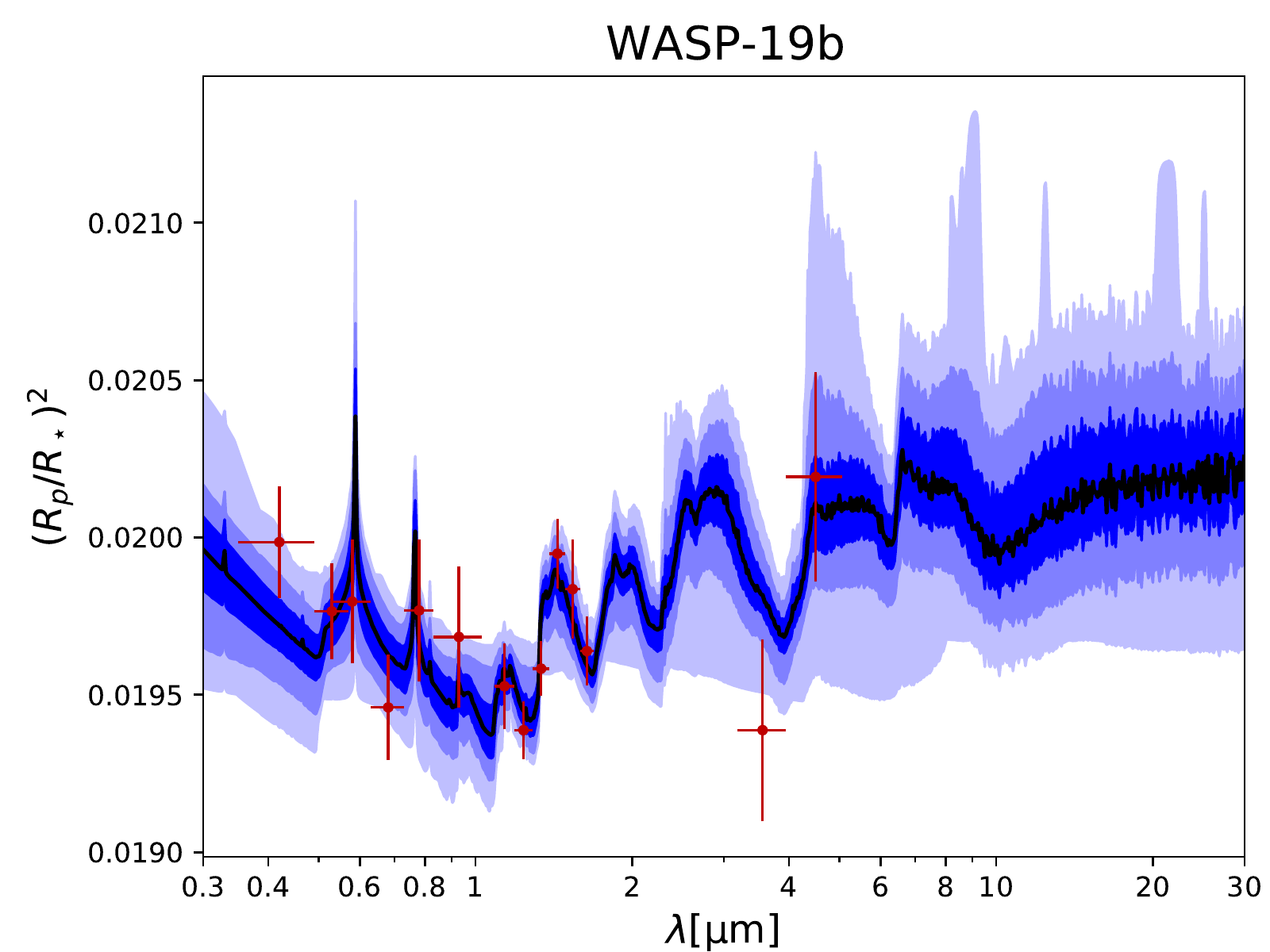}}}
\centerline{\resizebox{\hsize}{!}{\includegraphics{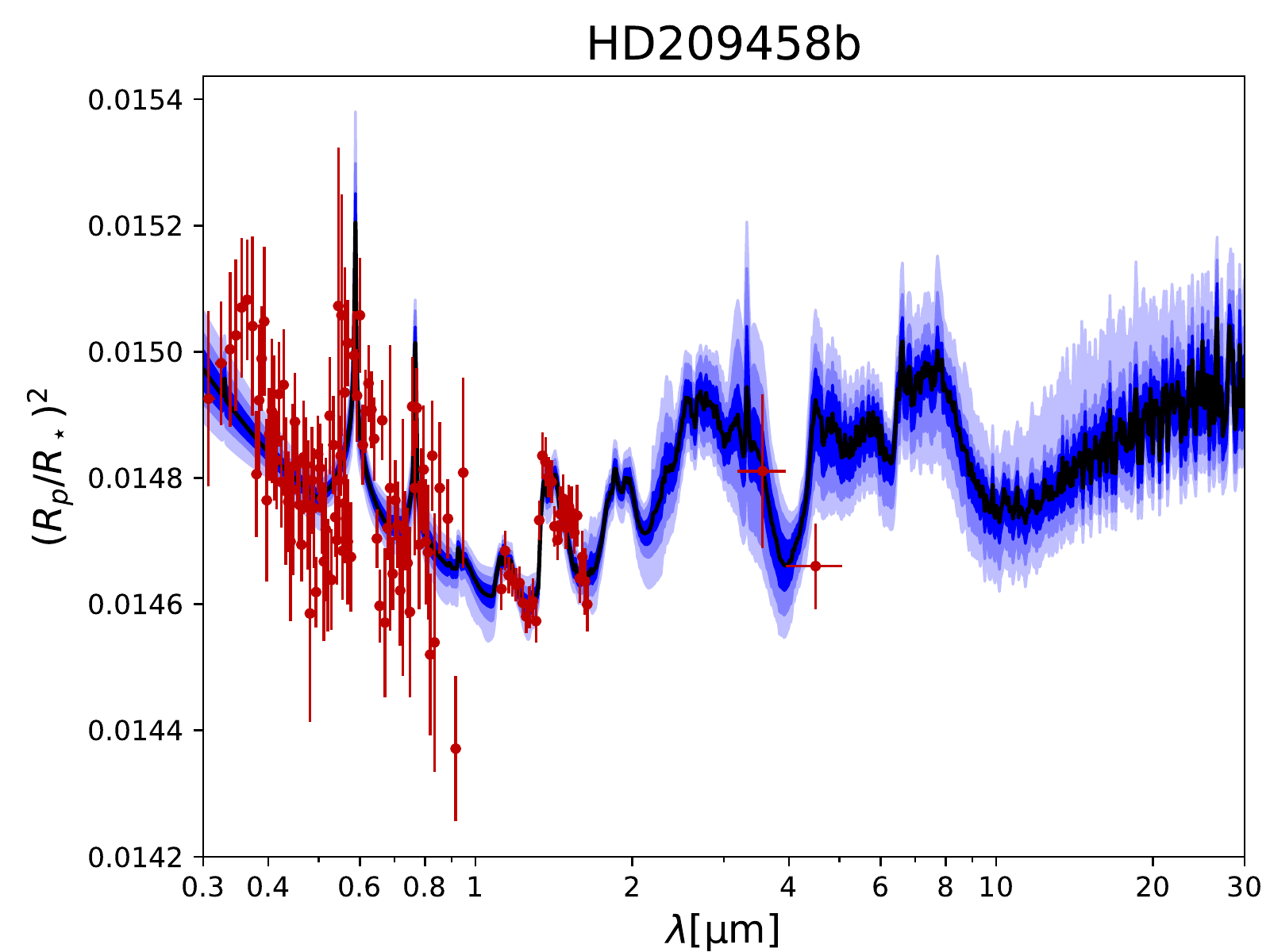}\includegraphics{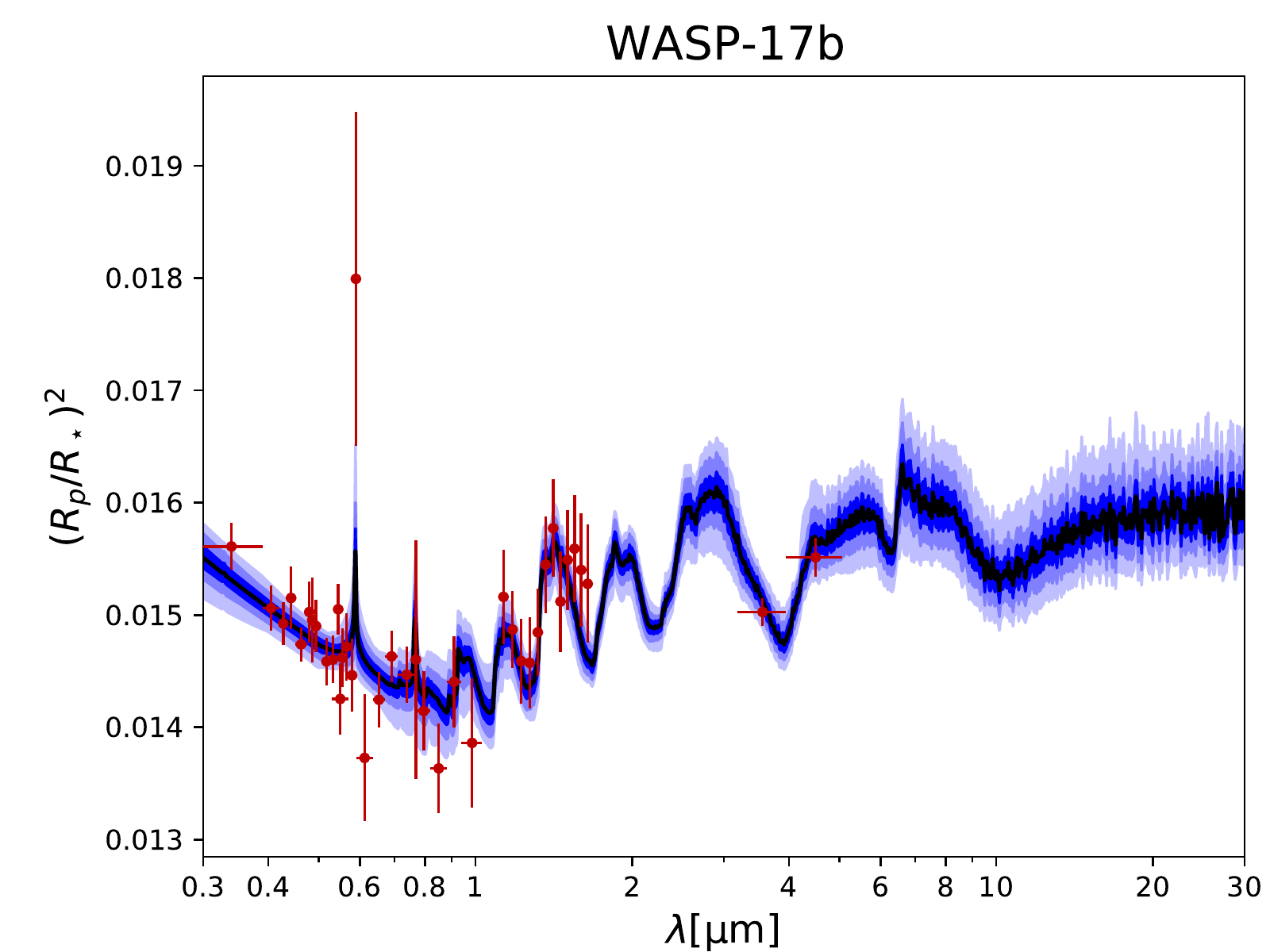}}}
\caption{Continued.}
\label{fig:spectra}
\end{figure*}

We can use the results from the model constrained retrieval to extrapolate into wavelength regions that will become accessible by future missions like JWST and Ariel. The retrieval gives the range of parameters possible within the constraints of the current dataset. We can use these model constraints to estimate the bounds in which we expect the observations with e.g. JWST to fall. For sources where the current dataset is not very constraining, the extrapolated spectra will have a broad range of possibilities. However, for the better constrained sources, we can make a prediction that is better constrained. This is of course under the assumption that our modelling is correct also extrapolated to longer wavelengths. Such an extrapolation dependents on the model used to make the retrieval fit.

In Fig.~\ref{fig:spectra} we show the extrapolated spectrum for all planets considered. The blue shaded areas show the 1, 2 and 3 $\sigma$ confidence intervals for the transmission depth. As can be seen, the region widens for the wavelengths further away from the current observational dataset. However, it is clear that for four planets with high retrieved values for $K_{zz}$ (HAT-P-12b, HD189733b, WASP-12b, WASP-6b), solid state features due to clouds are to be expected around 10~$\mu$m. The features here visible are caused by condensate particles of solid state materials, mainly SiO$_2$[s] and MgSiO$_3$[s]. We conclude here that for studies of cloud mineralogy the sources HD189733b is the most promising candidate. We previously mentioned the sensitivity of the HD189733b fit to small changes in the retrieval setup. In all cases, the mineral features in the JWST predicted spectrum are very significant. HAT-P-12b, WASP-6b and WASP-12b are other candidates showing cloud mineral features but with much less confidence. For HAT-P-12b we see that there are two options for the cloud features, carbon rich or oxygen rich clouds, and these can be distinguished spectroscopically in the 10 micron wavelength range. Finally, in the predicted spectrum for WASP-39b, slight hints can be seen for mineral features. However, detection of these require very high signal to noise spectra and determining the feasibility to detect those requires detailed instrument simulations. Overall, our results suggest that cloud mineral features are expected for sources with $\Delta Z_{UB-LM}/H_{eq}>3$.

\begin{figure}[!tp]
\centerline{\resizebox{\hsize}{!}{\includegraphics{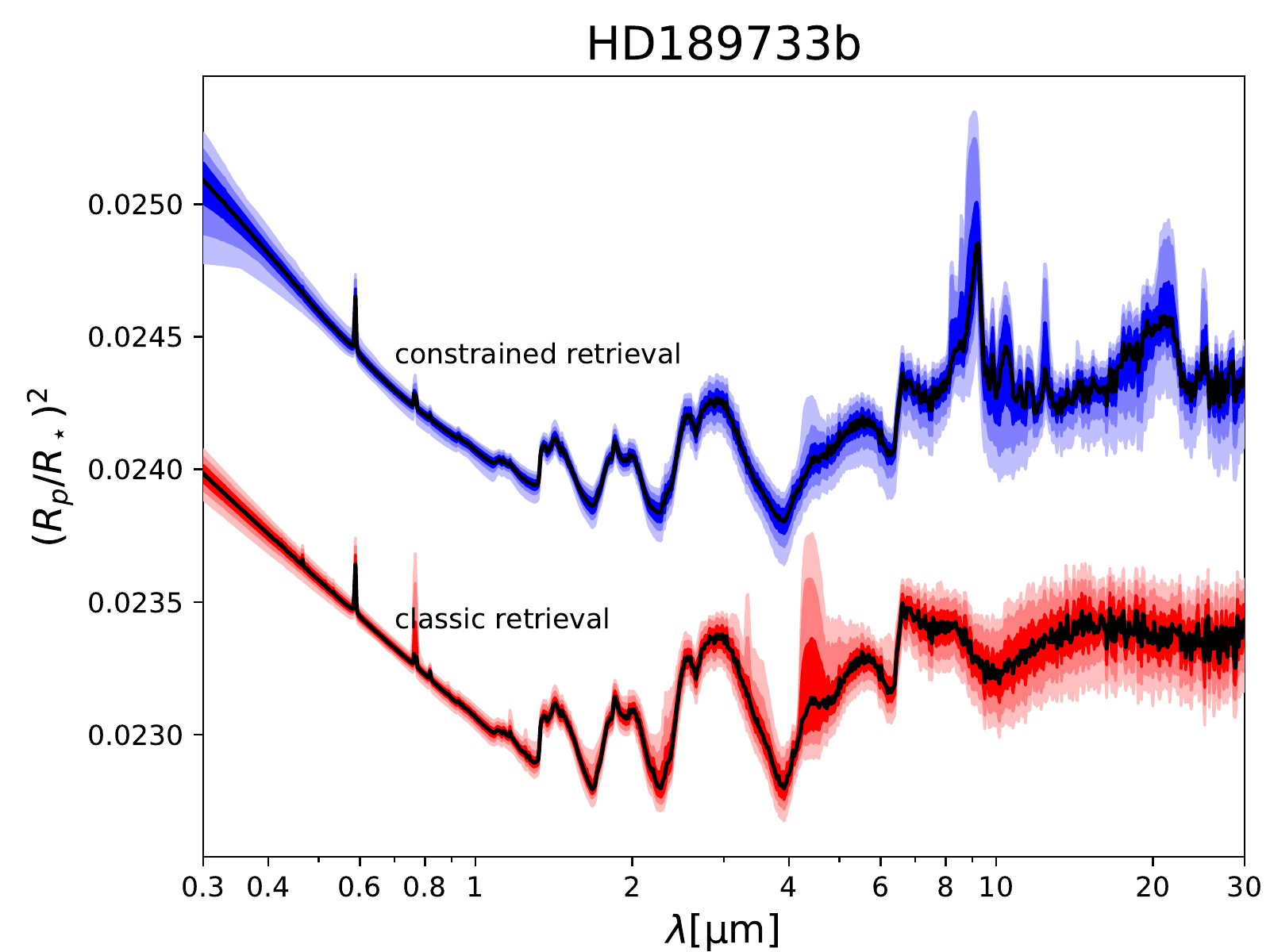}}}
\caption{Extrapolated spectrum into the JWST wavelength range for HD189733b. Here we compare the extrapolation using the classic retrieval and the model constrained retrieval. The classic retrieval result is shifted by 1000 ppm down for visual comparison.}
\label{fig:spectra compare}
\end{figure}

The retrieval results, and by this also the extrapolation to other wavelengths, depend on the underlying model assumptions. This is illustrated in Fig.~\ref{fig:spectra compare}. Here we show the extrapolation of the retrieval results for the classic retrieval and the model constrained retrieval into the JWST wavelength range. There are several noticeable differences. First, the classic retrieval does not contain information about the solid state features, and thus logically can also not predict their presence in the mid infrared. Second, it can be seen that the prediction from the model constrained retrieval are sometimes better constrained. For example, the classic retrieval has no information on the presence of some molecules like CO or CO$_2$, so their abundance can, in this context, be anything. The model constrained retrieval uses chemical information to predict the abundances of molecules of which no observational constraint can yet be made. Thus the prediction in this wavelength range is better constrained. This clearly illustrates the limitations of retrieval in general; a prediction or retrieval result is only valid within the context of the model assumptions. This holds for both the classic and the model constrained retrieval.

\section{Conclusions}
\label{sec:conclusions}

We present the ARCiS, the artful modelling code for exoplanet science. This code contains many physical concepts and can thus help us couple observations to theory. Even though chemistry and cloud formation are computed from physical concepts, the code is flexible and fast enough to be used in a retrieval framework. The code is applied to an existing dataset of 10 transmission spectra for exoplanets. From the retrieval analysis we conclude:
\begin{itemize}
\item The metallicity derived from the model constrained retrieval and the C/O ratio both do not appear to correlate with the H$_2$O abundance as derived from the classic retrieval approach using a parameterised homogeneous atmosphere.
\item We find one source that appears to have a C/O ratio significantly higher than solar (HAT-P-12b) and one with a value significantly lower than solar (WASP-39b).
\item The atmospheric diffusion strengths, which determines the cloud formation efficiency in our model, correlates strongly with the spectral index, $\Delta Z_{UB-LM}/H_{eq}$, for the sources within our sample.
\item We predict the presence of mineral features in the JWST spectra of some of the sources. These are the sources which observationally have a spectral index $\Delta Z_{UB-LM}/H_{eq}>3$. HD189733b is the most promising of these.
\end{itemize}

Overall the artful modelling approach as presented here provides a promising pathway to comparing theory and observation in a robust manner. In the future we will investigate incorporating disequilibrium chemistry and establish a physical link between the elemental abundances and the planet formation process.

\begin{acknowledgements}
The research leading to these results has received funding from the European Union's Horizon 2020 Research and Innovation Programme, under Grant Agreement 776403. We would like to express our gratitude to the referee, Ryan MacDonald, for an excellent review of the paper that significantly improved the quality and clarity.
\end{acknowledgements}

\emph{Data availability:} All data used and created in this paper is available online under DOI \href{https://doi.org/10.5281/zenodo.3935296}{10.5281/zenodo.3935296}.

\bibliographystyle{aa}
\bibliography{biblio}

\appendix

\clearpage
\section{Pressure temperature profiles for all retrieval sources}
\label{app:PT}

Here we plot the pressure temperature profiles obtained from the classic and the model constrained retrievals.

\begin{figure*}[!tp]
\centerline{\resizebox{\hsize}{!}{\includegraphics{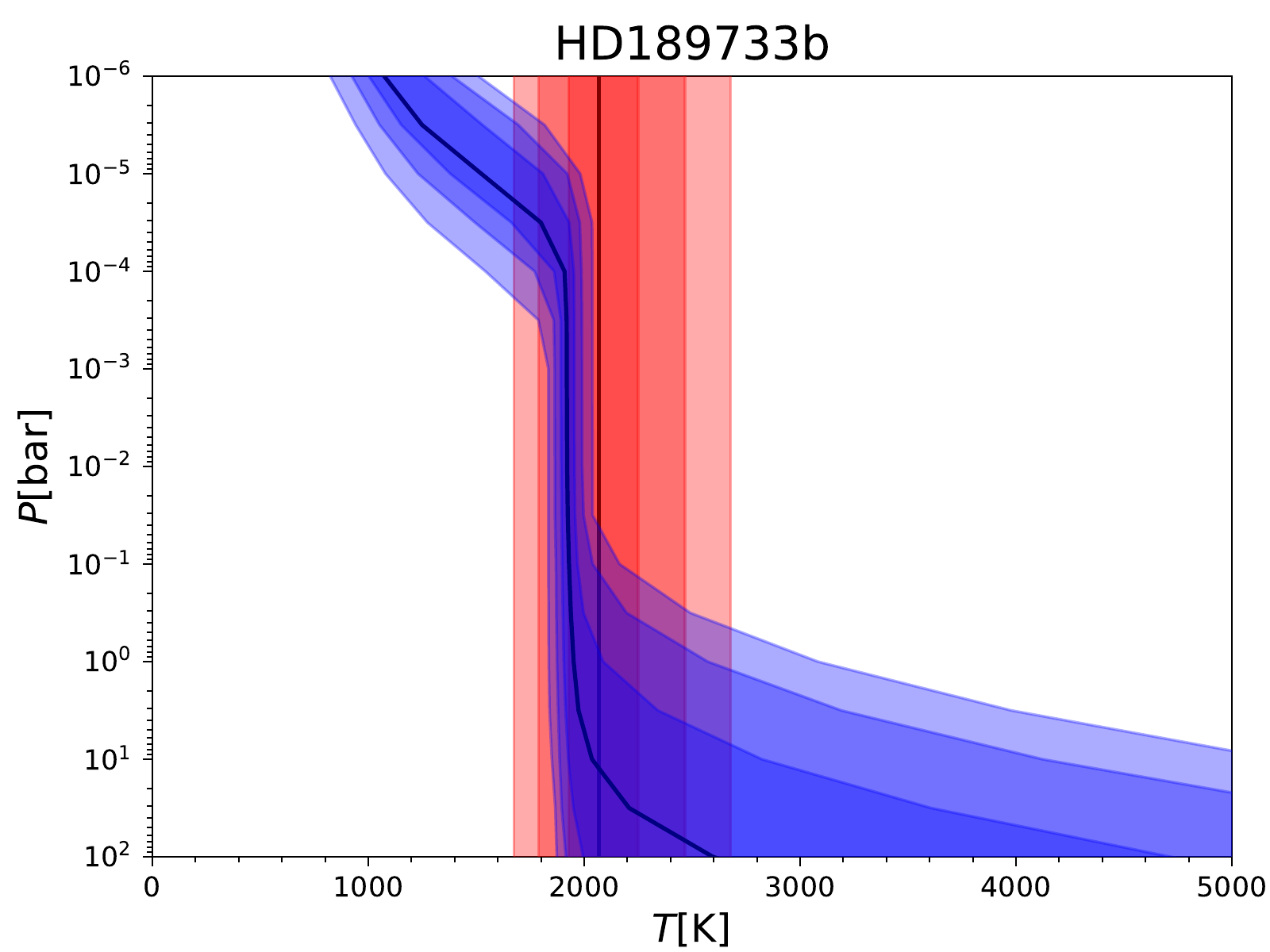}\includegraphics{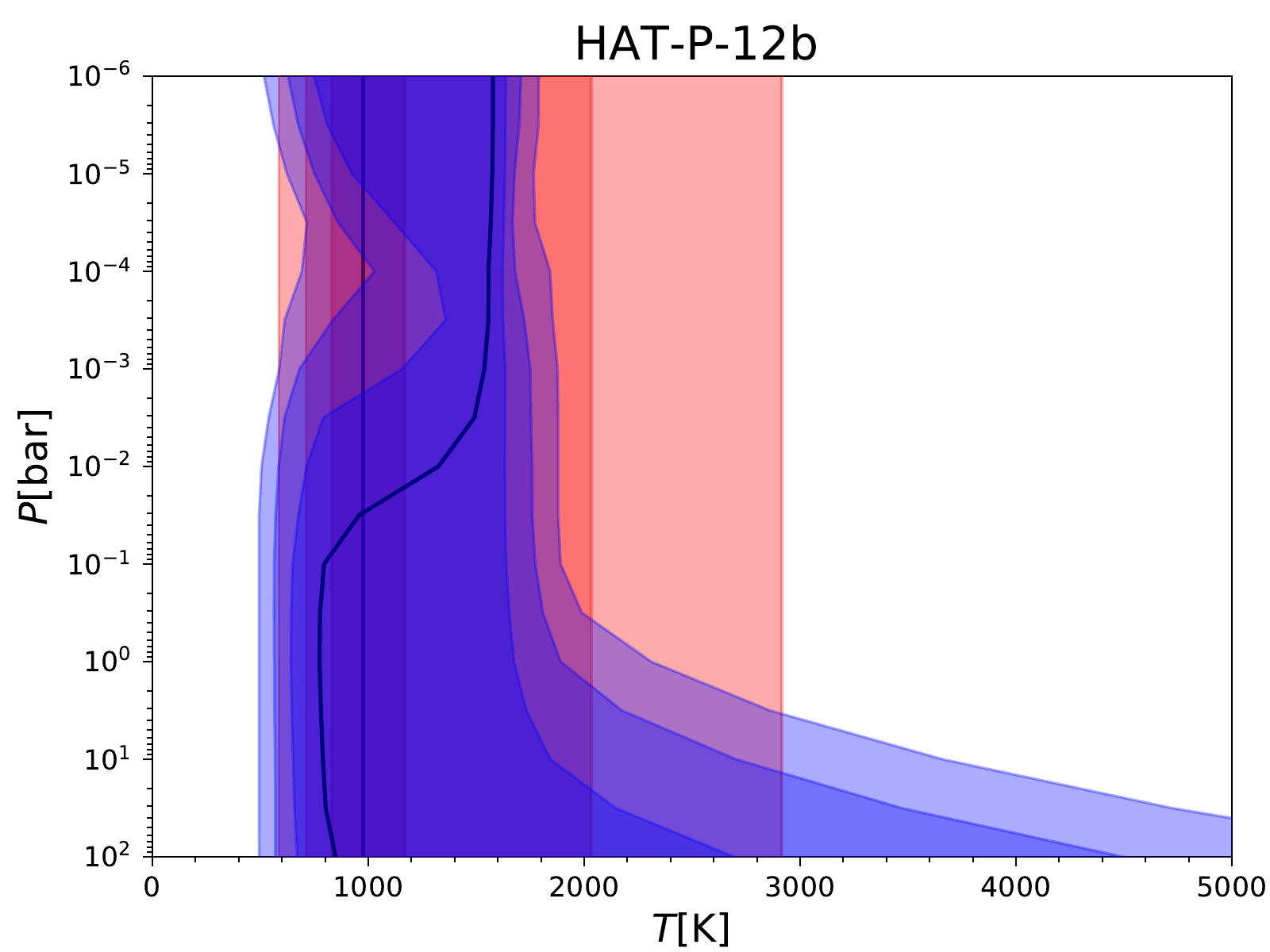}}}
\centerline{\resizebox{\hsize}{!}{\includegraphics{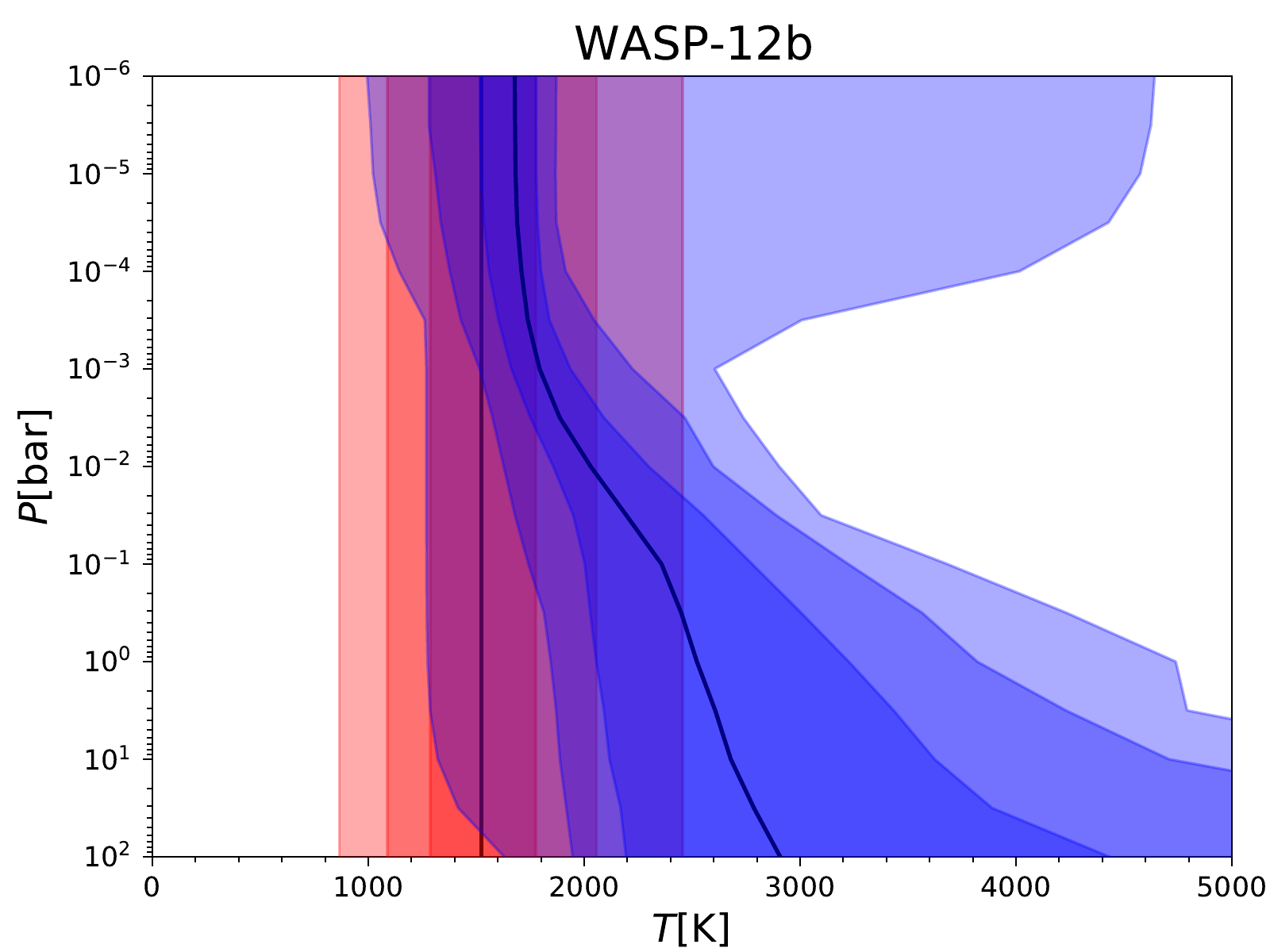}\includegraphics{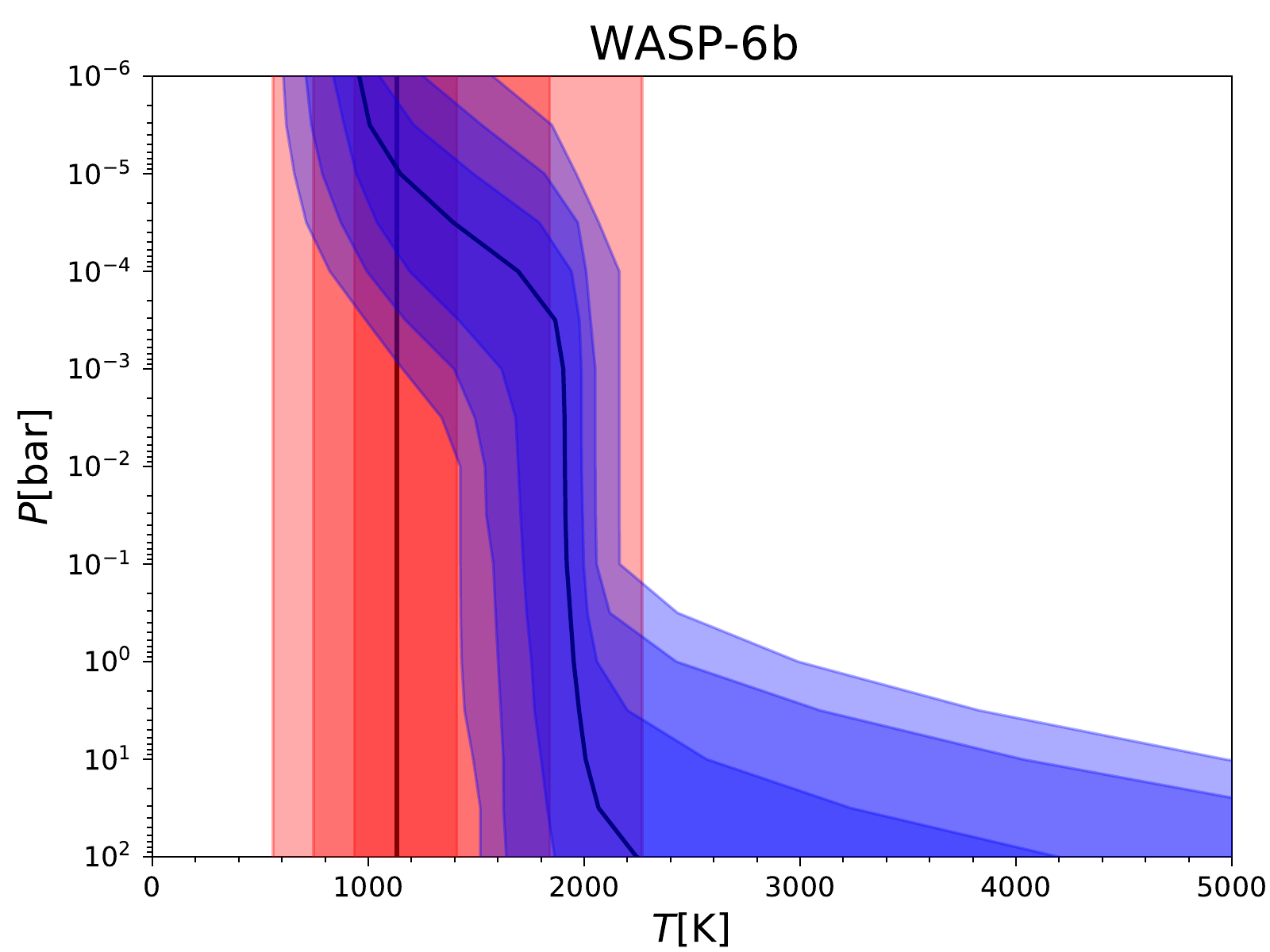}}}
\centerline{\resizebox{\hsize}{!}{\includegraphics{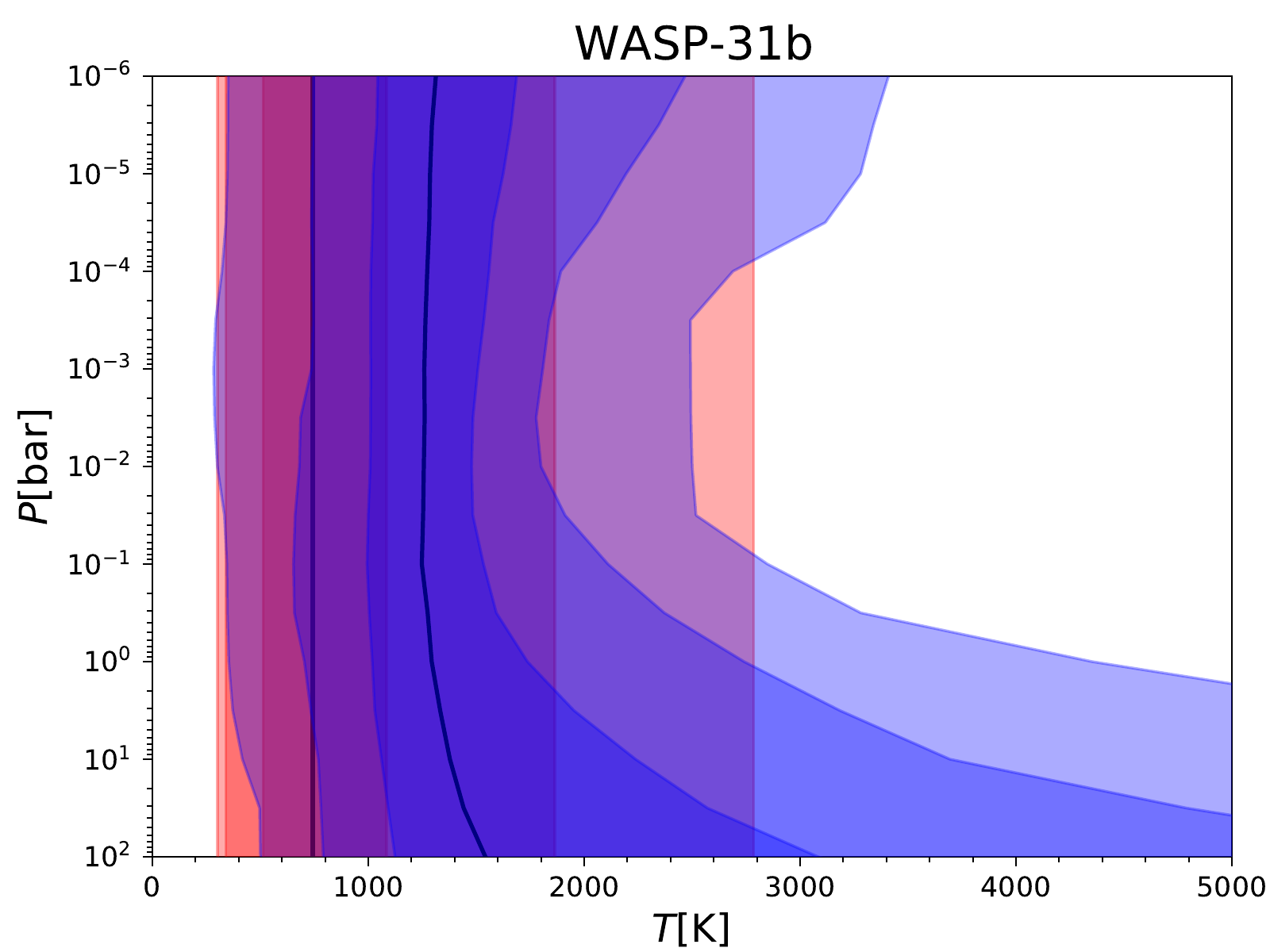}\includegraphics{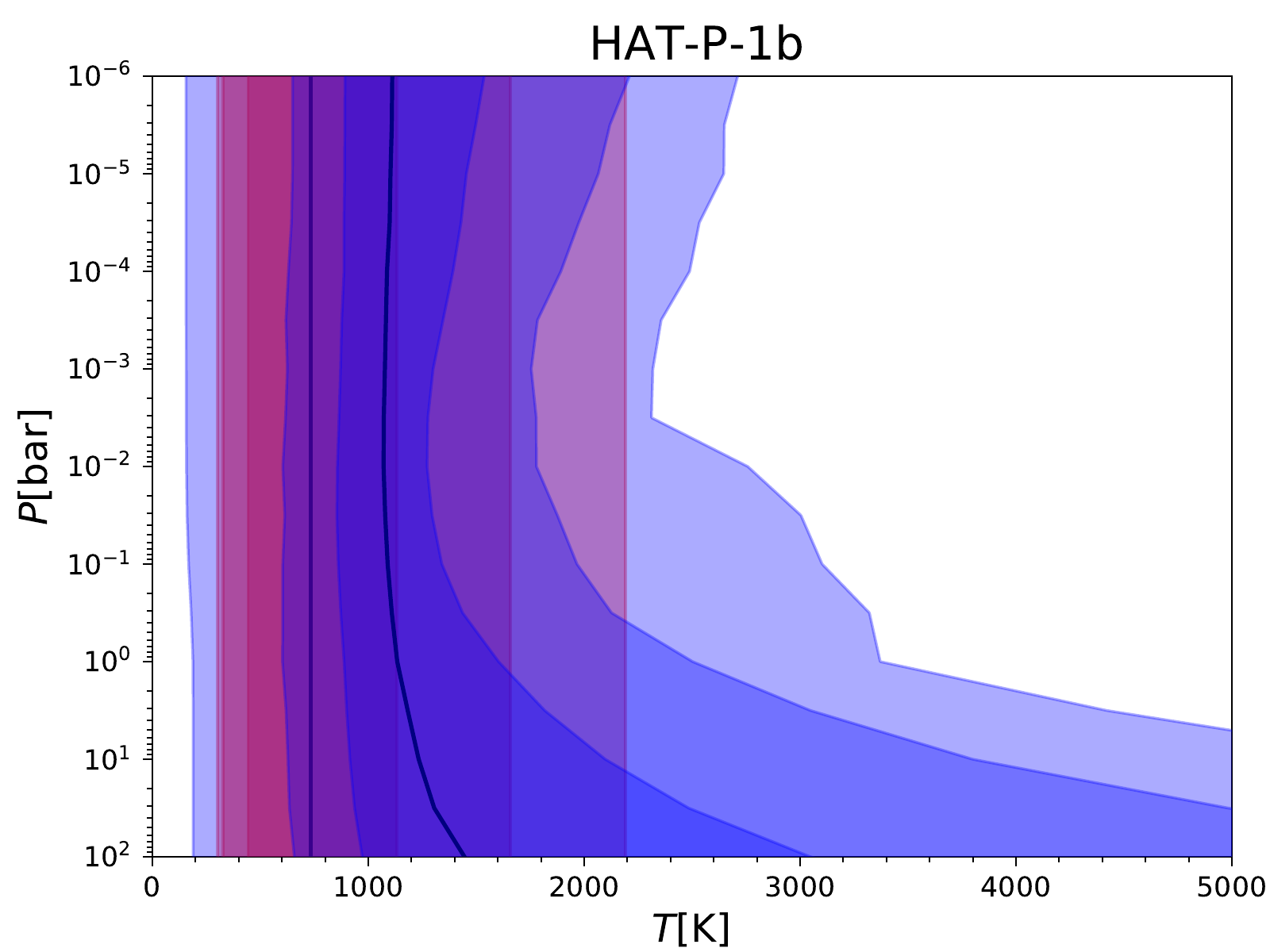}}}
\caption{Pressure temperature structures for all 10 planets considered. The blue area indicates the 1, 2 and 3 sigma limits from the model constrained retrieval, while the red areas indicate the structure for the classic retrieval.}
\label{fig:PT}
\end{figure*}
\begin{figure*}[!tp]
\ContinuedFloat
\centerline{\resizebox{\hsize}{!}{\includegraphics{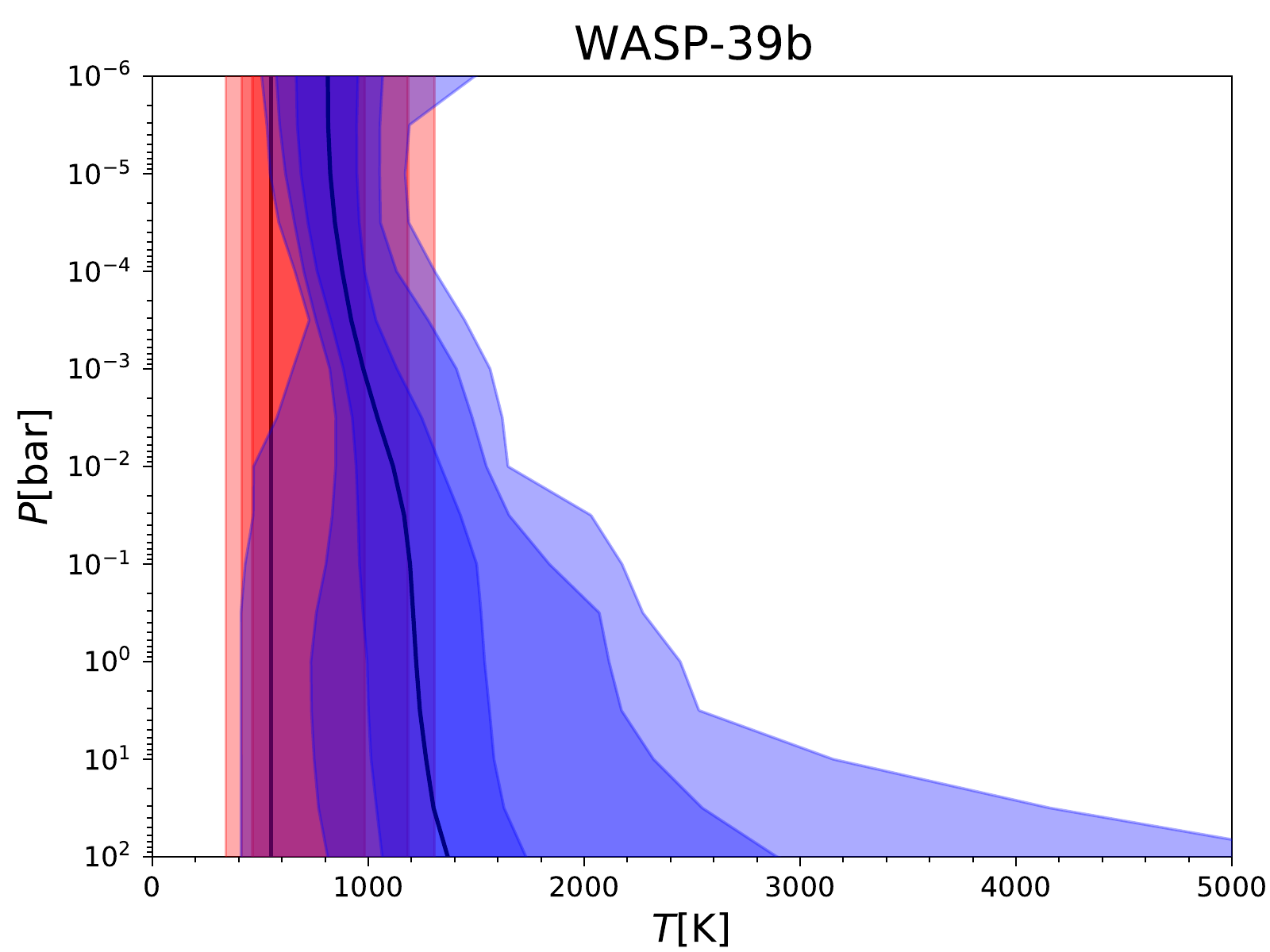}\includegraphics{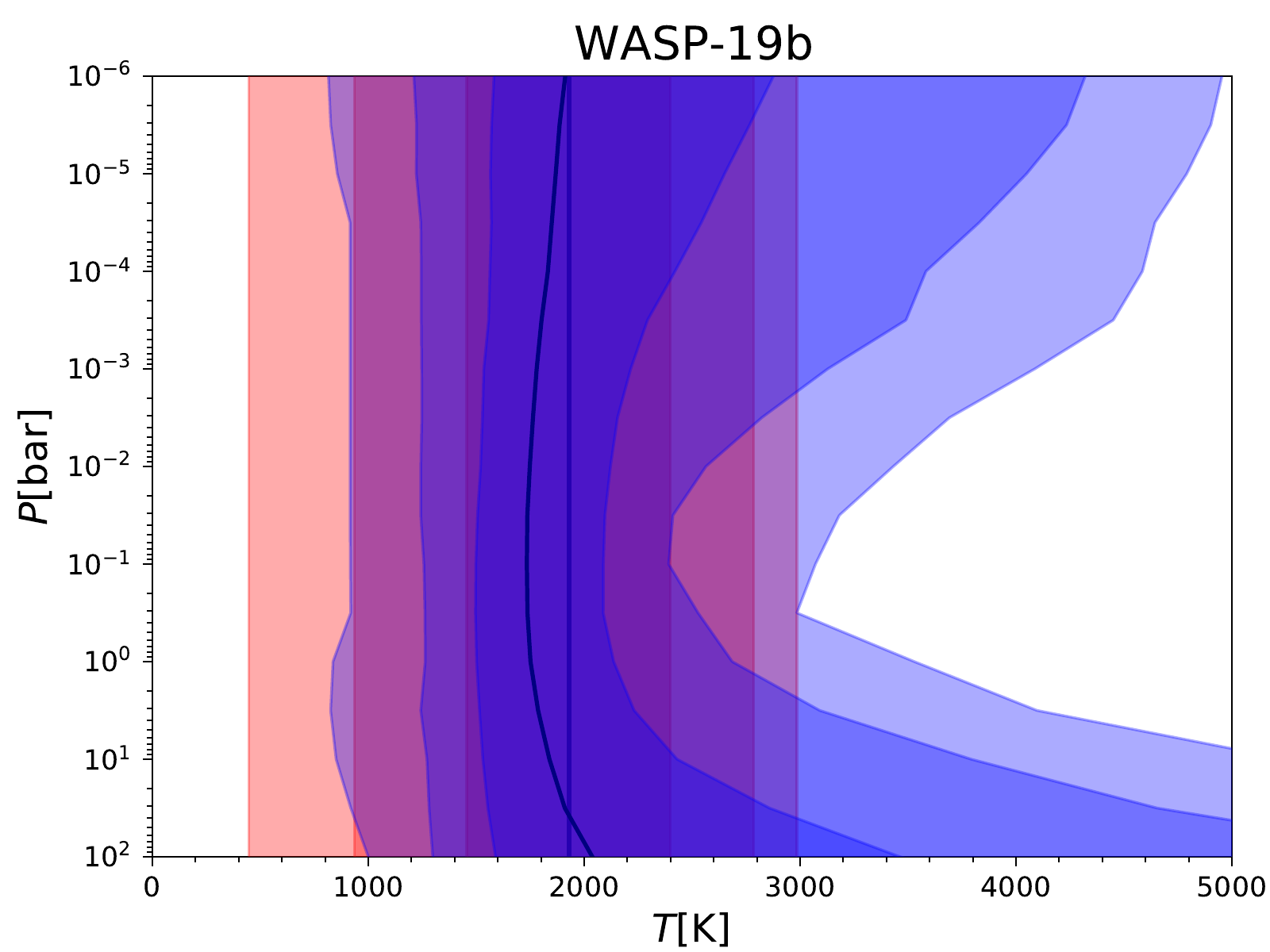}}}
\centerline{\resizebox{\hsize}{!}{\includegraphics{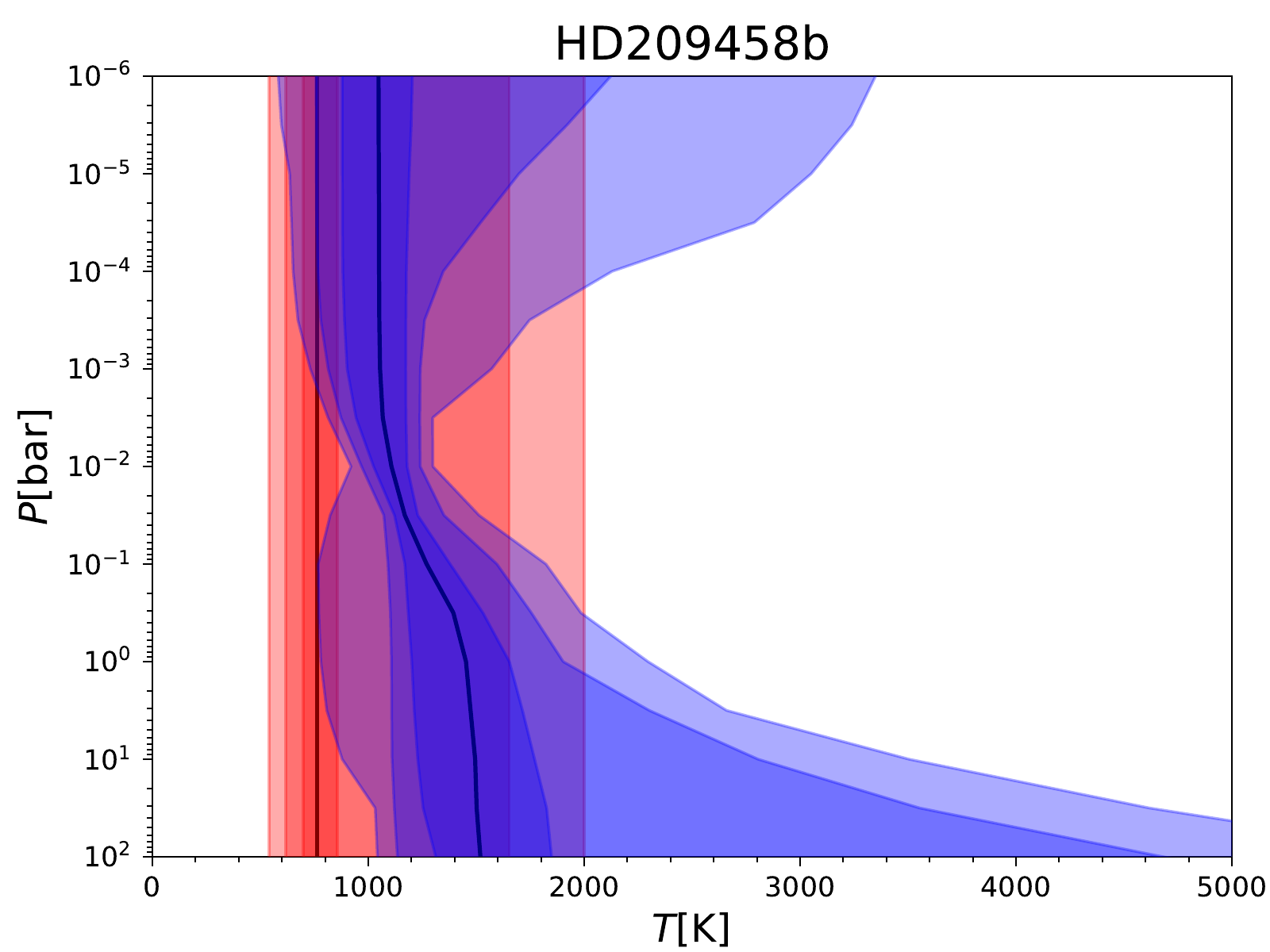}\includegraphics{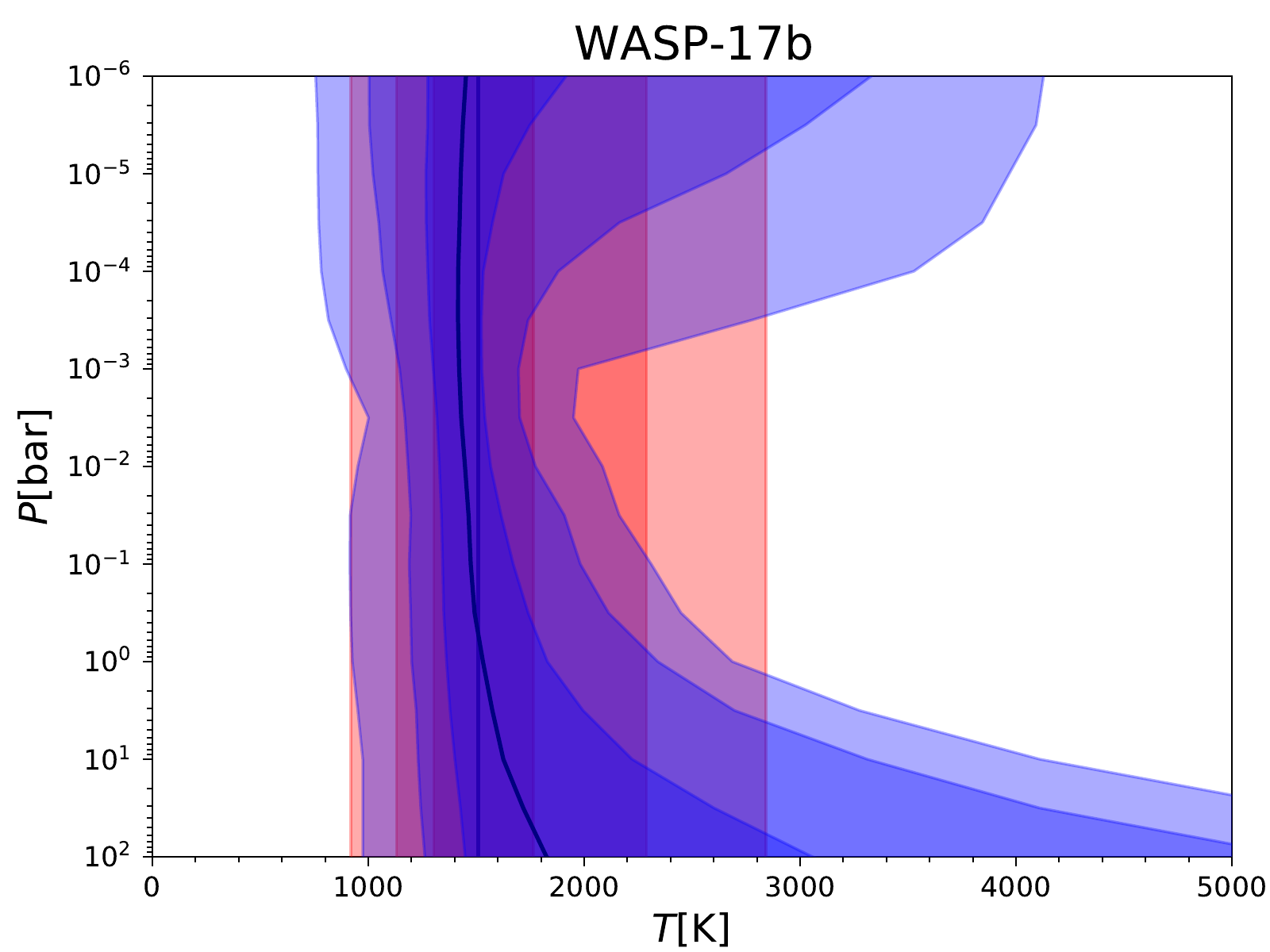}}}
\caption{Continued.}
\label{fig:PT}
\end{figure*}

\clearpage
\section{Posterior distributions}
\label{app:cornerplots}

\begin{figure*}[!tp]
\centerline{\resizebox{\hsize}{!}{\includegraphics{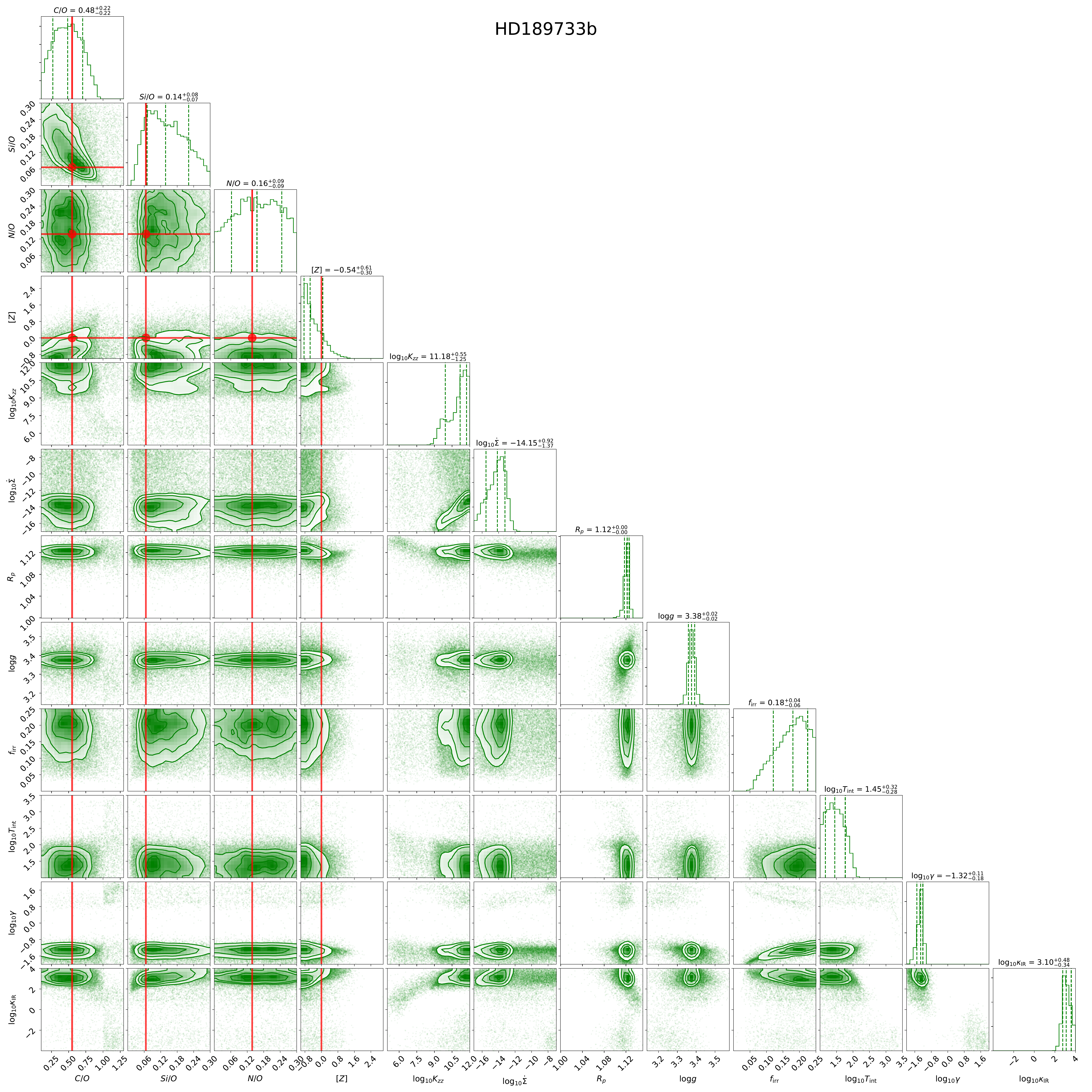}}}
\caption{Corner plots for the constrained retrieval for all 10 planets considered. The red lines indicate the solar elemental ratios.}
\label{fig:corner}
\end{figure*}
\begin{figure*}[!tp]
\ContinuedFloat
\centerline{\resizebox{\hsize}{!}{\includegraphics{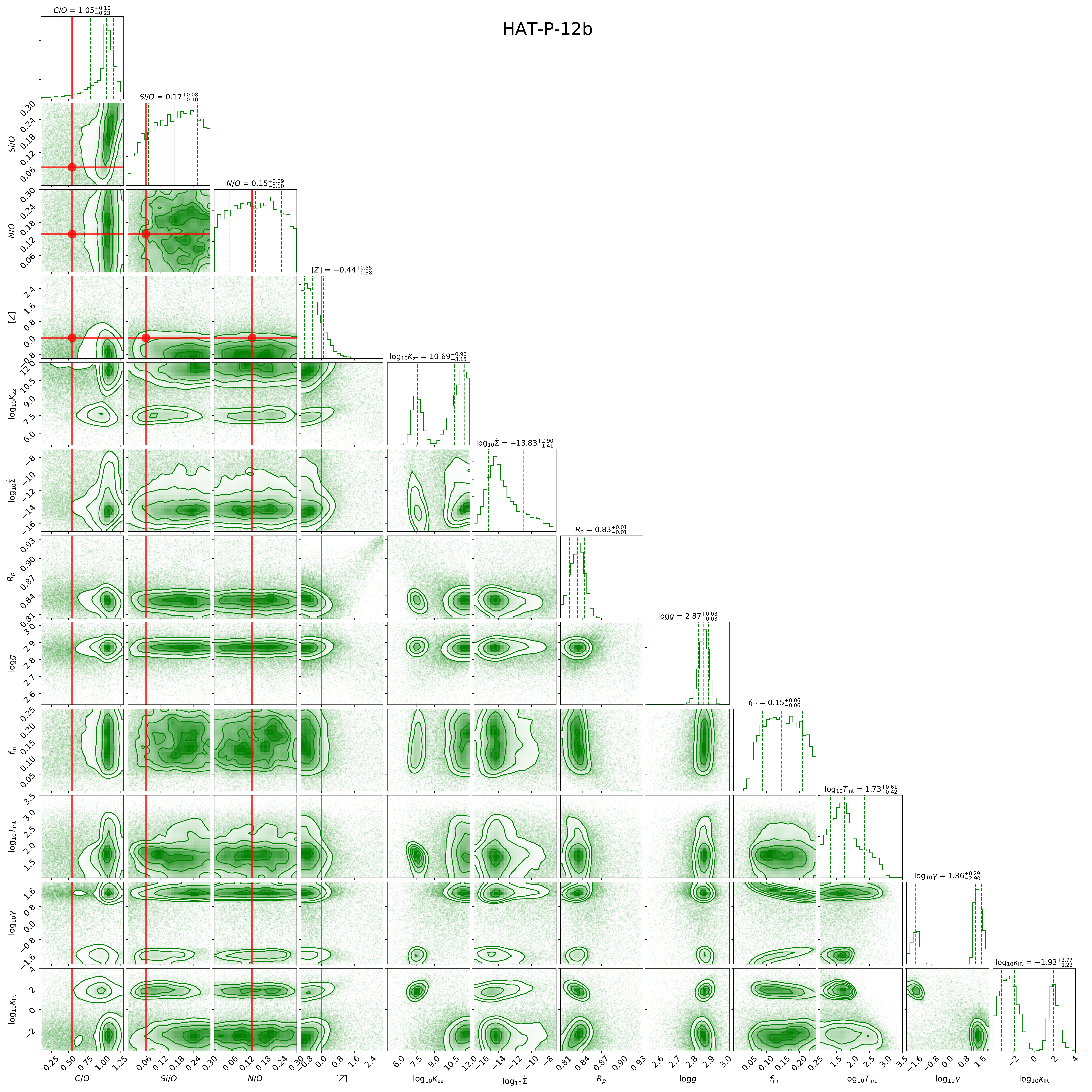}}}
\caption{Continued.}
\label{fig:corner}
\end{figure*}
\begin{figure*}[!tp]
\ContinuedFloat
\centerline{\resizebox{\hsize}{!}{\includegraphics{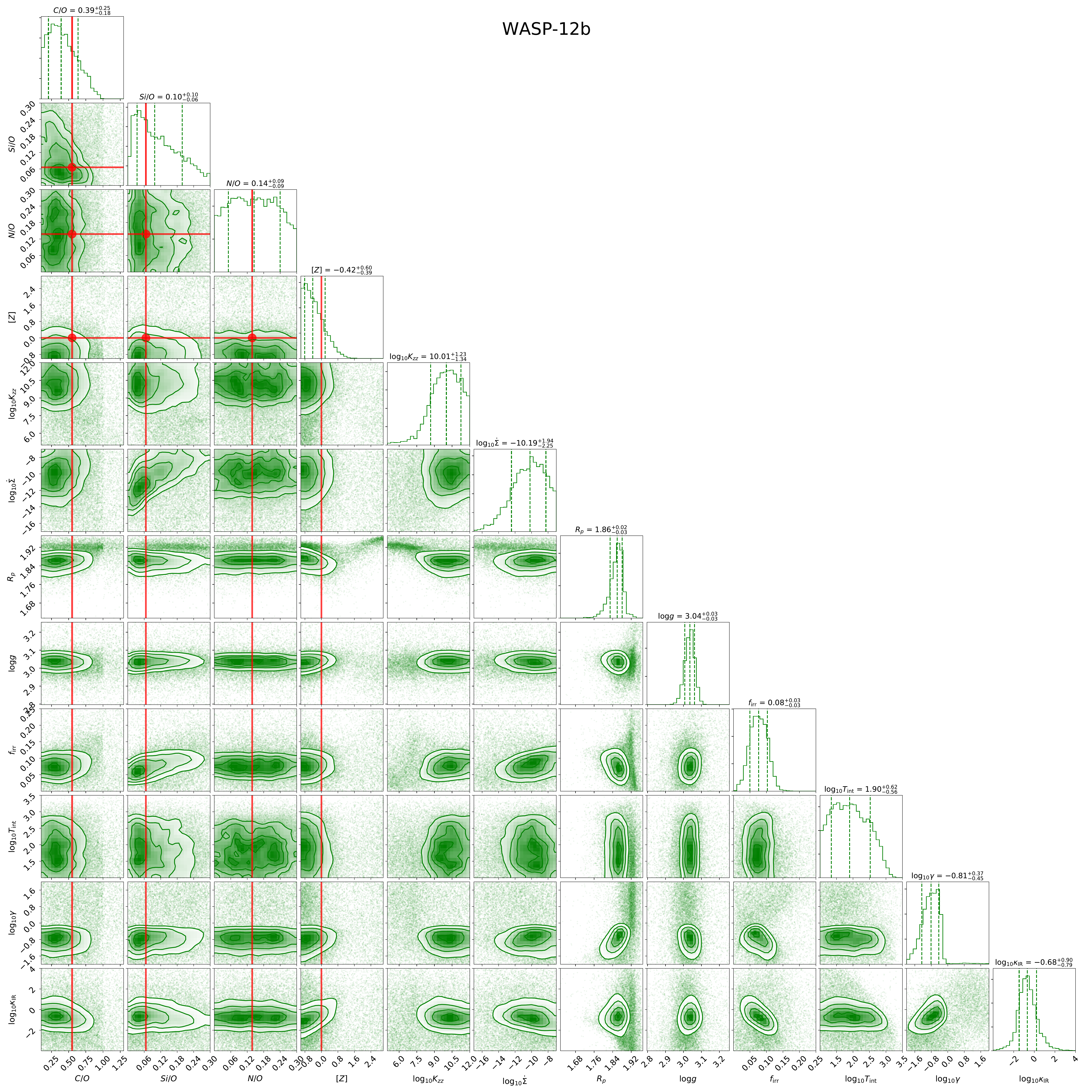}}}
\caption{Continued.}
\label{fig:corner}
\end{figure*}
\begin{figure*}[!tp]
\ContinuedFloat
\centerline{\resizebox{\hsize}{!}{\includegraphics{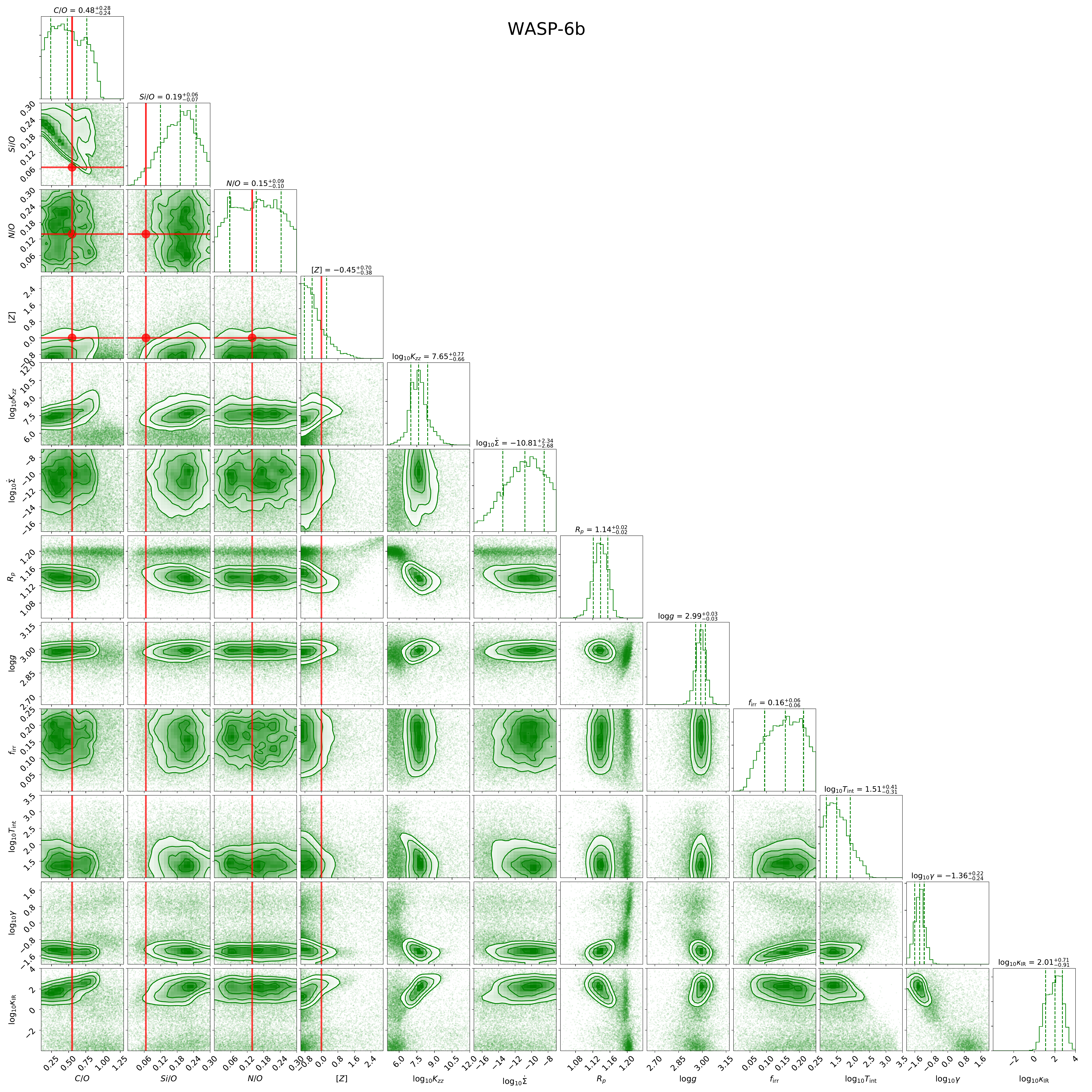}}}
\caption{Continued.}
\label{fig:corner}
\end{figure*}
\begin{figure*}[!tp]
\ContinuedFloat
\centerline{\resizebox{\hsize}{!}{\includegraphics{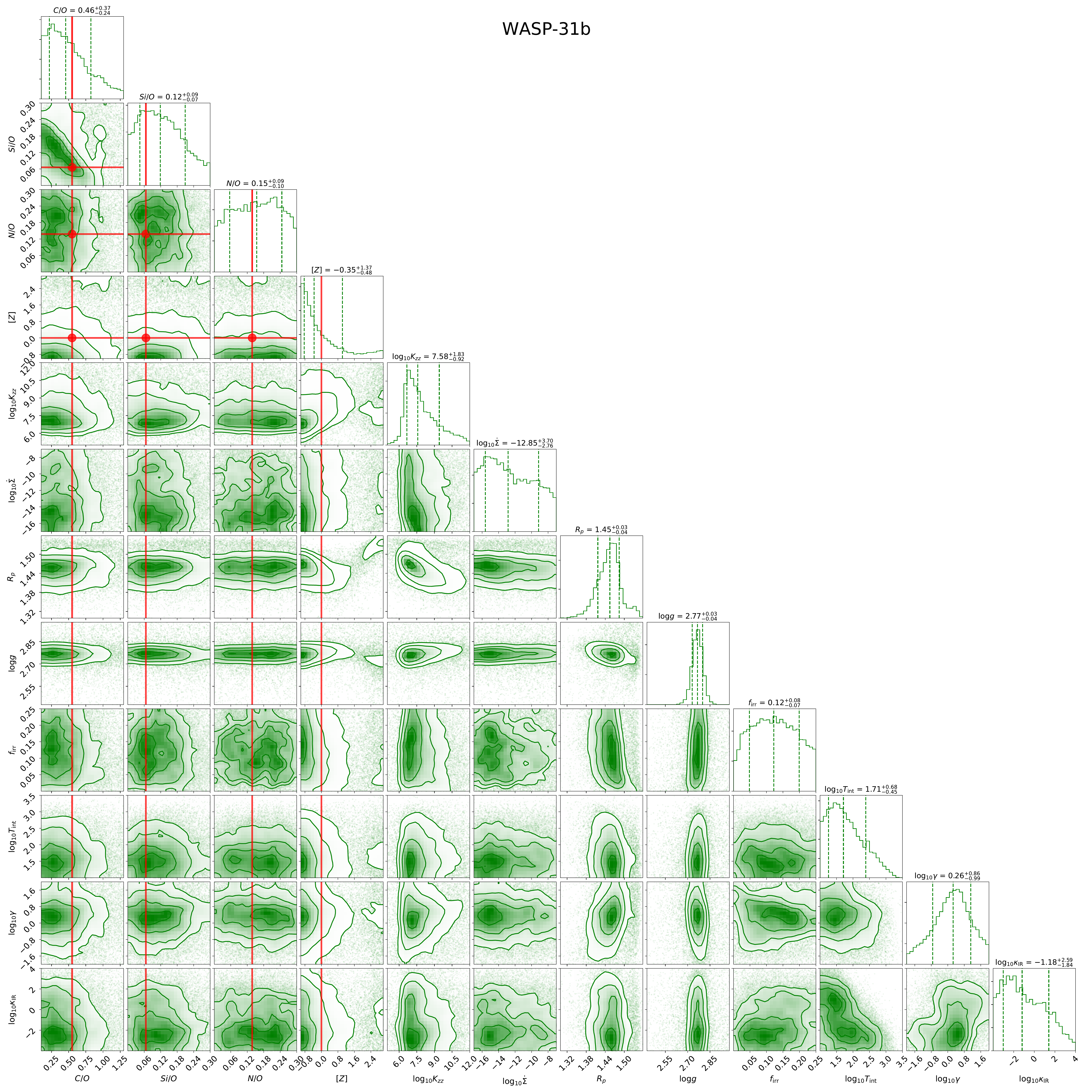}}}
\caption{Continued.}
\label{fig:corner}
\end{figure*}
\begin{figure*}[!tp]
\ContinuedFloat
\centerline{\resizebox{\hsize}{!}{\includegraphics{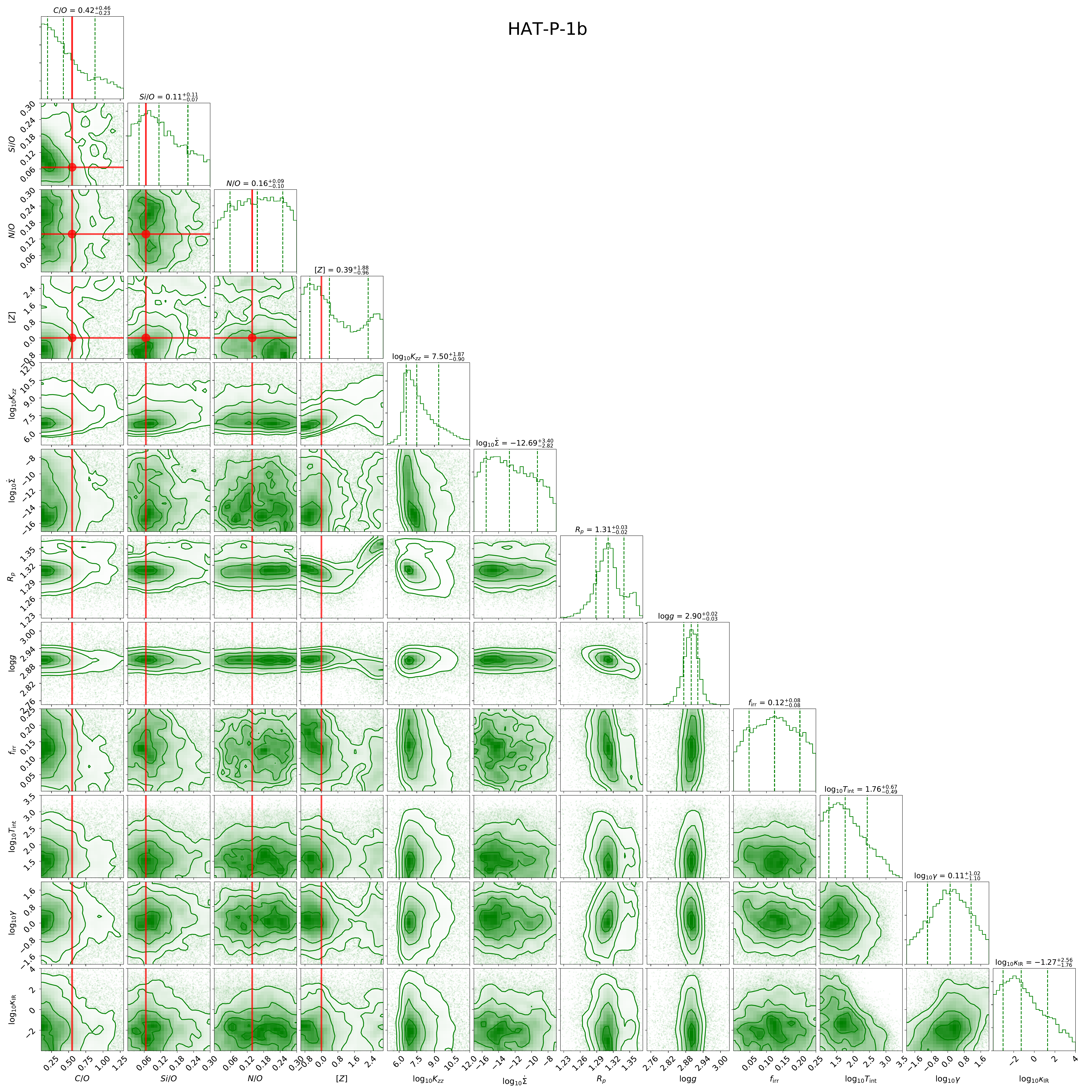}}}
\caption{Continued.}
\label{fig:corner}
\end{figure*}
\begin{figure*}[!tp]
\ContinuedFloat
\centerline{\resizebox{\hsize}{!}{\includegraphics{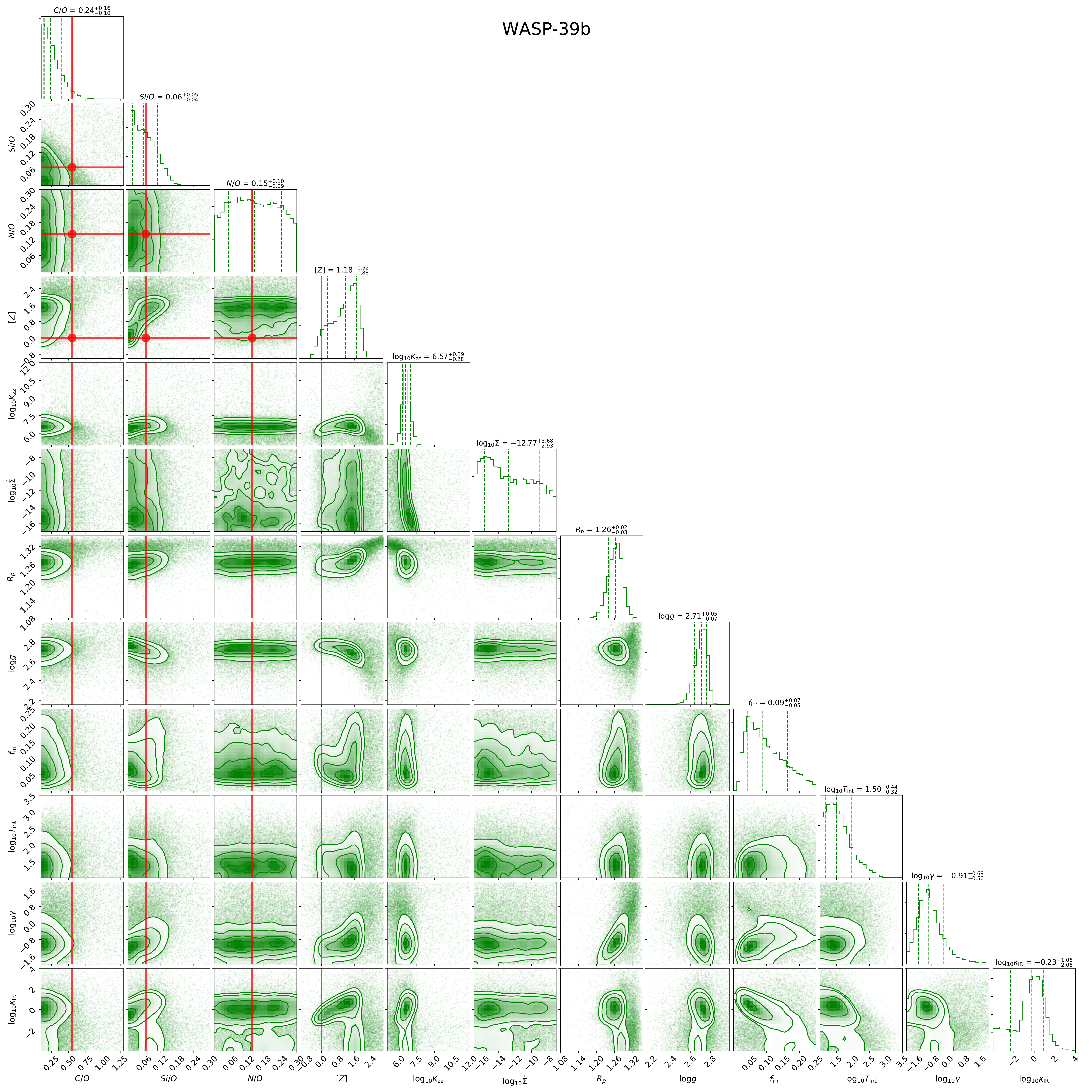}}}
\caption{Continued.}
\label{fig:corner}
\end{figure*}
\begin{figure*}[!tp]
\ContinuedFloat
\centerline{\resizebox{\hsize}{!}{\includegraphics{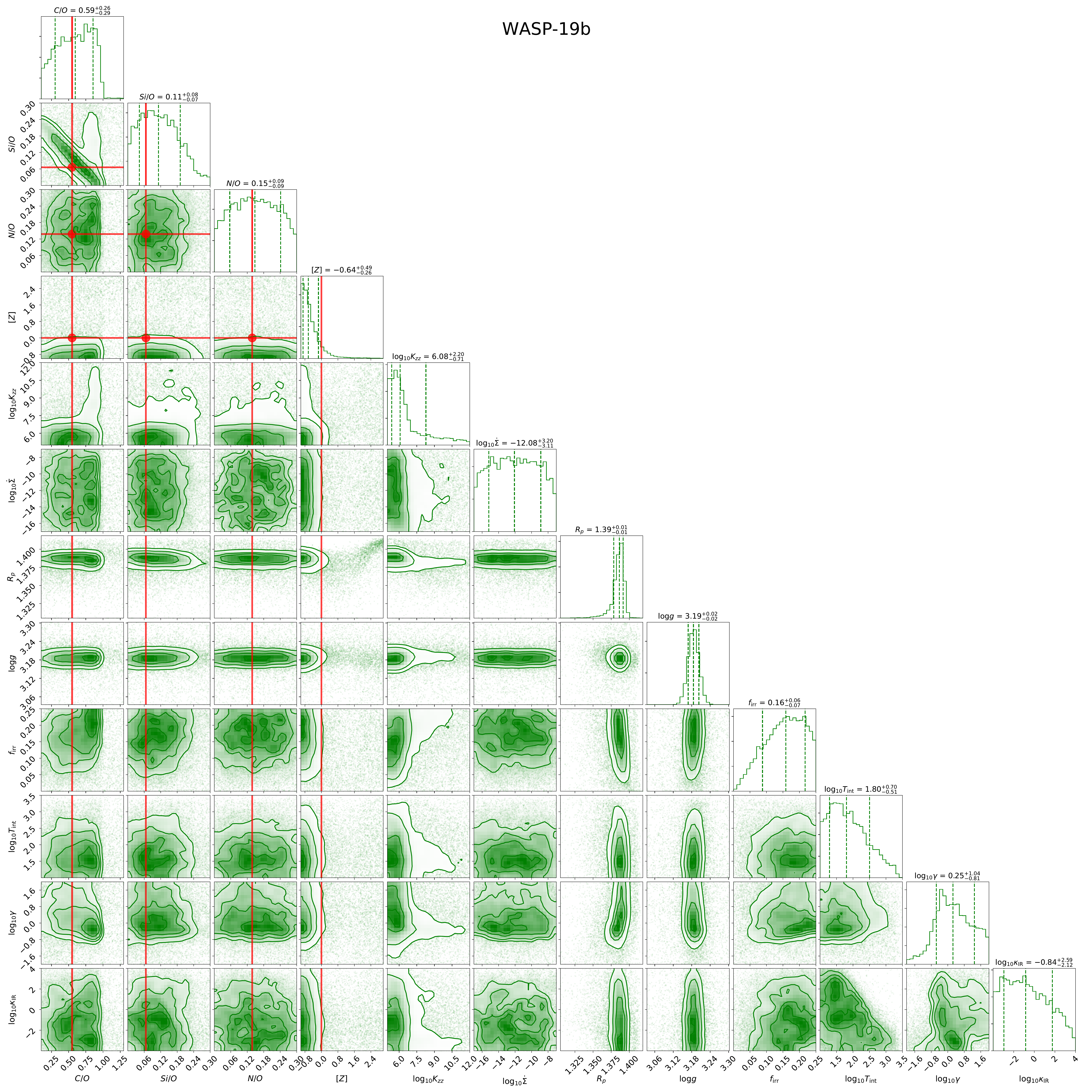}}}
\caption{Continued.}
\label{fig:corner}
\end{figure*}
\begin{figure*}[!tp]
\ContinuedFloat
\centerline{\resizebox{\hsize}{!}{\includegraphics{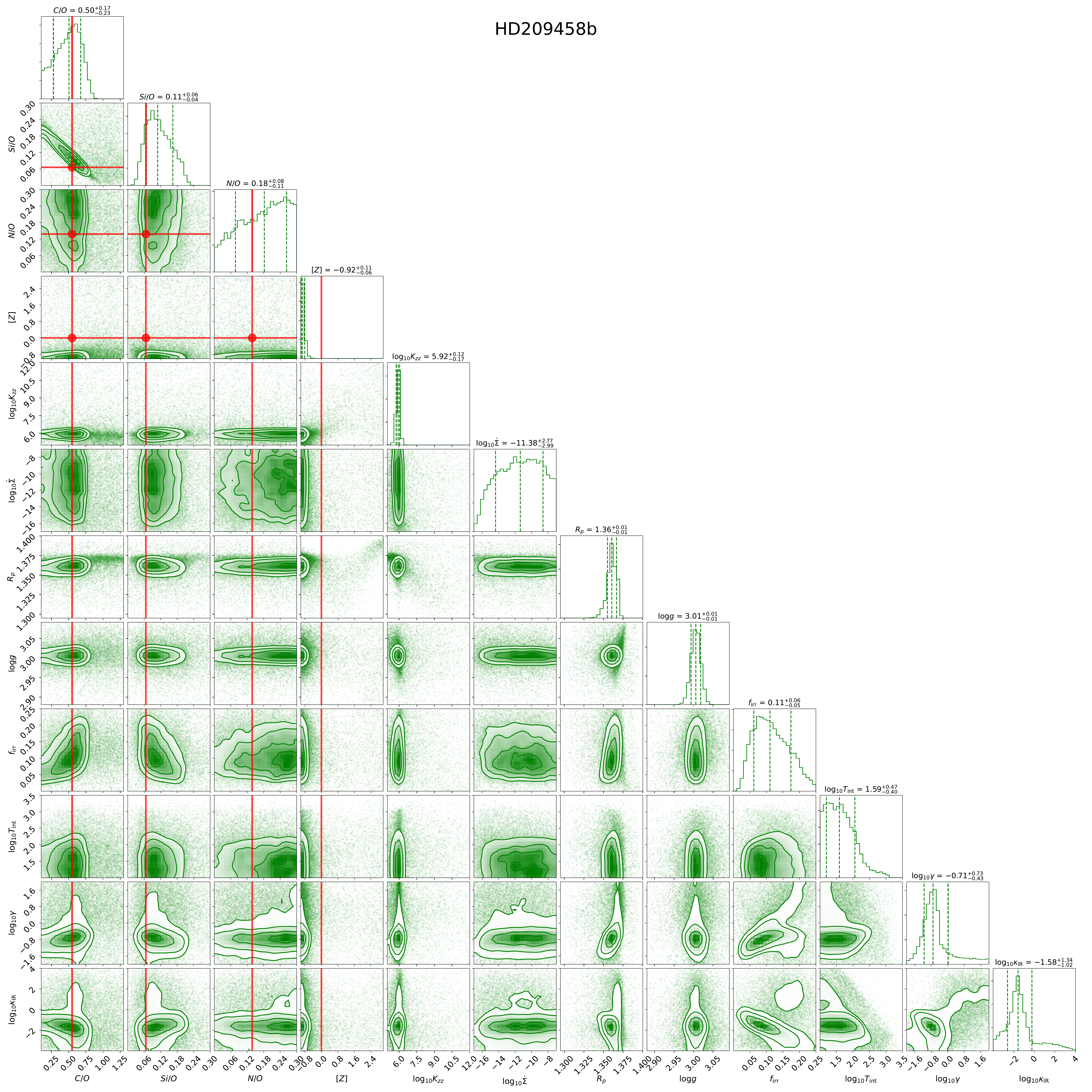}}}
\caption{Continued.}
\label{fig:corner}
\end{figure*}
\begin{figure*}[!tp]
\ContinuedFloat
\centerline{\resizebox{\hsize}{!}{\includegraphics{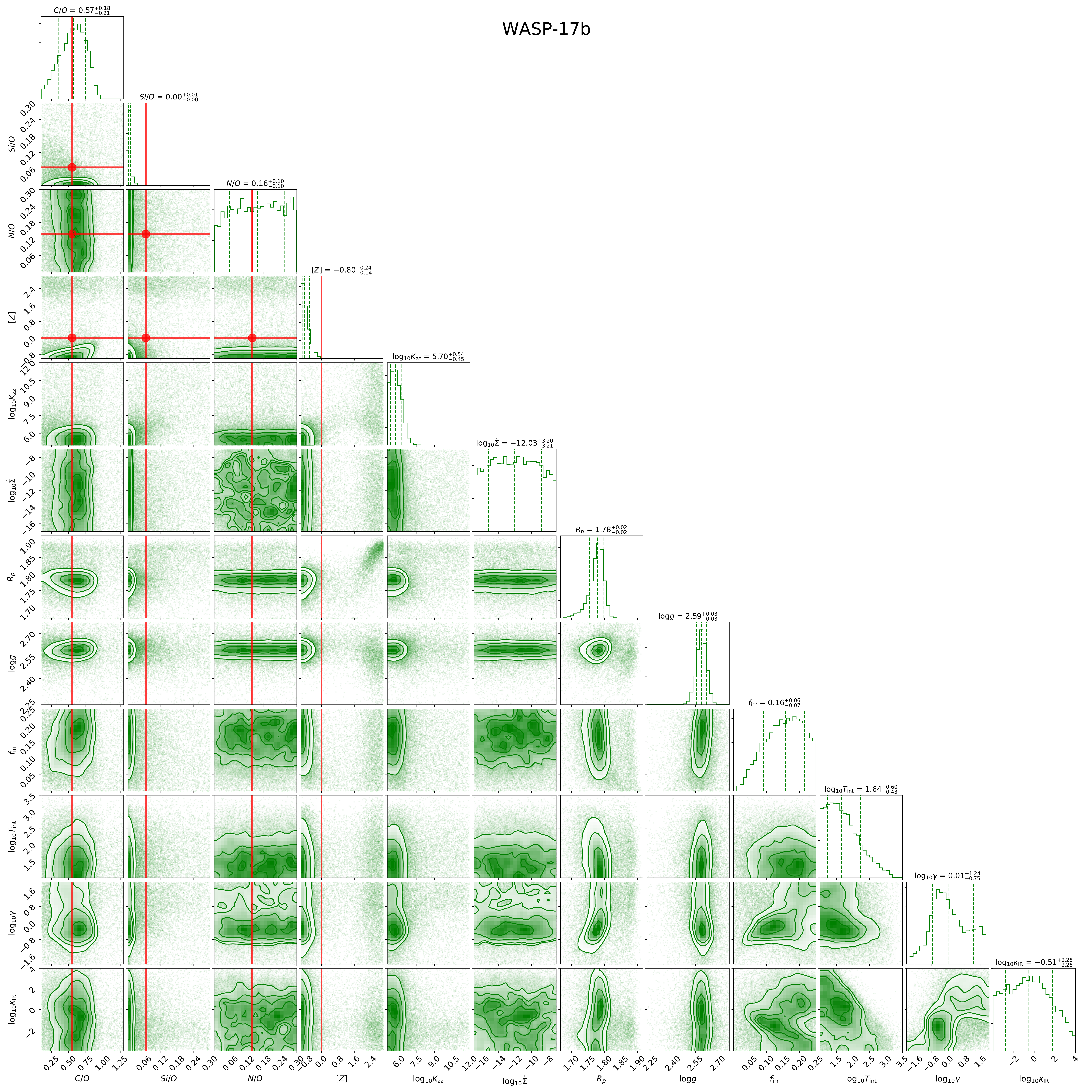}}}
\caption{Continued.}
\label{fig:corner}
\end{figure*}

\begin{figure*}[!tp]
\centerline{\resizebox{\hsize}{!}{\includegraphics{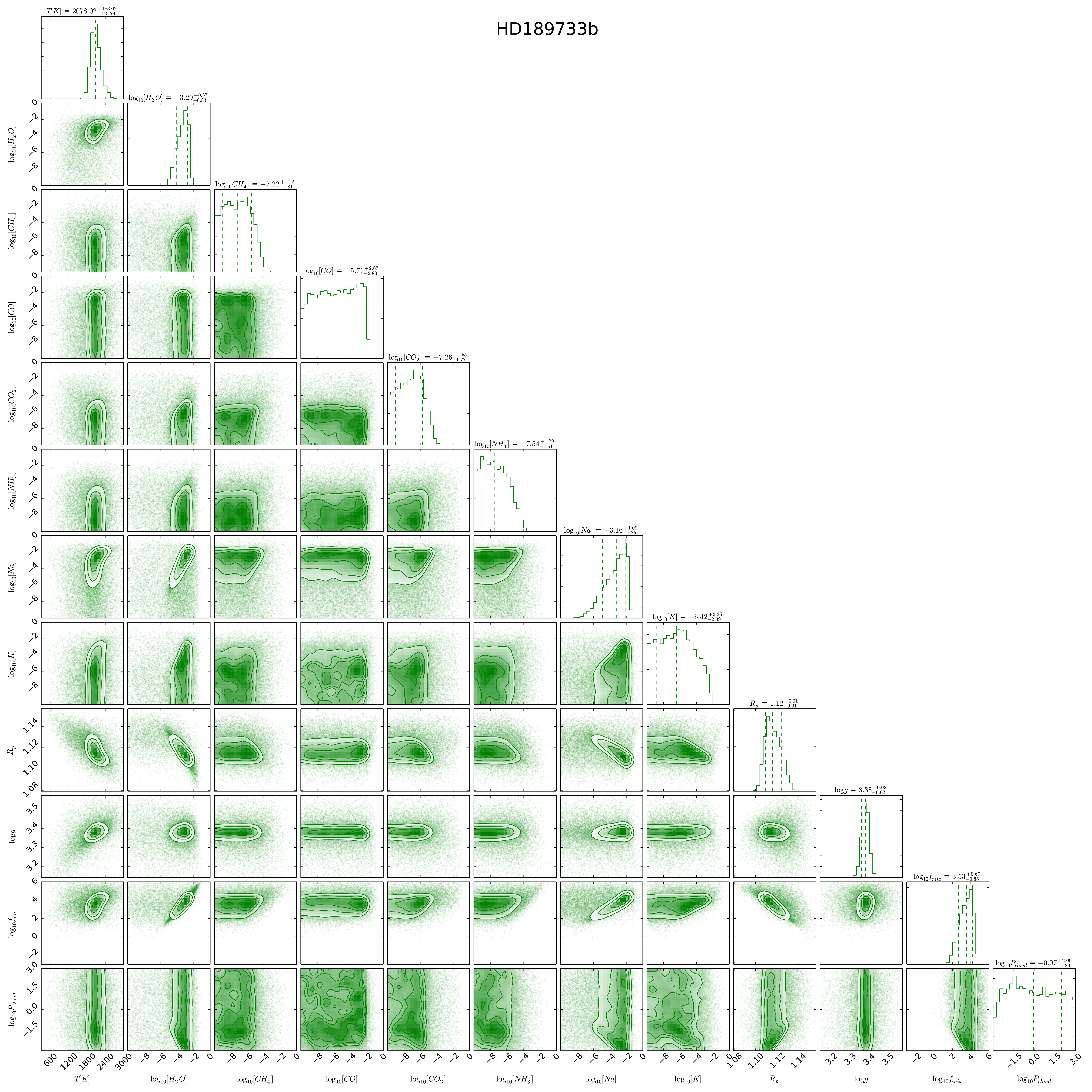}}}
\caption{Corner plots for the classic retrieval for all 10 planets considered.}
\label{fig:cornerclassic}
\end{figure*}
\begin{figure*}[!tp]
\ContinuedFloat
\centerline{\resizebox{\hsize}{!}{\includegraphics{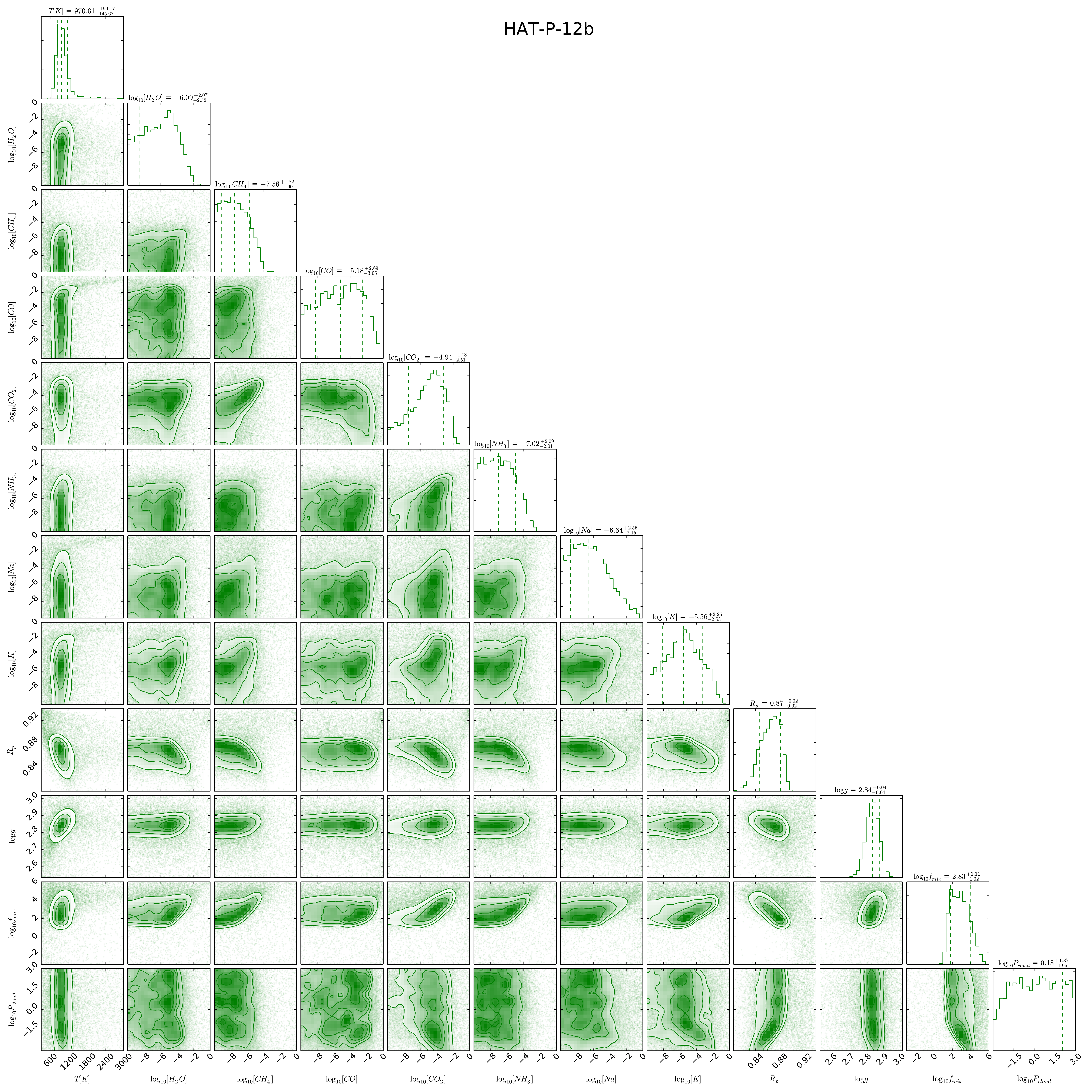}}}
\caption{Continued.}
\label{fig:cornerclassic}
\end{figure*}
\begin{figure*}[!tp]
\ContinuedFloat
\centerline{\resizebox{\hsize}{!}{\includegraphics{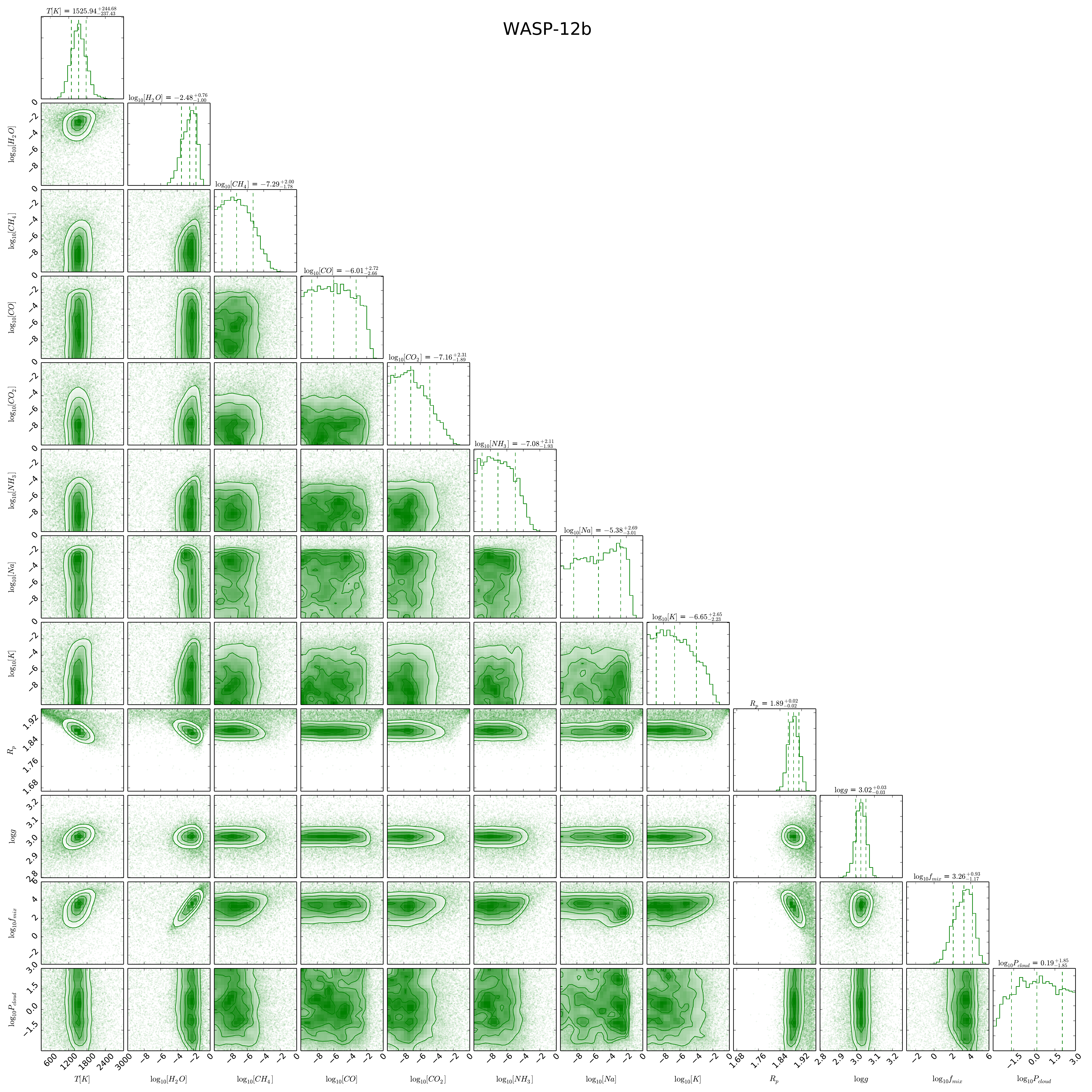}}}
\caption{Continued.}
\label{fig:cornerclassic}
\end{figure*}
\begin{figure*}[!tp]
\ContinuedFloat
\centerline{\resizebox{\hsize}{!}{\includegraphics{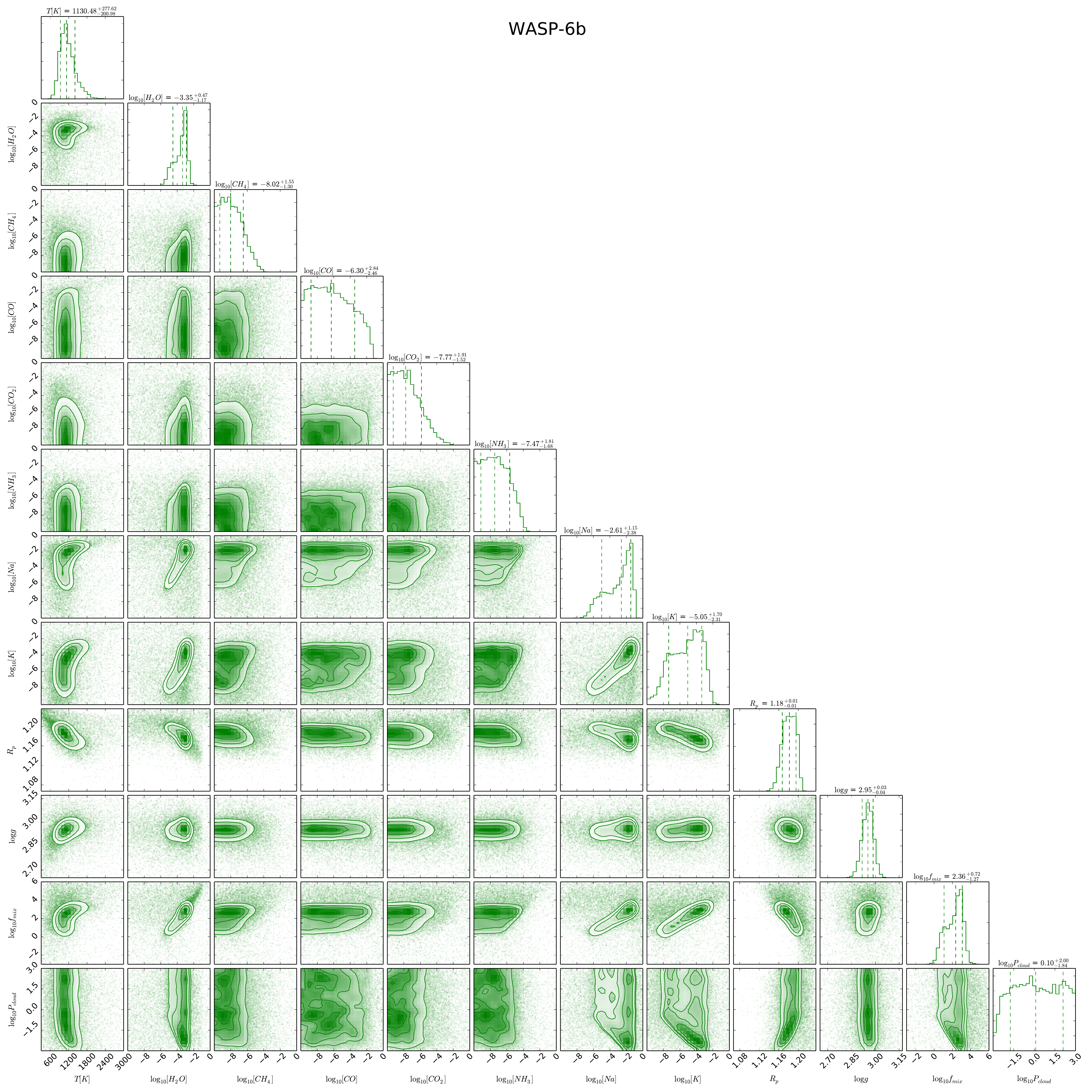}}}
\caption{Continued.}
\label{fig:cornerclassic}
\end{figure*}
\begin{figure*}[!tp]
\ContinuedFloat
\centerline{\resizebox{\hsize}{!}{\includegraphics{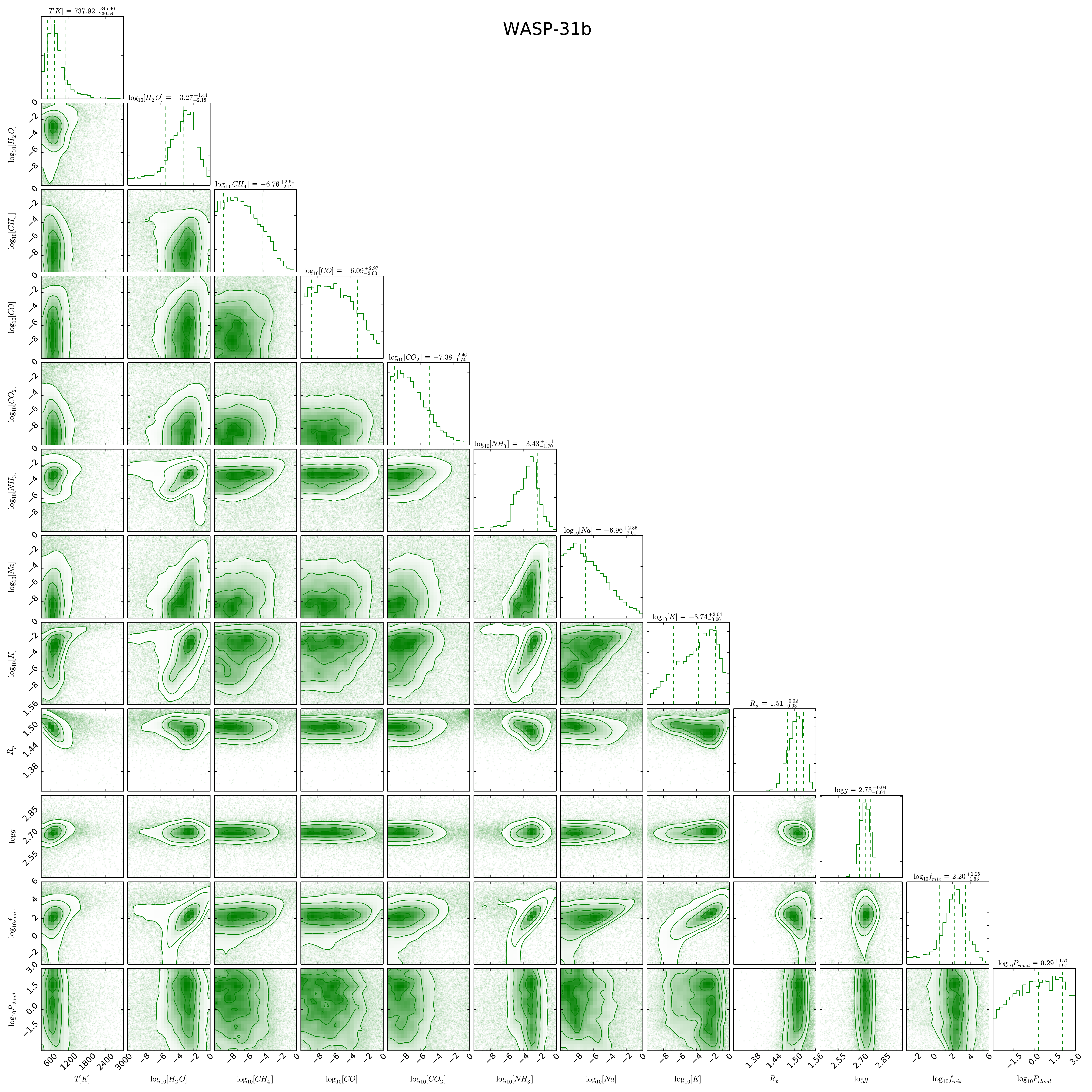}}}
\caption{Continued.}
\label{fig:cornerclassic}
\end{figure*}
\begin{figure*}[!tp]
\ContinuedFloat
\centerline{\resizebox{\hsize}{!}{\includegraphics{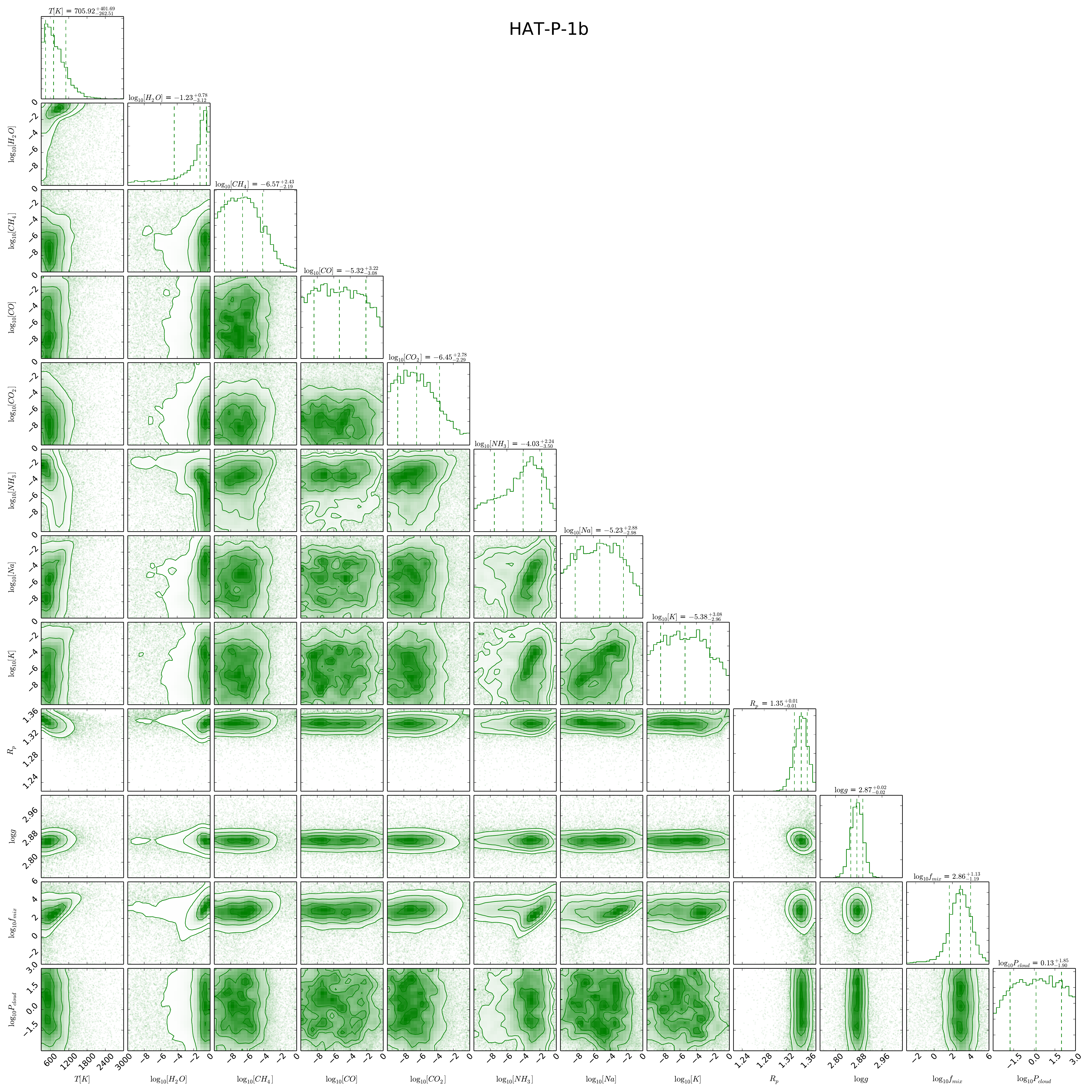}}}
\caption{Continued.}
\label{fig:cornerclassic}
\end{figure*}
\begin{figure*}[!tp]
\ContinuedFloat
\centerline{\resizebox{\hsize}{!}{\includegraphics{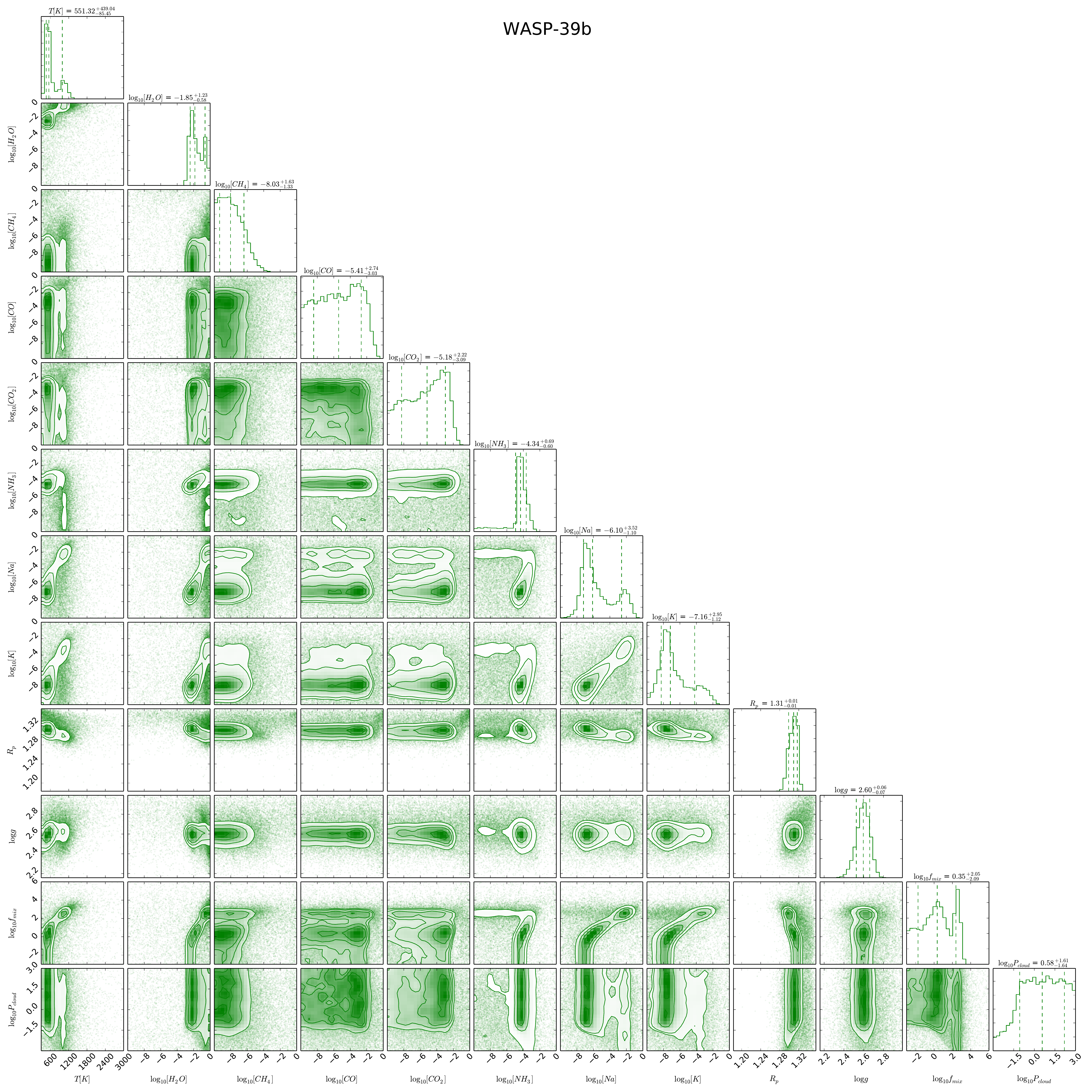}}}
\caption{Continued.}
\label{fig:cornerclassic}
\end{figure*}
\begin{figure*}[!tp]
\ContinuedFloat
\centerline{\resizebox{\hsize}{!}{\includegraphics{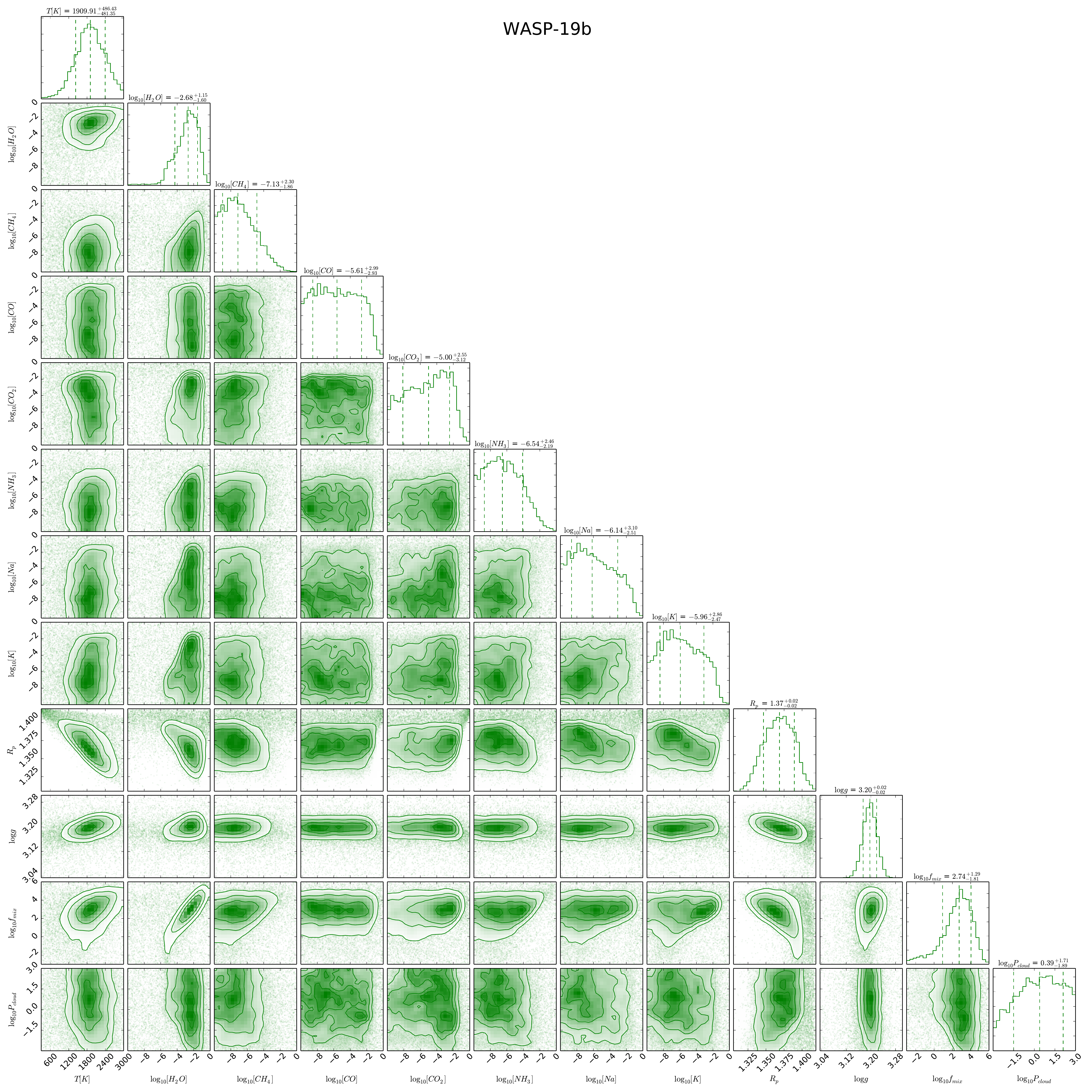}}}
\caption{Continued.}
\label{fig:cornerclassic}
\end{figure*}
\begin{figure*}[!tp]
\ContinuedFloat
\centerline{\resizebox{\hsize}{!}{\includegraphics{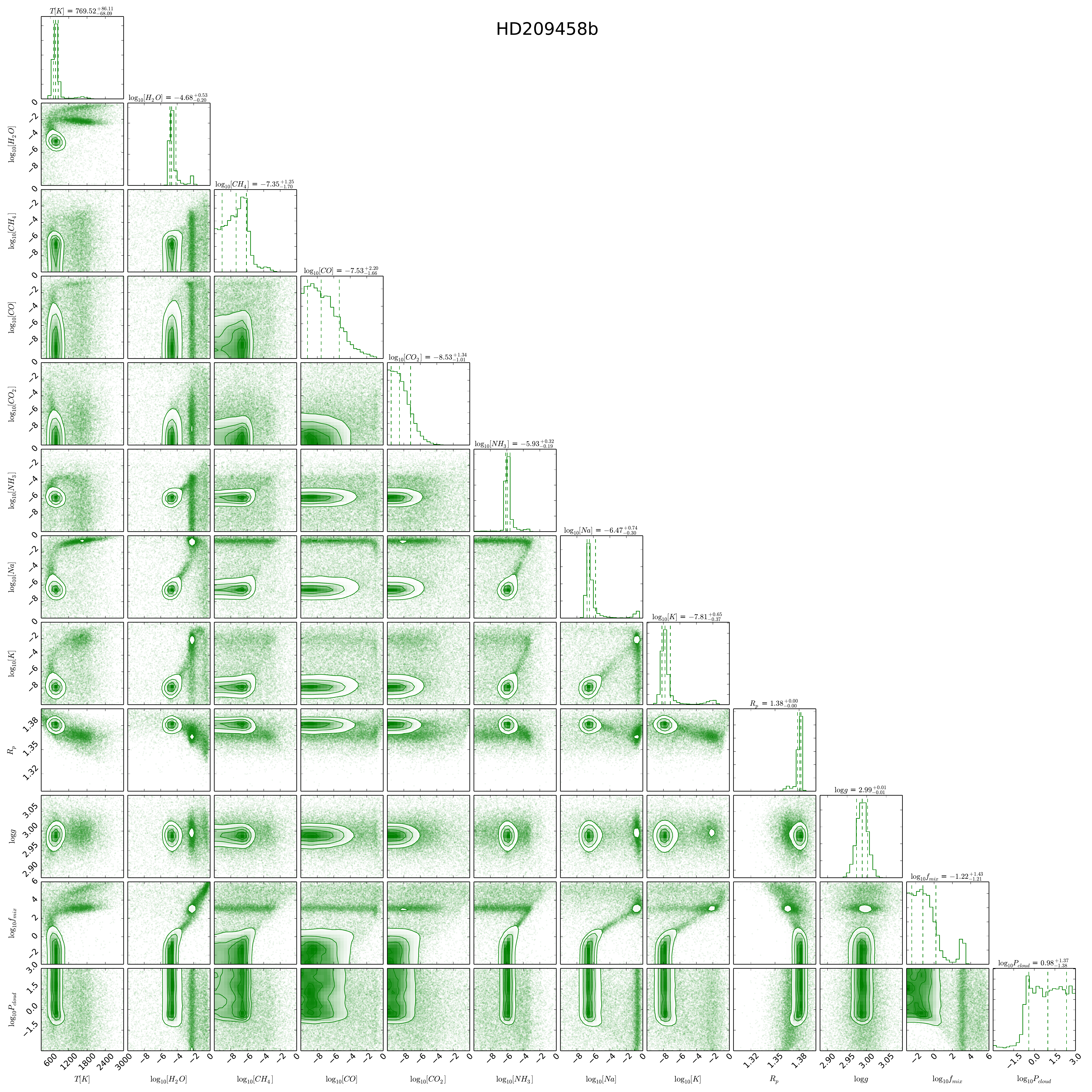}}}
\caption{Continued.}
\label{fig:cornerclassic}
\end{figure*}
\begin{figure*}[!tp]
\ContinuedFloat
\centerline{\resizebox{\hsize}{!}{\includegraphics{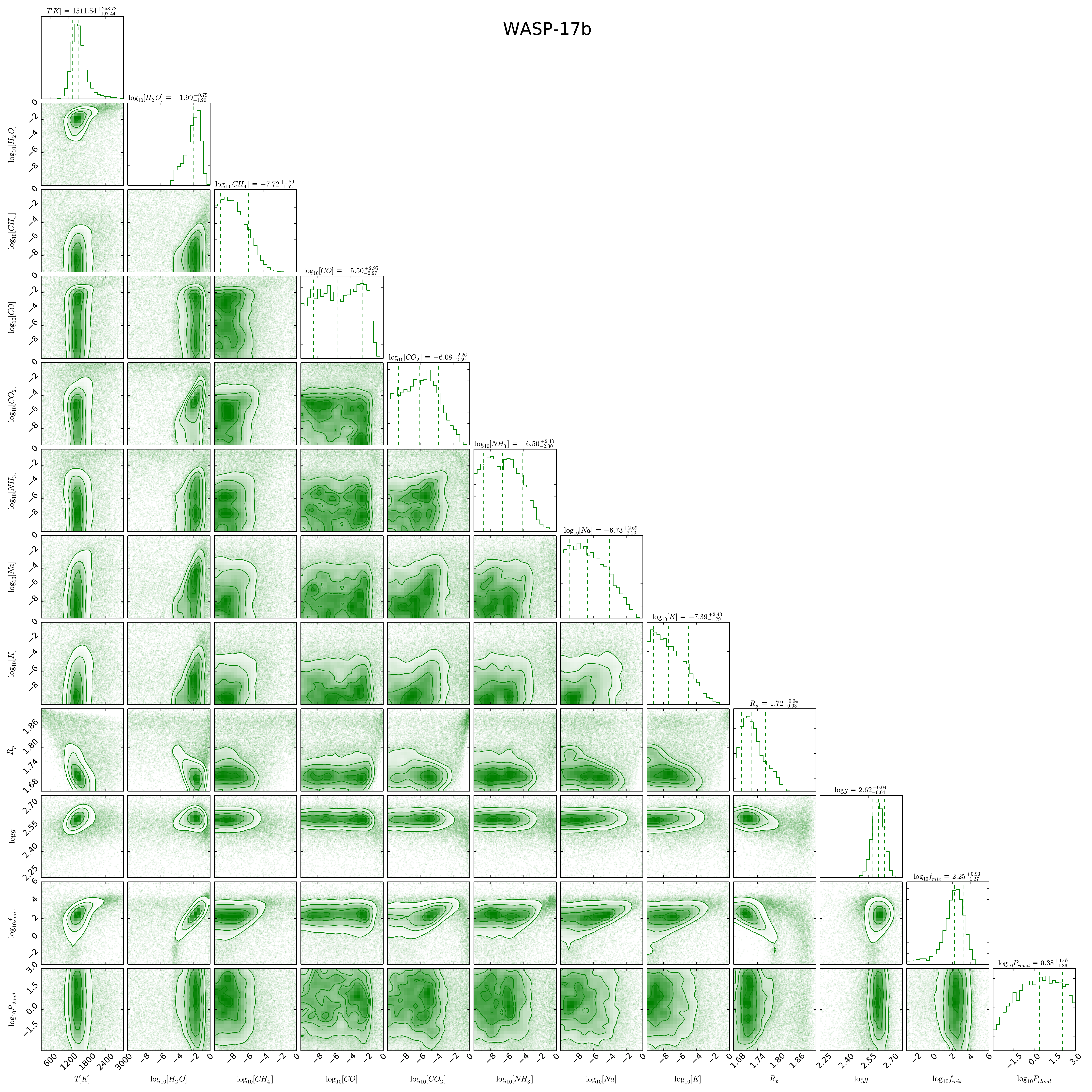}}}
\caption{Continued.}
\label{fig:cornerclassic}
\end{figure*}

\end{document}